# Competition between thermal pressurization and dilatant strengthening in laboratory ruptures

Caiyuan Fan[1], Gang Lin[1], Jérôme Aubry[2], Damien Deldicque[1], Carolina Giorgetti[1], Harsha S. Bhat[1], Alexandre Schubnel[1]

[1]Laboratoire de Géologie, École Normale Supérieure/CNRS UMR 8538, PSL University, Paris, France,

[2]Université Savoie Mont Blanc, Université Grenoble Alpes, CNRS, IRD, Université Gustave Eiffel, Le Bourget du Lac, France

Corresponding author: Caiyuan Fan (fan@geologie.ens.fr)

## CRediT

| | |
|---|---|
| **Conceptualization:** | C. Fan, C. Giorgetti, H. S. Bhat, A. Schubnel |
| **Methodology:** | C. Fan, G. Lin, J. Aubry, D. Deldicque, C. Giorgetti, H. S. Bhat, A. Schubnel |
| **Resources:** | H. S. Bhat, A. Schubnel |
| **Data curation** | C. Fan, G. Lin, D. Deldicque |
| **Investigation:** | C. Fan |
| **Writing – original draft:** | C. Fan |
| **Writing – review & editing:** | C. Fan, G. Lin, C. Giorgetti, H. S. Bhat, A. Schubnel |
| **Supervision:** | C. Giorgetti, H. S. Bhat, A. Schubnel |
| **Funding acquisition:** | C. Fan, H. S. Bhat, A. Schubnel |

## Key Points

- Direct experimental observations of pore pressure evolution during triaxial stick-slip rupture
- Co-seismic pore pressure shifts from rise to drop, interpreted as a transition from thermal pressurization to dilatant strengthening
- Fault structural evolution governs rupture dynamics through evolving shear zone hydromechanical properties

## Abstract

In fluid-saturated faults, thermal pressurization (TP) is predicted to induce rapid weakening, whereas dilatant strengthening (DS) can counteract this process through pore-space expansion. However, experimental constraints on their competition remain limited. Here, we investigate these processes using triaxial stick–slip experiments with direct on- and off-fault pore pressure (Pp) measurements on saturated, saw-cut, thermally cracked Westerly granite under effective confining pressures of 30–60 MPa. Results reveal a systematic rupture transition from co-seismic Pp rise (TP-type) to Pp drop (DS-type) with increasing shear strain, accompanied by a fast–slow–fast slip sequence, with slip velocities differing by up to three orders of magnitude. An undrained, adiabatic TP model reproduces the measured Pp rise evolution during early-stage fast events and indicates progressive shear zone widening that reduces TP efficiency. With continued slip, increasing cumulative plastic porosity reflects enhanced dilatancy during DS-dominated sequences. Microstructural observations reveal gouge development and a near-fault damage zone, providing independent evidence for this structural evolution. This evolution modifies fault stiffness and explains the observed transition in slip behavior. Despite contrasting Pp responses, fracture energy scales similarly with slip for both TP- and DS-type events, suggesting comparable rupture energetics. Overall, this study provides direct experimental evidence of TP–DS transitions, demonstrating that TP governs early-stage weakening but diminishes as dilatancy progressively strengthens with shear zone widening and near-fault damage. These results highlight that fault structure and its evolution plays a key role in controlling rupture dynamics in natural faults.

## Plain Language Summary

Earthquakes occur when shear stress exceeds fault strength on a preexisting fault, triggering rapid slip and dynamic rupture due to transient frictional weakening. In fluid-rich faults, pore pressure ($P_p$) plays a primary role in controlling fault strength by modulating the effective normal stress ($\sigma_n - P_p$). During rapid slip, frictional heating can cause fluid pressure to increase through thermal expansion, thereby weakening the fault. This process, known as thermal pressurization (TP), is widely considered a key mechanism for large earthquakes, although it remains poorly constrained by laboratory observations. TP may be counteracted by inelastic dilatancy occurring on the fault surface and in the surrounding near-fault region (e.g., the damage zone). Dilatancy reduces pore pressure, increases effective normal stress, and may inhibit rupture, a process referred to as dilatant strengthening (DS). This competition raises a fundamental question: under what conditions does TP dominate over DS? In this study, we present continuous measurements of on-fault pore pressure during spontaneous laboratory ruptures. Our results show that the evolution of fault structure plays a key role in controlling whether rupture exhibits TP- or DS-dominated behavior. These findings provide new experimental insights into fault hydromechanics and highlight the importance of near-fault inelastic deformation during earthquake rupture.

## 1. Introduction

During seismic slip, a significant fraction of mechanical work is dissipated within the fault zone, via frictional heating, and fracture energy, on and off-fault. Because the thermal expansion of fluids is orders of magnitude larger than that of rocks, localized frictional heating will induce pore-fluid expansion at high slip rates, resulting in a local increase of pore pressure (*Pp*). In impermeable fault zones, this will produce enhanced frictional weakening, a process generally referred to as thermal pressurization (TP) which has long been proposed as a major dynamic weakening mechanism during earthquakes (Sibson, 1973; Lachenbruch, 1980; Mase & Smith, 1987; Rice, 2006). On the opposite site of the spectrum, shear-induced dilatancy associated with on- and off-fault inelastic deformation will generate additional porosity, which may result in a drop of *Pp* prior, during or after rupture (Brace et al., 1966; Scholz, 1968; Brantut, 2020; Lin et al., 2026; Brantut et al., 2026). This reduction in *Pp* will result in fault strengthening, a mechanism often referred to as dilatant strengthening (DS) (Rice, 1975). TP and DS therefore represent competing source terms governing the evolution of *Pp* within a fluid saturated fault zone, which will control most aspects of rupture dynamics (Segall et al., 2010). Theoretical derivations of both mechanisms have emerged within a unified thermo–hydro–mechanical framework, taking into account the evolving balance between the respective rates of frictional heating and shear-induced dilatancy, thermal and hydraulic diffusion processes as well as shear zone structure (Rice, 2006; Suzuki & Yamashita, 2006).

Numerical studies have demonstrated that the effects of TP (often neglecting dilatancy) could strongly promote dynamic earthquake rupture (Andrews, 2002; Noda et al., 2009; Noda & Lapusta, 2010; Urata et al., 2008, 2012, 2014, 2015), while limiting temperature rise induced by frictional heating (Bizzarri & Cocco, 2006a, 2006b). TP efficiency has been shown to depend on rupture characteristics (whether crack or pulse-like) as well as fault roughness (Schmitt et al., 2011; Badt & Tal, 2025). TP is also expected to alter the nucleation phase, as a TP-induced *Pp* rise (Schmitt et al., 2011) may help promote unstable rupture when shear heating alone is not efficient enough (Segall & Rice, 2006). Structural observations and permeability measurements performed on the Median Tectonic Line fault zone in Japan (Wibberley & Shimamoto, 2005) were first shown to be consistent with the occurrence of TP during earthquakes, later confirmed by several fault zone analysis (Faulkner et al., 2010). Geological signatures consistent with TP, such as fluidization of comminuted material and thermally modified fluid inclusions, have also been documented (Ujiie et al., 2010). Post-seismic gouge fabric analyses further suggest a role for TP during fault seismic slip (Kuo et al., 2022). Finally, at larger scales, TP can reasonably explain the observed scaling of fracture energy with earthquake slip over more than eight orders of magnitude (Viesca & Garagash, 2015), while dynamic rupture simulations also demonstrate that TP can reproduce fracture-energy–slip scaling consistent with natural earthquakes (Perry et al., 2020).

On the other hand, numerical studies incorporating shear-induced dilatancy have demonstrated that DS stabilize fault motion, promoting stable or slow slip (Segall & Rice, 1995; Suzuki & Yamashita, 2009; Segall et al., 2010; Liu & Rubin, 2010). Quasi-dynamic models incorporating both TP and DS further suggest that rupture behavior can evolve, showing slow slips driven by DS at low stress evolving into fast ruptures dominated by TP at high stresses (Suzuki & Yamashita, 2007; Segall & Bradley, 2012). Field observations also provide the evidence of *Pp* drop by dilatancy during and after earthquakes. Groundwater monitoring following major earthquakes commonly shows rapid water-level declines followed by prolonged recovery in near-fault wells, such as 2018 Mw 6.3 Hualien earthquake (R. Hung et al., 2024). Modeling suggests that fault-zone damage and inelastic dilation increased permeability and specific storage, driving fluid depressurization (R. Hung et al., 2024, 2026). Similar groundwater-level drops were reported after the 2016 Mw 7.0 Kumamoto earthquake (Hosono et al., 2019), despite large slip and high slip rates inferred from rupture models (Hao et al., 2017).

Despite strong theoretical, numerical and observational support of the occurrence of TP during earthquakes, experimental evidence for TP remains limited. Most laboratory observations come from high-velocity rotary shear experiments on saturated gouge, where extreme frictional work produces large temperature increases along the slip surface (Yao et al., 2023; C.-C. Hung et al., 2025), while *Pp* increases are typically inferred indirectly (Acosta et al. 2018), measured outside the shear zone and often smaller than theoretical predictions. On the contrary, numerous laboratory experiments provide direct evidence for dilatancy. Rate-and-state velocity-stepping shear experiments document shear-induced dilation inferred from normal displacement, porosity change, or pore pressure reduction due to transient undrained conditions (Morrow & Byerlee, 1989; Marone et al., 1990; Marone, 1998; Samuelson et al., 2009a; Rathbun & Marone, 2010; Ashman & Faulkner, 2023; Scuderi et al., 2025). Undrained shear experiments on saturated gouge consistently show fault strengthening associated with dilation (Lockner & Byerlee, 1994; Affinito et al., 2025). Affinito et al. (2025) also further explored gouge shearing under drained and partially undrained velocity-stepping conditions, quantifying the detailed response of shear-induced dilatancy. Together, these studies demonstrate that dilatancy is a general and robust response during shear acceleration within gouge layers and fault zones. In addition, direct measurements of $Pp$ drop during triaxial experiments further confirm that dilatancy can actively counteract TP (Brantut, 2020; Proctor et al., 2020; Aben & Brantut, 2021; Lin et al., 2026; Brantut et al., 2026). Taken together, laboratory studies indicate that both thermal pressurization and dilatancy operate during fault slip, typically either emphasizing on thermal pressurization under imposed high-velocity frictional sliding or dilatant strengthening during low-velocity and/or quasi-static deformation.

This raises the question of whether $Pp$ rises or drops before, during, and after spontaneous frictional ruptures, such as laboratory stick-slip experiments, which provide a useful analogue of natural earthquake cycles (Brace & Byerlee, 1966). Tracking on-fault $Pp$ evolution across sequences of such events allows direct quantification of fluid-pressure responses under controlled laboratory conditions, where TP and DS may coexist, compete, and evolve within the same fault system. Here, we present laboratory observations of a systematic transition from TP–dominated to DS-dominated behavior during spontaneous triaxial stick–slip events, monitored via direct on- and off-fault $Pp$ measurements. This transition is accompanied by a fast–slow–fast spectrum of rupture styles, highlighting a coupled evolution between fluid-pressure processes and slip dynamics. Our results demonstrate that rupture behavior cannot be attributed to a single weakening mechanism, but instead is fundamentally controlled by fault structure and its evolution, including shear zone thickening and near-fault inelastic deformation.

# 2. Experimental Methodology

## 2.1 Sample preparation

Experiments were conducted on saw-cut cylindrical samples of Westerly granite, approximately 85 mm in length and 40 mm in diameter. Intact cylindrical samples were first heat-treated at 450 °C for two hours to induce thermal cracking and enhance permeability. The thermally-cracked samples were first cut to the target length, with both ends ground parallel to ensure axial alignment, and were then cut at an angle of 30° relative to the vertical axis to introduce a pre-existing planar fault. Both sides of the saw-cut fault surface were manually polished using #180 sandpaper, followed by diamond suspensions (ESCIL, France) with progressively decreasing particle sizes of 10, 3, and 1 μm. During polishing, a small amount of diamond suspension was applied to one surface, and the two sample halves were pressed together and reciprocally slid. This procedure produced smooth, parallel, and well-mated fault surfaces with reproducible roughness suitable for frictional experiments. The initial fault-surface roughness was characterized using a Keyence VHX-500 digital microscope. Root-mean-square height ($H_{RMS}$) variations ranged from 2 to 5 μm (Table 1). Fluid pressure transducers (Lin et al., 2024, following the design of Brantut & Aben, 2021) were installed through ports within the rubber sleeve, in direct contact with the rock-specimen in order to monitor internal $Pp$ during the experiments (Figure 1). Four transducers (#1–4) were positioned along the edge of the fault plane (±0.5 mm relative to the fault surface), to monitor on-fault $Pp$ evolution. Two additional transducers (# 5–6) were installed in the bulk of the foot and hanging walls, in order to monitor off-fault $Pp$. All 6 $Pp$ transducers were systematically calibrated during the first stage of each experiment (see Supporting information Text S1 for details). Eight acoustic emission (AE) sensors were installed around the fault plane and on the sample bulk to monitor microseismic activity during the experiments. In this study, AE data are used qualitatively to distinguish between seismic (fast) and aseismic (slow) slip behavior and are not analyzed further.

Table 1. Experimental sample information and initial physical parameter.

| Sample No. | $P_c$ | $P_f$ | $P_c - P_f$ | $H_{RMS}$ | $k$ | $\alpha_{hy}$ | $\beta$ |
|---|---|---|---|---|---|---|---|
| | MPa | MPa | MPa | μm | m$^2$ | m$^2$/s | MPa$^{-1}$ |
| WGF8 | 70 | 25 | 45 | ~3.4 | $1 \times 10^{-19}$ | $1.9 \times 10^{-6}$ | $5.3 \times 10^{-5}$ |
| WGF9 | 75 | 45 | 30 | ~3.4 | $9 \times 10^{-21}$ | $1.5 \times 10^{-7}$ | $6 \times 10^{-5}$ |
| WGF10 | 85 | 25 | 60 | ~2.6 | $4 \times 10^{-21}$ | $1.0 \times 10^{-7}$ | $4 \times 10^{-5}$ |
| WGF7 | 90 | 45 | 45 | ~5.3 | $8 \times 10^{-21}$ | $6.0 \times 10^{-7}$ | $1.4 \times 10^{-5}$ |

Remark: $P_c$: confining pressure. $P_f$: boundary fluid pressure. $P_c - P_f$: effective (boundary) confining pressure. $H_{RMS}$: initial roughness. $k$: initial permeability. $\alpha_{hy}$: initial hydraulic diffusivity, as measured during the calibration process of $Pp$ transducers (see Supporting information Text S1 for details). $\beta$: Storage capacity, calculated as $\beta = \alpha_{hy}/k\eta$ with the fluid viscosity $\eta$ equal to $10^{-3}$ MPa/s.

**2.2 Experimental set-up**

Samples were deformed within a triaxial high-pressure apparatus (Figure 1a) installed at the Laboratoire de Géologie of École Normale Supérieure (Schubnel et al., 2005; Fortin et al., 2007), identical to the apparatus described by Lin et al. (2026). Confining pressure ($P_c$) and differential stress ($\sigma_1 - P_c$) were applied using two independent servo-controlled syringe pumps, with maximum capacities of 300 MPa and 717 MPa (for a 40 mm diameter sample) respectively. Fluid pressure was independently controlled at the top and bottom of the sample using two micro-volumetric syringe-pumps with a maximum pressure of 100 MPa. Deionized water was used as pore fluid medium, and silicon oil as confinement. Axial loading was imposed at a constant displacement rate of 0.3 μm s$^{-1}$. Displacement was measured by an external linear variable differential transformer (LVDT) with a resolution of ±0.1 μm. In addition, a laser vibrometer mounted on the axial piston (Figure 1) measured dynamic piston velocity during stick-slip ruptures at an acquisition rate of 1.5 MHz, with a sensitivity of 50 mm s$^{-1}$ V$^{-1}$ and a maximum measurable axial velocity of 0.5 m s$^{-1}$. Stress, displacement, fluid pump pressure and volume, and pore pressure transducer data were recorded synchronously at a sampling rate of 100 Hz.

**2.3 Experimental procedure**

Each experiment followed the same three stages: 1) sample saturation and permeability measurement; 2) pore pressure transducer calibration and 3) and axial loading. Dry samples were initially loaded to a low confining pressure of 10 MPa and saturated by injecting fluid from the bottom at 5 MPa while allowing fluid to flow out from the top open to the atmosphere. Once saturation was achieved, the top of the specimen was connected to the fluid pump and maintained at the same pressure as the bottom. Confining pressure ($P_c$) and fluid pressure ($P_f$) were then increased to target experimental values (see $P_c$ and $P_f$ values in Table 1). At the target confining pressure, permeability was measured using the constant-flow method with a fixed fluid pressure gradient of 1 MPa (e.g., Fortin et al., 2011). Permeability was calculated from fluid pump flow rates using Darcy's law and measured twice with reversed flow directions to obtain a mean value. Initial permeabilities are reported in Table 1. One experiment conducted at an effective confining pressure of 45 MPa used a sample with an initial permeability approximately one order of magnitude higher ($\sim 10^{-19}$ m²) than the other samples ($< 10^{-20}$ m²), allowing assessment of the influence of initial permeability on stick-slip behavior. *Pp* transducers were then calibrated under constant confining pressure by applying stepwise changes in boundary fluid pressure and allowing sufficient time for internal pressure equilibration. Calibration procedures, uncertainty estimates (±1 MPa), and derivation of hydraulic diffusivity and storage capacity from the transient diffusion response are detailed in Supporting Information Text S1, with resulting values summarized in Table 1. Following calibration, axial loading was applied at a constant displacement rate of 0.3 µm $s^{-1}$. A total of four samples were tested under varying effective confining pressures ($P_c - P_f = 30,\ 45,\ 60$ MPa) with boundary fluid pressures (25 and 45 MPa). Complete experimental conditions are listed in Table 1.

**2.4 Data pre-processing and conventions**

In the following, we use the convention of compressive stresses being positive and dilatant strains negative. Fault slip is corrected following the same procedure as in Marty et al. (2023), i.e. using the total displacement projected onto the fault plane and assuming that the fault remains fully locked during elastic loading and:

$$D_\mathrm{f} = \frac{D_{LVDT} - D_E}{\cos\theta}$$

where $D_f$ is the slip along fault, $\theta$ is the angle between the fault-plane normal and the axial stress direction (i.e., 60°), and $D_E$ represents the elastic response of the system. Slip velocity along the fault is derived from the time derivative of $D_f$, after appropriate filtering and smoothing (Supporting Information Text S2). Dynamic piston velocity during stick-slip ruptures is independently measured using the laser vibrometer mounted on the axial piston.

For sake of simplicity, we only report the mean on-fault pore pressure $P_{\text{on-fault}}$, calculated as the average of transducers #1–4, and the mean off-fault *Pp* from transducers #5–6. The heterogeneity of *Pp* field will be discussed in Section 5.4. Shear stress ($\tau$), normal stress ($\sigma_\mathrm{n}$) and effective normal stress ($\sigma_\mathrm{n,eff}$) acting on the fault plane are calculated from the applied differential stress and confining pressure, accounting for the measured on-fault pore pressure $P_{\text{on-fault}}$:

$$\tau = \left(\frac{\sigma_1 - P_c}{2}\right)\sin 2\theta$$

$$\sigma_\mathrm{n} = \left(\frac{\sigma_1 + P_c}{2}\right) + \left(\frac{\sigma_1 - P_c}{2}\right)\cos 2\theta$$

$$\sigma_\mathrm{n,eff} = \sigma_n - P_\mathrm{on-fault}$$

The apparent effective friction coefficient of the fault is defined as $\mu = \tau/\sigma_\mathrm{n,eff}$. Finally, volumetric strain is estimated via fluid pump volume change (normalized by the total specimen volume) during the experiment.

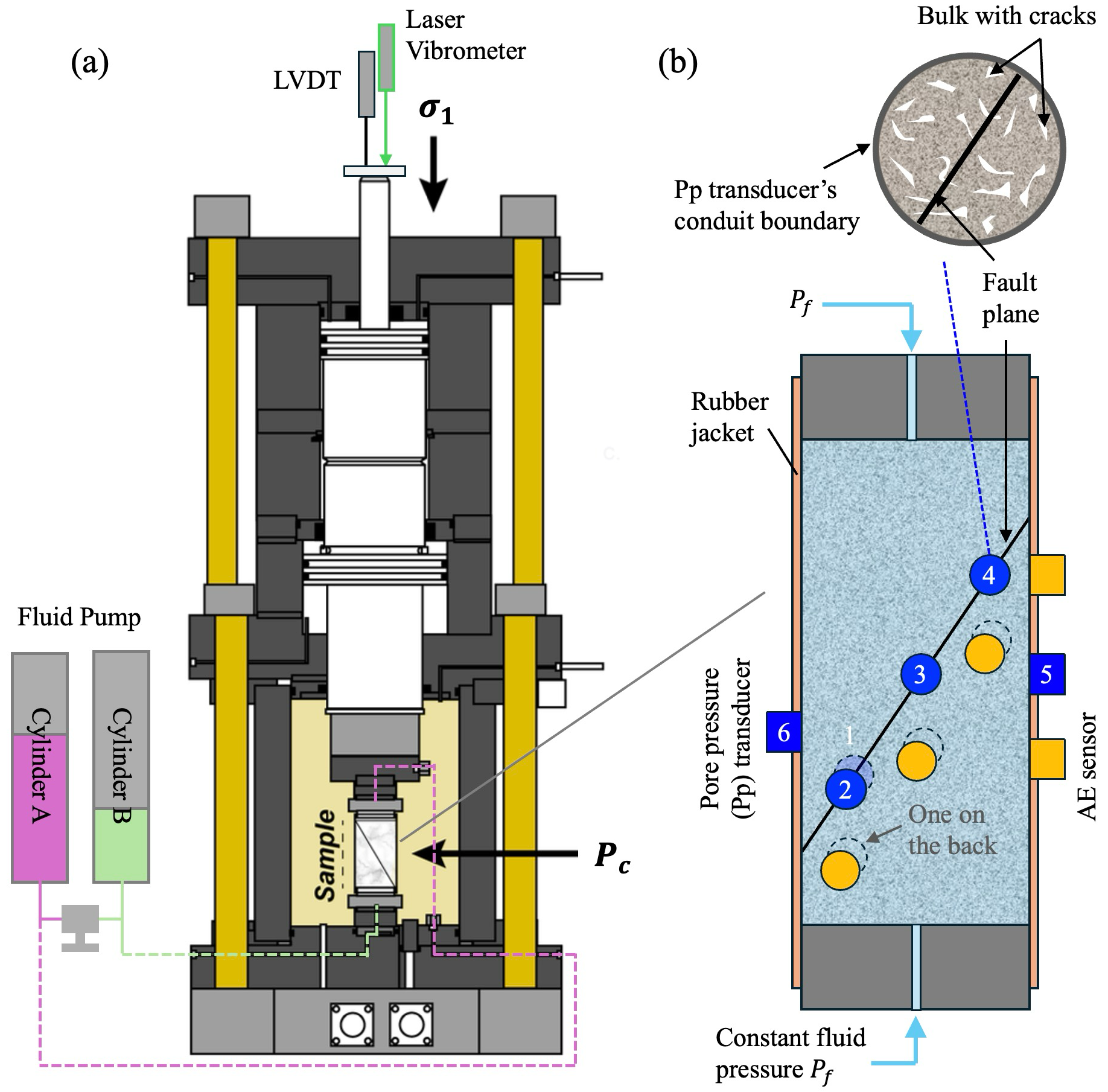

Figure 1. (a) Schematic of the triaxial loading cell with fluid pressure pump. Modified from Fortin et al. (2007), Schubnel et al. (2005) and Aubry (2019). (b) Saw-cut diagram of sample assemblage with *Pp* transducers and AE sensor.

# 3. Experimental Results

### 3.1 Stress, slip and pore pressure evolutions

The evolution of shear stress, cumulative fault slip, mean on- and off-fault pore pressure (*Pp*), and volumetric strain is shown in Figure 2 using experiment WGF9 ($P_c$ = 75 MPa, $P_f$ = 45 MPa) as a representative example. Similar results from the three additional experiments can be found within the Supplementary Figure S13-16.

During the initial loading stage, shear stress increased linearly with time, accompanied by slow and stable fault slip. Stick-slip events then occurred periodically, marked by abrupt shear-stress drops and discrete fault-slip increments. At the early stage of the stick-slip sequence, each stress drop was associated with a rapid co-seismic increase in mean on-fault *Pp*. During this period (stage A in Figure 2), mean on-fault *Pp* was consistently higher than off-fault *Pp*, which in turn exceeded the boundary fluid pressure ($P_f$) and established an internal fault overpressure state (Figure 2, inset). Following each event, *Pp* decayed post-seismically toward its pre-event level. The difference between on- and off-fault *Pp* during the first 3-4 events in Stage A reflected repeated co-seismic on-fault pressure increases combined with incomplete post-seismic recovery.

As the experiment progresses, the *Pp* response during stick-slip events evolved systematically, exhibiting a gradual transition from *Pp* rise to drop, with the amplitude of both the co-seismic stress and *Pp* drops increasing through the sequence. This transition from co-seismic *Pp* rise to drop has been consistently observed across all experiments and occurs at approximately 4500 s during WGF9. At later times (stage B during approximately 14000–18000 s), the *Pp* configuration reverses: mean on-fault *Pp* became lower than both off-fault and boundary pressures, indicating a fault under-pressure state and enhanced dilatancy-induced *Pp* drop localized near the fault-core. This transition from internal overpressure to under-pressure was also observed in experiments WG7 and WG10. On-fault *Pp* even approached near-zero values in experiment WGF10 (Figure S16). In contrast, WGF8 exhibited no pronounced background *Pp* evolution (Figure S15), higher diffusivity maintaining *Pp* on and off-fault close to the imposed boundary value throughout the experiment.

The total volumetric strain of the sample exhibited bulk compaction at the beginning (stage A), followed by the development of net dilatancy at later stages. This transition from background compaction to dilatancy, which was observed across all experiments, is broadly consistent with the evolution of internal *Pp* states (Figure 2) and with the transition from co-seismic *Pp* rise to drop. In samples with slightly higher permeabilities (WG7 and WG8), the correspondence between volumetric strain and *Pp* transitions were closely aligned. In WGF10 (see Figure S16), the volumetric-strain transition was delayed relative to the *Pp* response, consistent with the low hydraulic diffusivity (Table 1), and thus respectively smaller and higher characteristic diffusion length and time.

In summary, stress drop (thereby apparent friction) and associated slips generally evolved with event number, highlighting a systematic fault evolution in all 4 experiments (see Figure 2, Supplementary Figure S13-17). Based on the sign of the co-seismic on-fault *Pp* change, stick-slip events were classified into events with *Pp* rise (hereafter referred to as TP events) and events with *Pp* drop (hereafter referred to as DS events). DS events occurred at higher shear stress levels and were associated with larger stress drops and larger slips than TP events. The transition from TP-type to DS-type was generally coupled with a fast–slow–fast spectrum of slip velocities (see next sections), as inferred from LVDT-derived fault displacement, in combination with audibility and wave-based laser velocity records (Figure S18). The characteristics of TP events, DS events and of the transition phase are presented in the following sections.

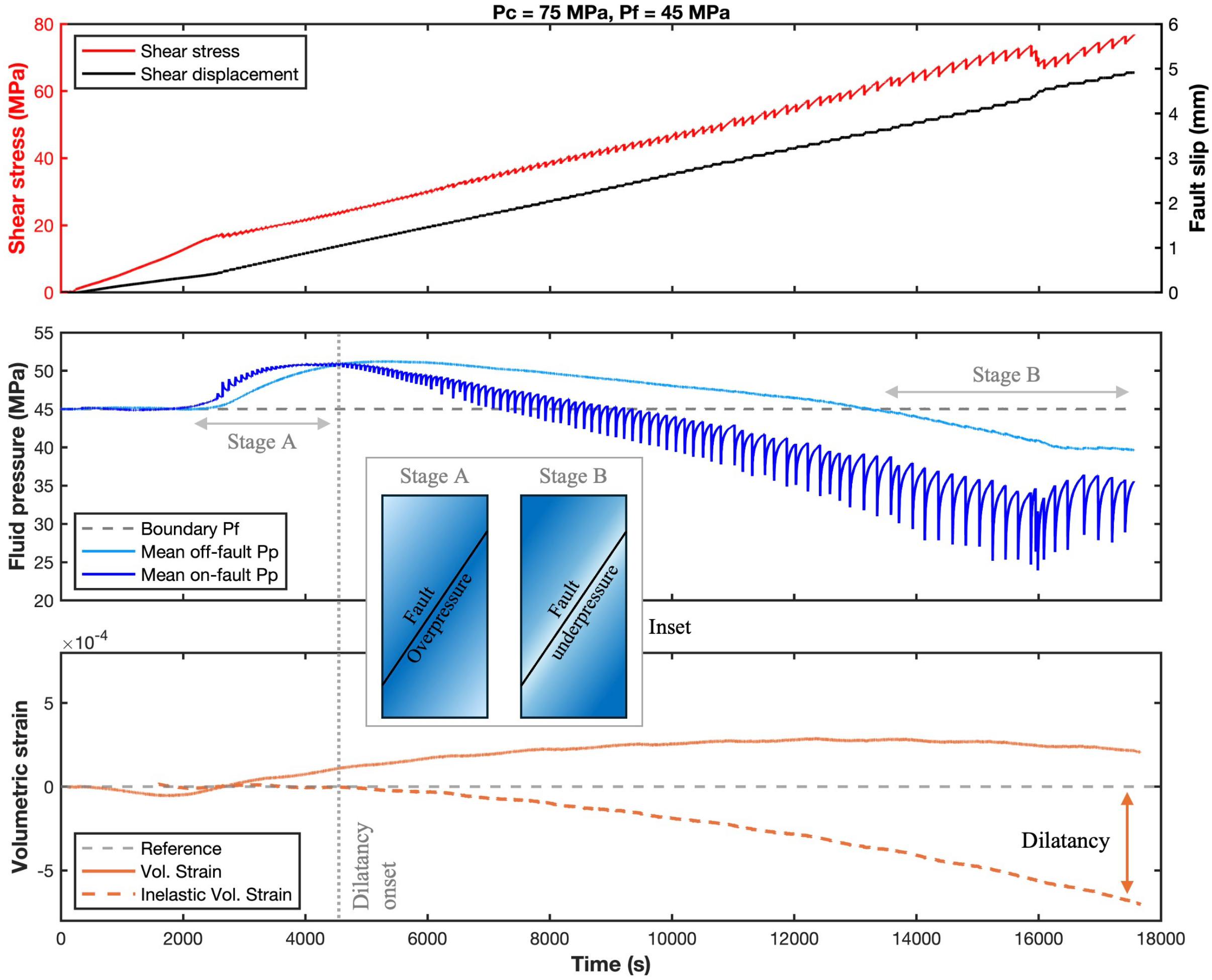

Figure 2. Evolution of shear stress, cumulative fault slip, mean on- and off-fault pore pressure (*Pp)*, and volumetric strain during experiment WGF9 conducted at $P_c - P_f = 30$ MPa. Mean on-fault *Pp* is averaged from transducers #1–4 and mean off-fault *Pp* from transducers #5–6. Volumetric strain is calculated from fluid pump volume change normalized by the sample volume, and the inelastic volumetric strain (dilatancy) is obtained by removing the linear elastic trend. The dilatancy onset is marked and roughly corresponds to the transition from co-seismic *Pp* rise to drop. The inset panel schematically illustrates the internal pore pressure (*Pp*) distribution within sample WGF9 during the experiment, where darker blue indicates higher *Pp* than lighter blue. During the early stage A (2000–4500 s), an internal overpressure state is observed, with on-fault *Pp* exceeding off-fault *Pp* and the boundary pressure $P_f$. In contrast, during the late stage B (14000–18000 s), an internal under-pressure state develops. The boundary fluid pressure $P_f$ remained constant at 45 MPa throughout the experiment.

### 3.2 TP-events with on-fault pore pressure rise

Figure 3a displays a zoom-in on the first four stick-slip events (events #1–4) in experiment WGF10, characterized by pronounced co-seismic rises in mean on-fault *Pp* and therefore classified as TP events. Among them, the first event exhibited the largest shear-stress drop (~2.2 MPa), the largest co-seismic on-fault *Pp* increase (~7 MPa), and the largest slip (~0.05 mm). Subsequent events displayed progressively smaller stress drops and *Pp* rises, indicating a systematic decrease of rupture intensity. Following each event, on-fault *Pp* decayed toward its pre-event level during the post-seismic stage, although recovery remains incomplete before the onset of the next rupture. Event #4 exhibited substantially reduced stress drop and fault slip, characterized by the absence of audible signals and detectable acoustic emissions, indicating a transition toward aseismic slip. Mean off-fault *Pp* is also shown for comparison. While off-fault *Pp* exhibits only a weak background increase, the difference between on- and off-fault *Pp* increases through successive TP events, reflecting increasingly localized *Pp* rise near the fault. Off-fault *Pp* also shows a small, rapid increase during rupture (Figure S19).

A detailed view of event #1 is shown in Figure 3b. The shear-stress drop consisted of a short dynamic drop followed by a brief overshoot (due to machine stiffness and fast-reloading, see Lockner et al., 2017) and a second, more gradual stress decrease. Fault slip increased rapidly during the dynamic phase and accumulated slowly afterwards. Slip velocity exhibits a transient pulse during the dynamic stage, reaching a peak value of ~1.2 mm $s^{-1}$, the highest among TP events. Note here that the slip velocity estimate using the LVDT is a lower bound only, the laser vibrometer generally giving velocities two orders of magnitudes higher (See Figure 8a). The on-fault *Pp* signal displayed a similar temporal structure, with a sharp pulse of pressurization characterized by a peak rate of ~180 MPa $s^{-1}$. The dynamic response of event #1 is further resolved at higher temporal resolution in Figure 3d. At the sampling interval of 0.01s, the on-fault *Pp* rise was synchronous with shear-stress drop, indicating that *Pp* pressurization was directly associated with the rupture process. Notably, *Pp* continued to increase by an additional ~1.5 MPa during the subsequent stage of gradual stress drop and stable slip following the dynamic phase. A precursory on-fault *Pp* increase was also observed prior to the dynamic rupture (Figure 3c). More than 50s before the main stress drop, *Pp* progressively deviated upward from the background trend, reaching an increase of ~1.1 MPa, while far-field shear stress remained nearly constant or decreased only slightly (~0.01 MPa). This precursory pressurization (Figure 3c) highlights the possible role played by TP during the nucleation phase.

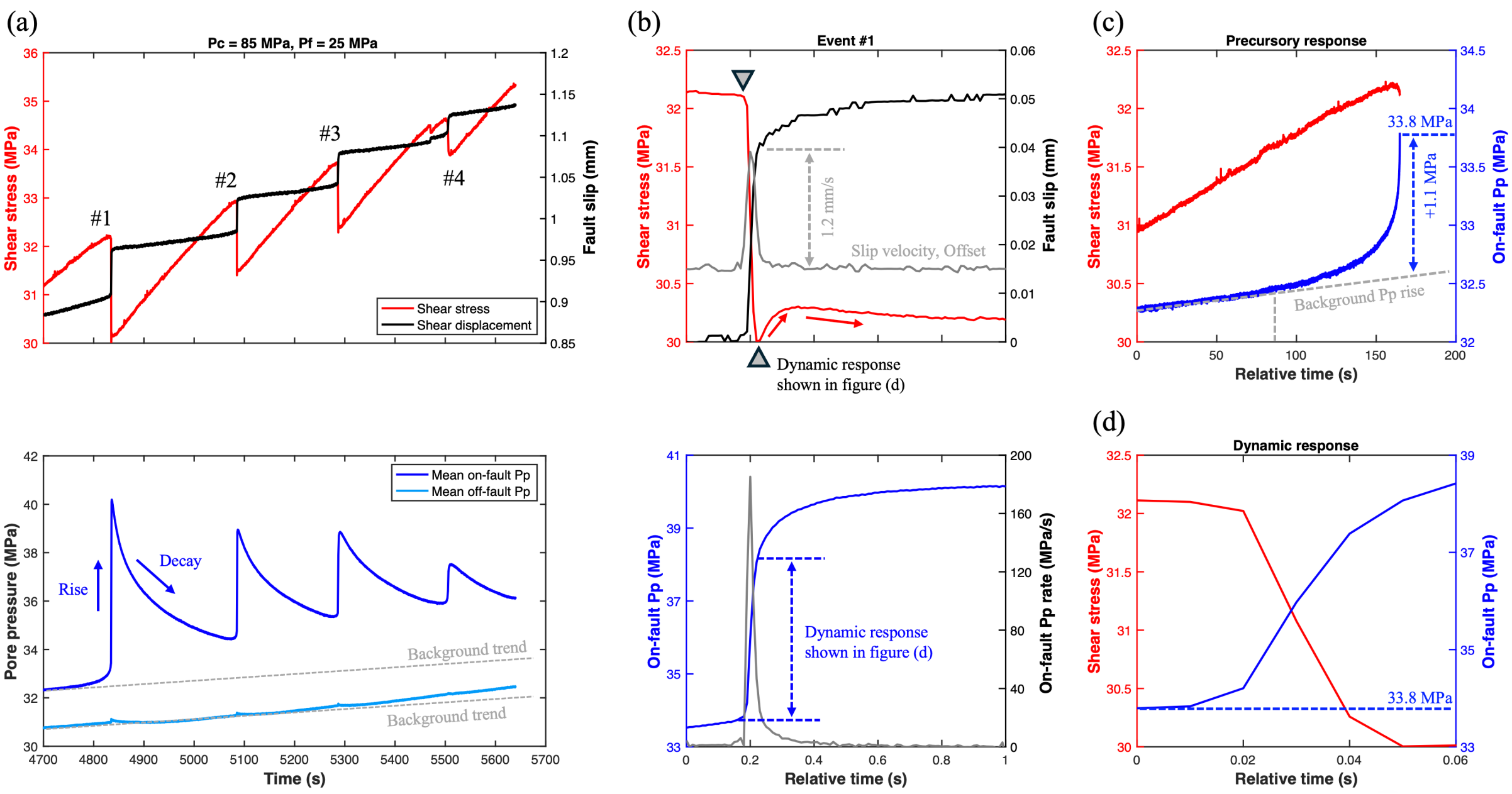

Figure 3. TP events with co-seismic on-fault pore pressure (*Pp*) rise in experiment WGF10 conducted at $P_c - P_f = 60$ MPa. (a) Zoom-in on four TP events (events #1–4) showing shear stress, cumulative fault slip, and mean on- and off-fault *Pp*. (b) Co-seismic stage of event #1 over a 1s window, with time and fault slip offset for clarity. Slip velocity and on-fault *Pp* rate are first derivatives of fault slip and on-fault *Pp*, respectively. Gray triangles indicate the dynamic stress-drop interval. (c) Precursory on-fault *Pp* response prior to the dynamic rupture of event #1, illustrating a gradual deviation from the background trend more than 50s before the main stress drop. (d) Dynamic response of event #1 at a temporal resolution of 0.01s, showing the synchronization between shear-stress drop and on-fault *Pp* rise. The time interval shown in panel (d) starts where the precursory response shown in panel (c) ends.

### 3.3 DS events with on-fault pore pressure drop

Figure 4a displays three DS events (#108–110 of experiment WGF9) characterized by co-seismic on-fault *Pp* drop. For all three events, on-fault *Pp* dropped abruptly during rupture, followed by a prolonged post-seismic recovery that remained incomplete before the onset of the next event, resulting in a decreasing background *Pp* level. Mean off-fault *Pp* is shown in Figure 4a for comparison. At this stage, off-fault *Pp* is lower than the imposed boundary pressure (Pf = 45 MPa) but remains substantially higher than on-fault *Pp*, reflecting strong pressure gradients localized near the fault. Off-fault *Pp* also exhibits a small co-seismic increase during rupture (Figure S19). A detailed view of event #110 is shown in Figure 4b. Similarly to TP events, the shear-stress drop consists of a single dynamic drop and corresponding slip, followed by a brief overshoot (due to machine stiffness and fast-reloading, see Lockner et al., 2017). A slight post-seismic stable sliding phase was also observed after DS events. Slip velocity consisted of a single sharp pulse, reaching a peak of ~1.5 mm $s^{-1}$. Note again that the slip velocity estimate using the LVDT is a lower bound only, the laser vibrometer generally giving velocities two orders of magnitudes higher (See Figure 8a). The on-fault *Pp* drop was synchronous with the dynamic slip and released nearly all *Pp* instantaneously, with a peak drop rate of ~380 MPa $s^{-1}$. As observed for TP events, shear-stress and on-fault *Pp* drops are perfectly synchronized, at our temporal resolution of 0.01 s (Figure 4d). Finally, a precursory on-fault *Pp* signal is observed as a precursor to dynamic rupture (Figure 4c). Indeed, on fault *Pp* starts decreasing 10s before rupture, while far-field shear stress continues to increase slowly toward failure. This precursory behavior is consistently observed in DS events, highlighting the possible role played by DS during the nucleation stage.

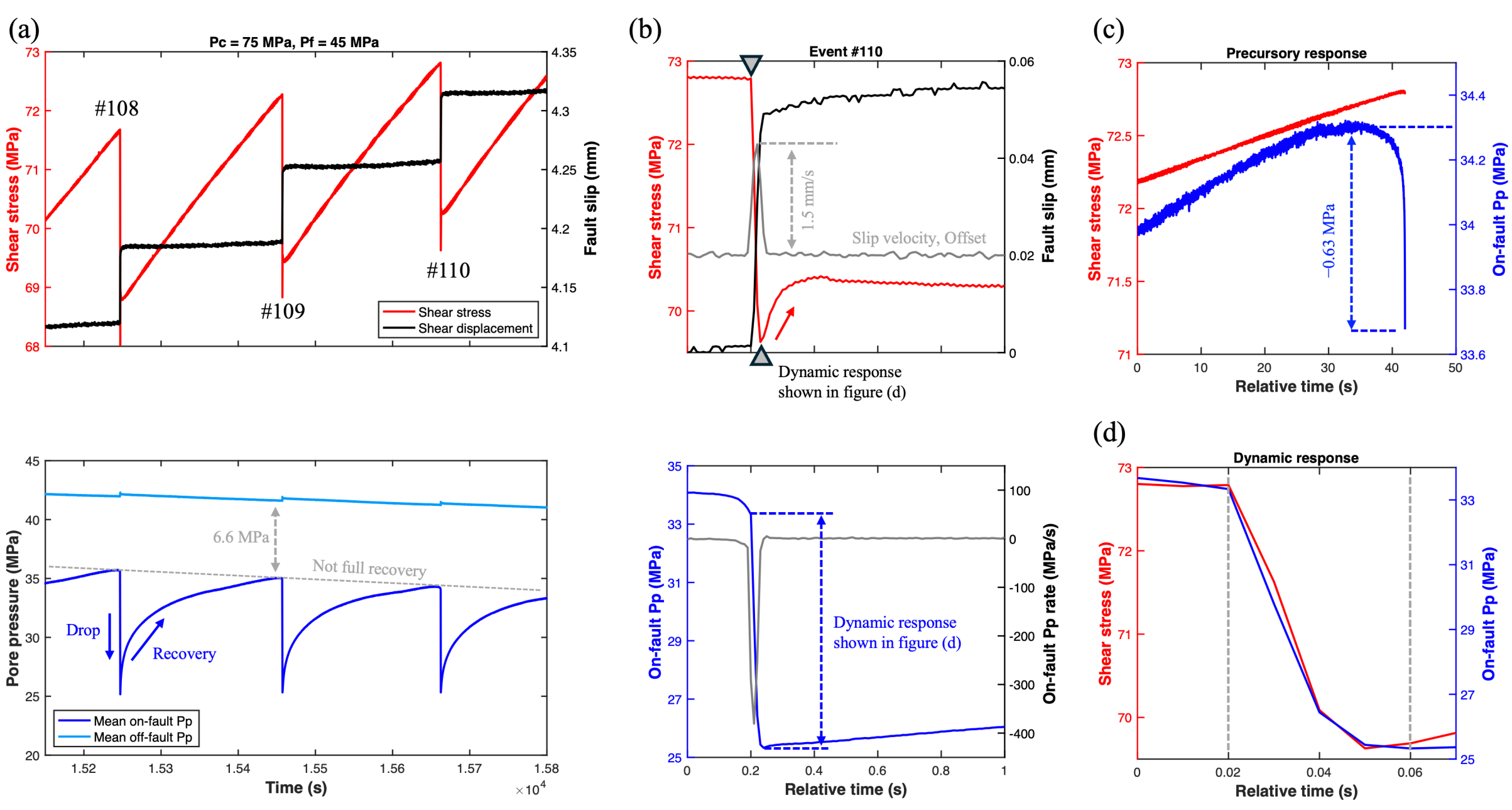

Figure 4. DS events with co-seismic on-fault pore pressure (*Pp*) drop in experiment WGF9. (a) Zoom-in on three DS events (events #108–110) showing shear stress, cumulative fault slip, and mean on- and off-fault *Pp*. (b) Co-seismic stage of event #110 over a 1s window, with time and fault slip offset for clarity. Slip velocity and on-fault *Pp* rate are first derivatives of fault slip and on-fault *Pp*, respectively. Gray triangles indicate the dynamic stress-drop interval. (c) Dynamic response of event #110 resolved at a temporal resolution of 0.01s, illustrating the synchronization between shear-stress and on-fault *Pp* drops. (d) Precursory on-fault *Pp* response prior to the dynamic rupture of event #110, showing a progressive *Pp* decrease more than 10s before the main stress drop. The time interval shown in panel (d) starts where the precursory response shown in panel (c) ends.

### 3.4 Transition from TP to DS

The transition phase between TP to DS events is characterized by complex *Pp* responses, generally accompanied by slow stick-slip events (~0.01 mm $s^{-1}$) for samples of lower permeabilities (WGF7, 9, 10) and only fast events for WGF8 of higher permeability. Events #7–13 of experiment WGF10 are displayed in Figure 5a. Events #7–10 exhibit small stress drops accompanied by co-seismic on-fault *Pp* rise and represent the late-stage TP-type behavior. Event #11 marks a transition, exhibiting both stress and co-seismic *Pp* drops, together with a large post-seismic *Pp* increase. Following this event, the system evolves into a sequence of DS-dominated events.

Representative slow events (#7, #9, and #13) are displayed in Figures 5b–d. Compared to fast events, slow events exhibit smaller stress drops (~0.7 MPa), longer durations (several seconds), limited co-seismic slip (~0.02 mm), and lower slip velocities (~0.02 mm $s^{-1}$), approximately one to three orders of magnitude lower than those of fast events (>0.5 mm $s^{-1}$, measured by the LVDT). On-fault *Pp* responses are characterized by lower rates and amplitudes. For example, event #7 shows a gradual *Pp* rise with a peak rate of ~0.06 MPa $s^{-1}$ and a total increase of ~0.9 MPa, with *Pp* continuing to rise after stress drop, indicating a time-lag between stress drop and the evolution of on-fault *Pp*. Event #9 exhibited a similar mechanical evolution to event #7 but displays an initial small *Pp* drop prior to the subsequent rise. Comparable *Pp* behavior is also observed in event #10. Subsequent slow events (e.g., event #13) exhibit *Pp* drops that terminate before the end of the stress drop and slip, indicating that *Pp* recovery initiates while rupture is still ongoing. In contrast, fast DS events described in Section 3.3 exhibit a near-synchronous termination between *Pp* and shear-stress drops.

Overall, mixed *Pp* responses during the transition phase are consistently observed, including small initial drops followed by rises, progressively larger co-seismic drops, and diminishing post-seismic over-recovery. These behaviors define a transitional spectrum connecting fast rupture events characterized by co-seismic *Pp* rise and those characterized by co-seismic *Pp* drop and are consistently observed across all experiments, including the high-permeability experiment WGF8 despite the absence of slow events.

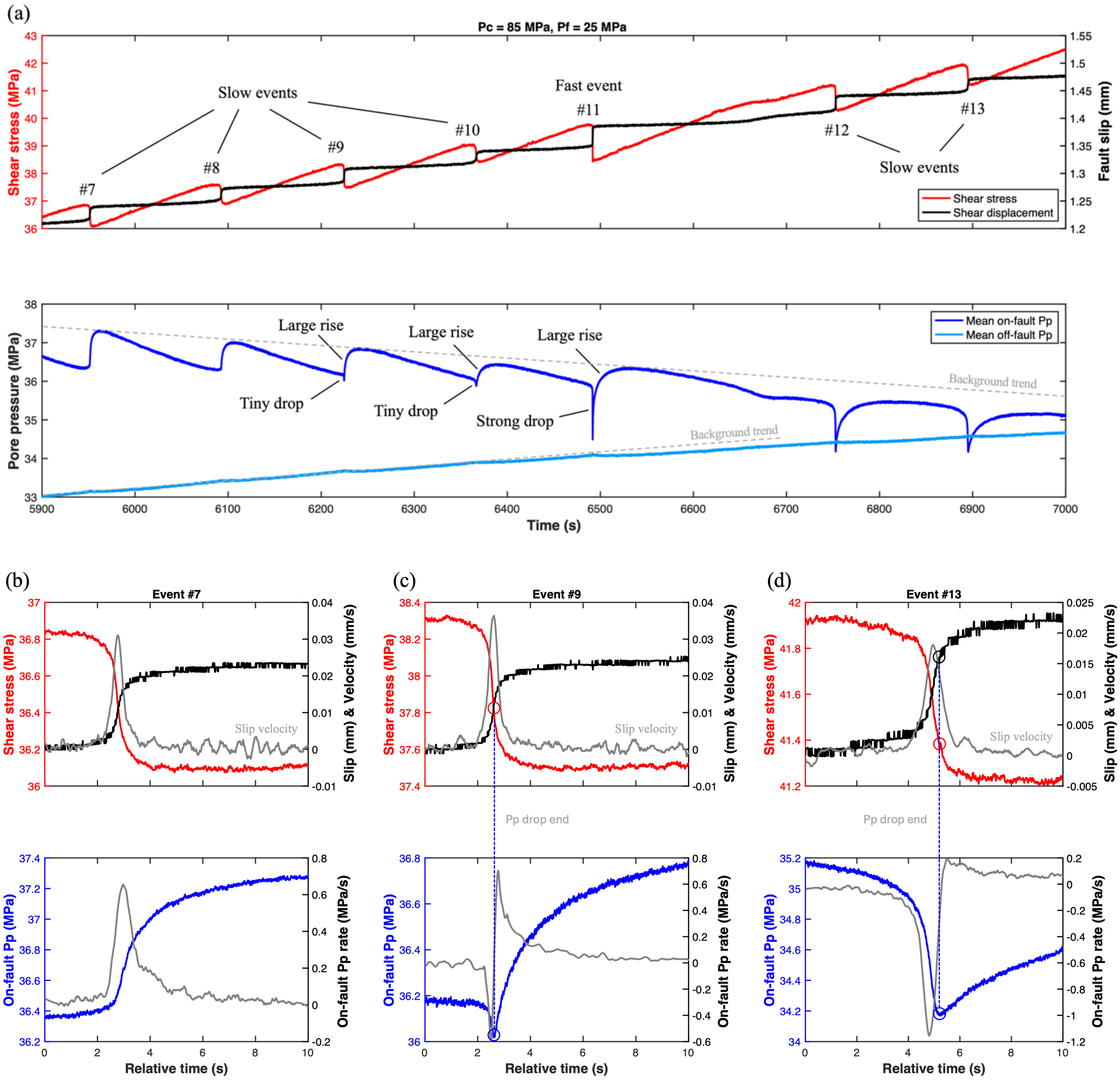

Figure 5. Transition from TP to DS events in experiment WGF10. (a) Evolution of shear stress, cumulative fault slip, and mean on- and off-fault pore pressure (*Pp*) during events #7–13. Except for event #11 (~0.58 mm $s^{-1}$), all events show much lower slip velocities (~0.02 mm $s^{-1}$) and are thus referred to as slow events (velocity not shown). (b–d) Co-seismic stages of representative slow events (#7, #9, and #13), respectively. Time and fault slip are offset for clarity. Slip velocity and on-fault *Pp* rate are first derivatives of fault slip and on-fault *Pp*, respectively. Colored circles with dashed blue lines highlight the relative temporal offset between *Pp* evolution, shear stress drop and fault slip during the transition phase.

### 3.5 Microstructural observation

After the experiments, the bottom block of sample WGF10 was observed using scanning electron microscopy (SEM) to investigate fault-zone microstructures, in both surface and cross-sectional views (Figure 6). The fault surface (Figure 6a-b) exhibits gouge patches forming topographic asperities relative to surrounding flatter regions. Zoomed-in observations show that these gouge patches are characterized by the presence of micron and submicron grains (Figure 6b). Filaments indicating frictional melting are observed on the fault surface (Figure 6c), similar to those reported by Acosta et al. (2018), suggesting strong frictional heating during rupture.

To investigate damage patterns, several cross-sectional images parallel to the shear direction and covering nearly the entire fault length (~75 mm over a maximum fault length of 80 mm) were analyzed. High-density clusters of smaller cracks are locally observed (Figure 7a). Long cracks parallel or at low angle to the fault surface are observed along most of the section (Figure 7b). Gouge patches can be here also observed in cross-sectional view (Figure 6d), indicating a gouge thickness of less than ~10 μm.

To quantify this damage, SEM images of consistent size and magnification were used to calculate crack density profiles (Figure 7b, see also Okubo et al. (2019) for a similar grid-based approach; methodological details are provided in Supporting Information Text S3). The resulting depth-averaged crack density (Figure 7c) captures the spatial distribution of cracks along the fault, showing several local peaks corresponding to zones of high-density crack clusters (Figure 7a). In addition, near the fault tip (right end), fractures are more extensively developed, resulting in very high crack density, approaching 1. On the other hand, the depth profiles show an overall decay in crack density with increasing distance normal to the fault. The background crack density is estimated at depths of 150–200 μm (~0.06), and the damage zone half-thickness is defined as the depth at which crack density decreases to this background level. Local depth profiles, averaged over ~1.5-mm-long fault segments, indicate that the damage zone half-thickness varies from ~10 μm to ~160 μm. Averaging over the full image length (~75 mm) yields a mean depth profile corresponding to a damage zone half-thickness of ~60 μm. A power-law fit describes the mean depth profile well, with an exponent of 0.80, indicating a gradual decay of damage with depth.

Overall, these observations of a narrow but spatially variable near-fault damage zone suggest the progressive generation of additional pore space, consistent with the progressive pore pressure decrease observed during the late-stage DS events (e.g., Stage B in Figure 2).

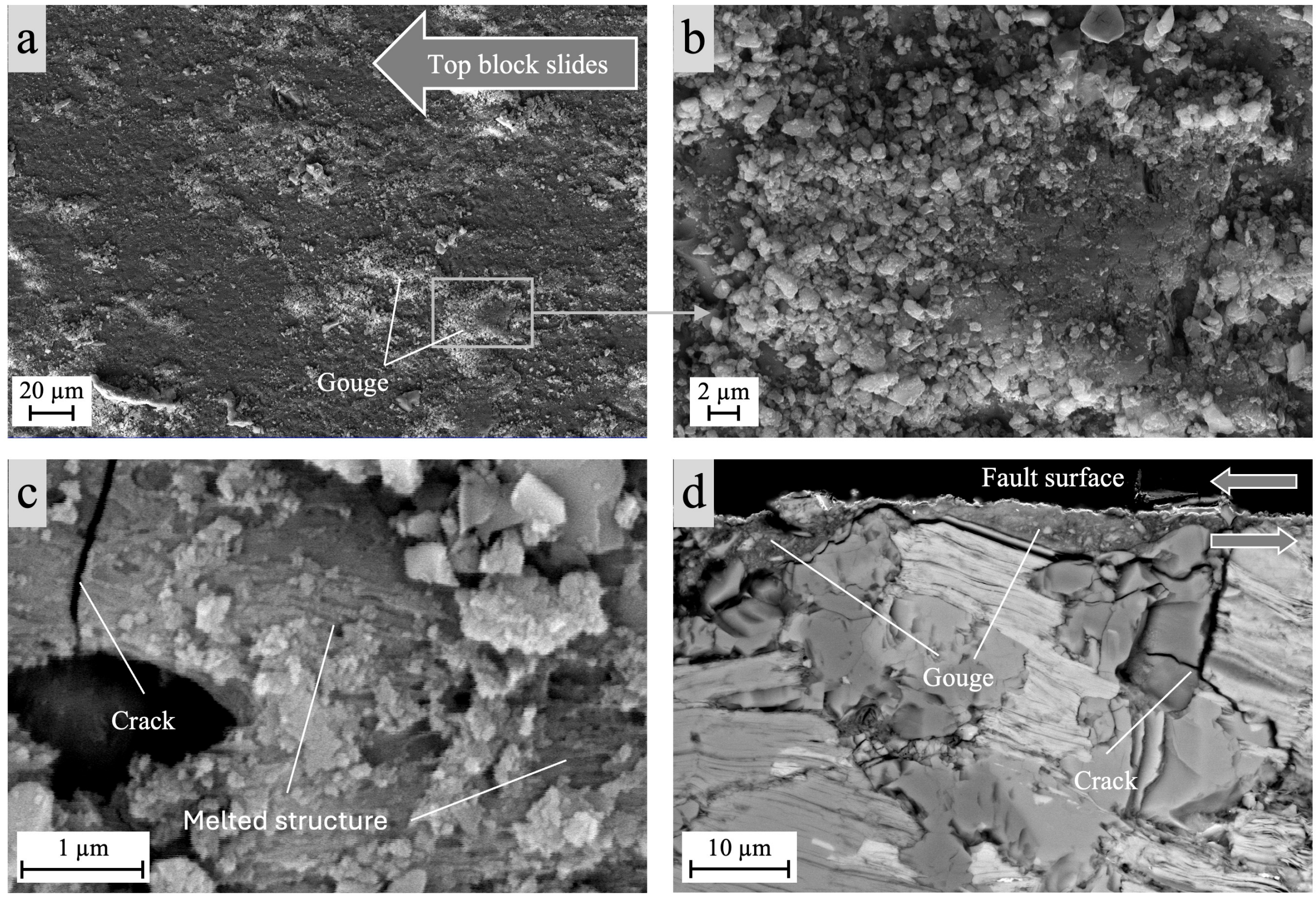

Figure 6. SEM observations of post-experiment sample WGF10 (bottom block). (a–b) Fault surface images, acquired in secondary electron mode, showing the development of gouge patches and associated grain-size reduction. (c) Localized melted structure indicative of frictional heating. (d) Cross-sectional image, acquired in backscattered electron mode, perpendicular to the fault plane and parallel to the slip direction, illustrating a near-fault zone with cracks and a thin gouge layer (<10 μm). Shear direction is indicated by arrows.

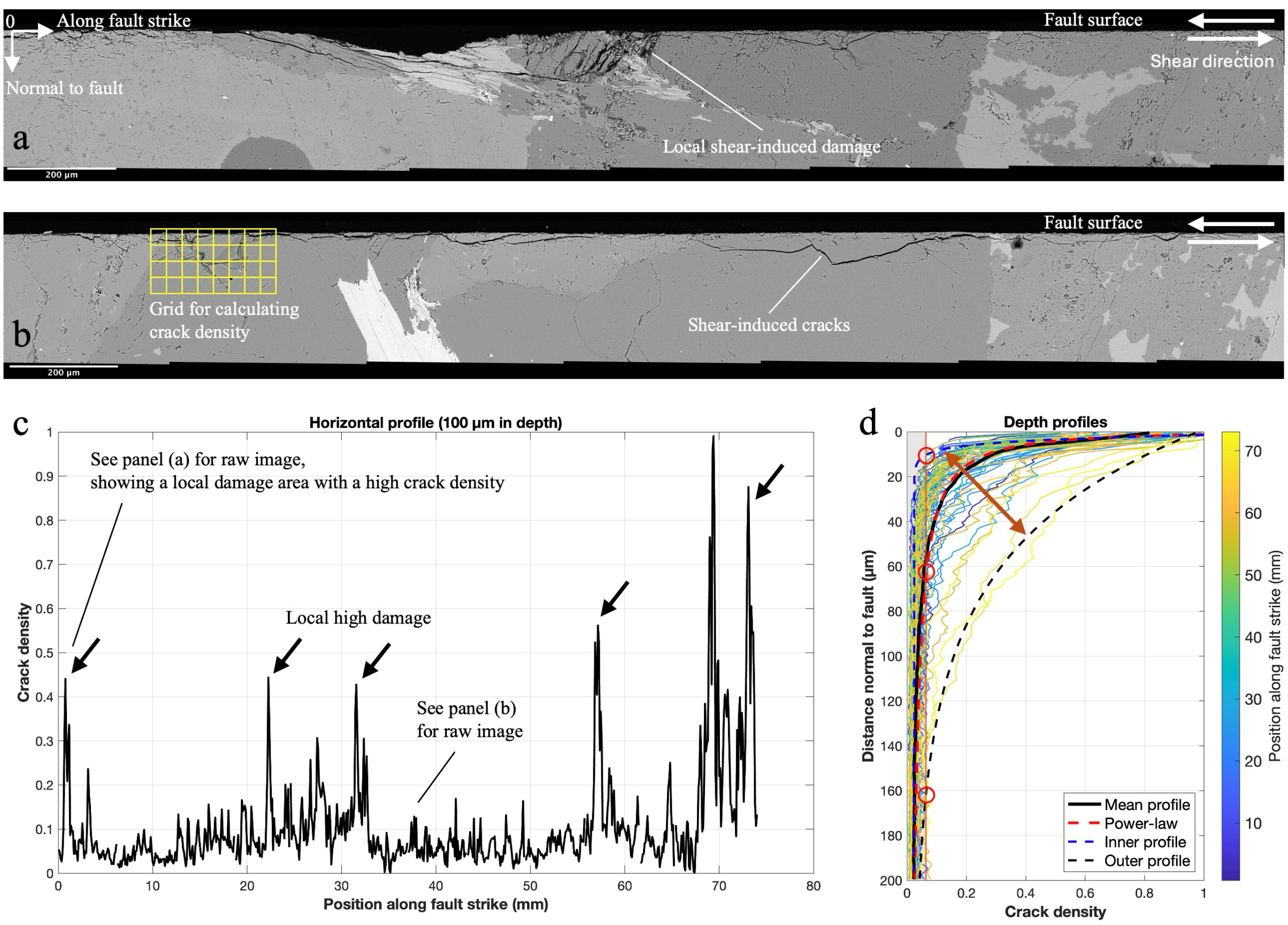

Figure 7. Cross-sectional SEM observations of post-experiment sample WGF10 (bottom block), perpendicular to the fault plane and parallel to the shear direction, illustrating a spatially variable near-fault damage zone. (a–b) Representative cross sections taken along the longest axis of the fault, showing both localized small crack clusters and long, continuous cracks. Shear direction is indicated by arrows. (c) Depth-averaged crack density along the fault (~75 mm), revealing strong along-fault heterogeneity and localized zones of high crack density, particularly near the fault tip (right end). (d) Depth profiles of crack density showing a mean damage zone half-thickness of ~60 μm. Spatially averaged profiles (color scale) indicate variability from ~10 μm to ~160 μm for damage zone half-thickness.

## 4. Analysis

### 4.1 Fast and slow stick-slip events and pore pressure response

The stress state and rupture characteristics of all stick-slip events for the all low-permeability samples tested at effective confining pressures of 30, 45, and 60 MPa are summarized on Figure 8. Figure 8a highlights a broad range of peak slip velocities among stick-slip events. Events with shear stress drops exceeding ~1 MPa were accompanied by higher velocities (>0.5 mm $s^{-1}$). For these, it is likely that the LVDT signal saturated (due to both aliasing and sampling frequency) and recordings using the laser vibrometer indicated peak slip velocities of > 0.1 m/s (see Figure S18). These results are broadly consistent with Passelègue et al. (2016a), Lockner et al. (2017), Hayward and Cox (2017), Aubry et al. (2018) and Acosta et al. (2018). Events with stress drops below ~1 MPa exhibited lower peak velocities (<0.5 mm $s^{-1}$), corresponding to aseismic (or too weakly seismic) events, undetectable with our laser vibrometer, which is consistent with Passelègue et al. (2019). For these, it is unlikely that the LVDT saturated since stress drop and slip occurred over several seconds, so that our results clearly emphasize a velocity contrast of several orders of magnitude between fast and slow events.

In all experiments, the shear stress increased over the course of the experiment (Figure 8b). First stick-slip events occurred at a peak friction of ~0.45, while the last ones at peak frictional stresses up to ~0.9 (for the lowest effective pressure experiment), which is broadly consistent with the Byerlee's range of values (Byerlee, 1978). While the fault strengthened, stress drops also evolved systematically following first a decrease and then an increase. Note here that shear stress drops and final slips are linearly correlated via (machine+sample) loading stiffness ~ 42 MPa/mm in terms of shear stress (see Figure S20). At $P_c - P_f = 30$ MPa, effective normal stress initially decreased with increasing shear stress due to background $Pp$ rise, during which shear stress drops progressively diminished to values < 1 MPa, and a transition from fast to slow stick-slips was observed. With continued loading, the reverse transition (from slow to fast) was observed as stress drops increased again. Similar observations were performed for samples deformed at $P_c - P_f = 45$ MPa and 60 MPa. Indeed, although no slow stick-slips were observed during the experiment performed on the higher permeability sample (WGF8, see Figure S21), the stress drops evolution was similar.

Co-seismic on-fault $Pp$ changes are displayed on Figures 8c. At early stages of each experiment, events exhibited positive co-seismic $Pp$ pulses, with amplitudes decreasing progressively through the sequence. At later stages, co-seismic $Pp$ pulses became negative, and the magnitude of the drop in $Pp$ increased with increasing shear stress drop. Only for the experiment at $P_c - P_f = 60$ MPa, for which $Pp$ drops were there largest, reaching up to ~15 MPa, co-seismic $Pp$ close to zero were observed. In summary, fast stick-slips could be accompanied both by co-seismic $Pp$ rise or drop, corresponding to fast TP and fast DS events, respectively, while slow stick-slips only occurred during the transition from TP to DS events. Note finally that for the experiment conducted on the higher initial permeability sample, a similar co-seismic stress–slip–$Pp$ relationships were observed, except slow events were absent and TP and DS ruptures transitioned directly between fast events. These results are presented in the Figure S21.

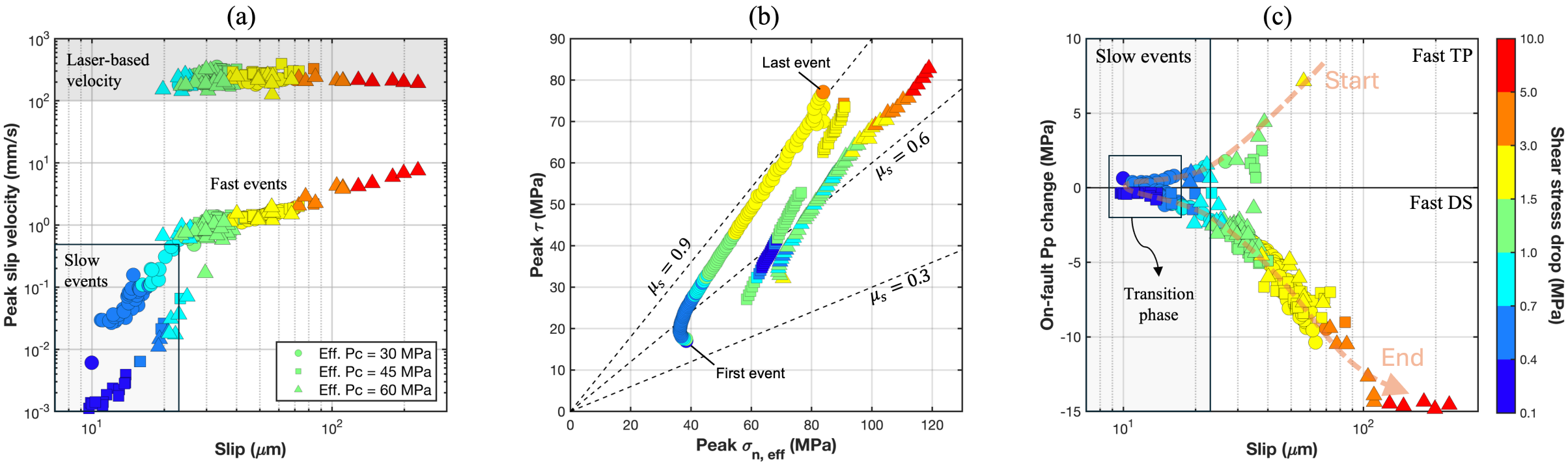

Figure 8. Stress state and co-seismic rupture characteristics of stick-slip events conducted at effective confining pressures of 30, 45, and 60 MPa on samples with similar initial low permeability. (a) Peak slip velocity versus total slip. Slip velocities exceeding 0.1 m s$^{-1}$ are derived from laser vibrometer measurements and correspond to fast events only. (b) Peak shear stress versus peak effective normal stress immediately prior to stress drop, illustrating the stress state at rupture onset. (c) Maximum co-seismic on-fault $Pp$ change during rupture. Symbol color indicates shear stress drop magnitude. The dashed orange arrow in panel (c) indicates the temporal

progression of events during experiments. Based on co-seismic $Pp$ change and slip velocity, events in panel (c) are classified as fast TP, fast DS, and slow events, with slow events occupying a transitional regime.

### 4.2 Evidence for co-seismic thermal pressurization

Co-seismic increases of on-fault pore pressure ($Pp$) were observed during the early stages of all experiments, initially associated with fast stick-slip events. These pressure rises occurred synchronously with rapid slip and shear stress drop (Figure 3), indicating a direct coupling between $Pp$ generation and frictional heating. In the following, we evaluate whether the observed co-seismic $Pp$ rise may be explained by thermal pressurization (TP).

TP operates within the micrometer-scale shear zone, whereas $Pp$ measurements sample a millimeter-scale rock volume defined by the effective diameter of the transducers (±0.5 mm). For an initial hydraulic diffusivity of $a_{hy} \approx 10^{-7}$ m$^2$ s$^{-1}$, the characteristic diffusion timescale across a half-width of 0.5 mm is ~2.25 s. Because permeability and diffusivity are expected to decrease with increasing differential stress during loading, this estimate represents a lower bound. Consequently, during fast events, with typical dynamic slip durations of <0.1 s, the measured $Pp$ responds under effectively undrained conditions at the measurement scale. For slow events, although the total duration can extend to ~2 s, the main slip occurs within a velocity-pulse-dominated phase, during which fluid diffusion remains limited. Therefore, an undrained approximation is applied to the primary slip phase for both fast and slow events. Based on this consideration, TP analysis is restricted to the main slip-velocity pulse, isolating the interval of concentrated frictional heating and minimizing the influence of post-seismic diffusion.

Within this interval, we apply a simplified undrained, adiabatic TP formulation (Rice, 2006), neglecting dilatancy, such that the $Pp$ change is directly proportional to the cumulative frictional work:

$$p_{(t)} - p_0 = \Lambda(T_{(t)} - T_0) = \frac{\Lambda}{\rho c w}\int_0^t \tau V \ dt \quad (1)$$

where $\Lambda$ is the thermal pressurization coefficient, $(T_{(t)} - T_0)$ is the temperature rise, $\tau$ is shear stress, $V$ is slip velocity, $w$ is the effective shear zone width, and $\rho c$ is the volumetric heat capacity (2.7 MPa K$^{-1}$). We adopt a constant thermal pressurization coefficient of $\Lambda = 0.5$ MPa K$^{-1}$, which lies within the range estimated for crustal fault rocks by Rice (2006) and is close to the value (0.468 MPa K$^{-1}$) used by Schmitt et al. (2015). Although $\Lambda$ may vary with thermodynamic conditions during slip, treating it as constant provides a first-order estimate of fault heating. Under this assumption, the inferred temperature rise scales as $\Delta T \propto 1/\Lambda$, whereas the inferred shear-zone thickness scales as $w \propto \Lambda$.

For early fast events, this formulation reproduces well the observed $Pp$ evolution (Figure 9a), indicating that undrained TP provides an excellent description of the co-seismic pressure rise. In contrast, for later events characterized by slower slip and longer rupture durations, the fit quality progressively degrades (Figure 9b) with the predicted and observed pore pressure responses being time-shifted (Figure 10a). This systematic time-shift indicates that the initial efficiency of TP has decreased in the earliest stages of rupture or potential dilatancy compensates TP. Diffusion may also contribute for the longer rupture phase. Applying this above analysis across all TP events reveals a consistent reduction in fitted $\Lambda/(\rho c w)$ with event sequence (Figure 10). The fitted $\Lambda/(\rho c w)$ decreases by ~70–90% relative to its initial value, implying a progressive reduction in $Pp$ generation per unit frictional work. Interpreted in terms of shear zone properties, this trend corresponds to an increase in the effective shear zone width, from ~0.05 mm during early fast events to ~0.4 mm during later events, nearly an order-of-magnitude thickening. Furthermore, the inferred peak temperature rise decreases from ~10 K during early fast events to ~0.2 K or less during later slow events. Consequently, combining with increasing shear zone width, frictional heat is distributed over a larger volume, buffering temperature rise. Because diffusion is neglected, this simplified analysis cannot reproduce post-seismic *Pp* decay. Numerical modeling of pressure and temperature diffusion to capture the full temporal evolution of *Pp* during the post-seismic phases demonstrates that our above estimates of *w* and ΔT should be regarded as an upper and lower bounds respectively (see Supporting Information Text S4 for details), due to the combined effects transducer design and sampling frequency. Nonetheless, our results show that, at first order, co-seismic *Pp* rise can be interpreted as an undrained response of the fault core dominated by TP.

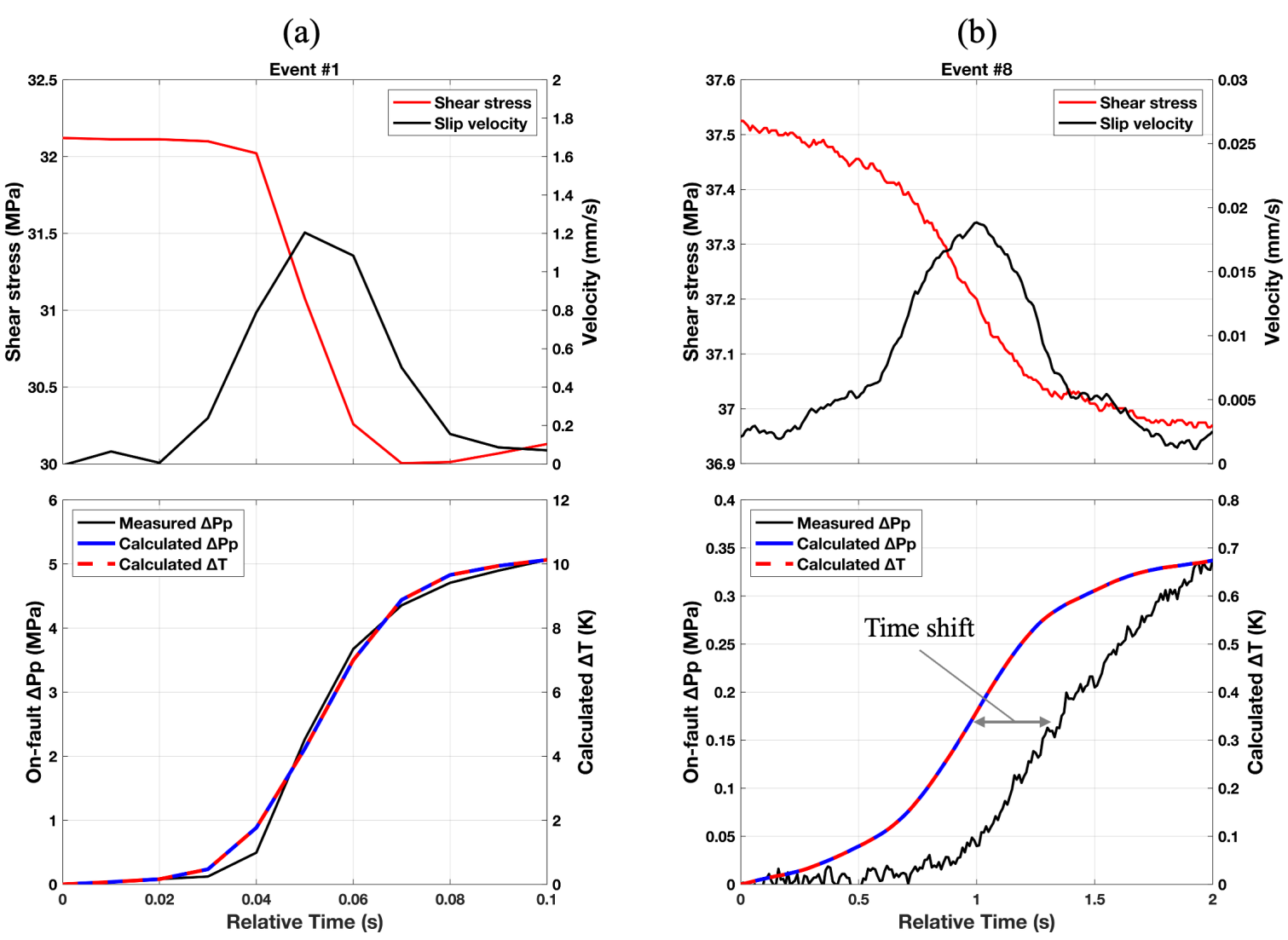

Figure 9. Fits of measured on-fault pore pressure (*Pp*) using a simplified thermal pressurization (TP) model neglecting diffusion and dilatancy for (a) the first fast event and (b) the last slow event exhibiting co-seismic Pp rise in experiment WGF10 under $P_c - P_f = 60$ MPa.

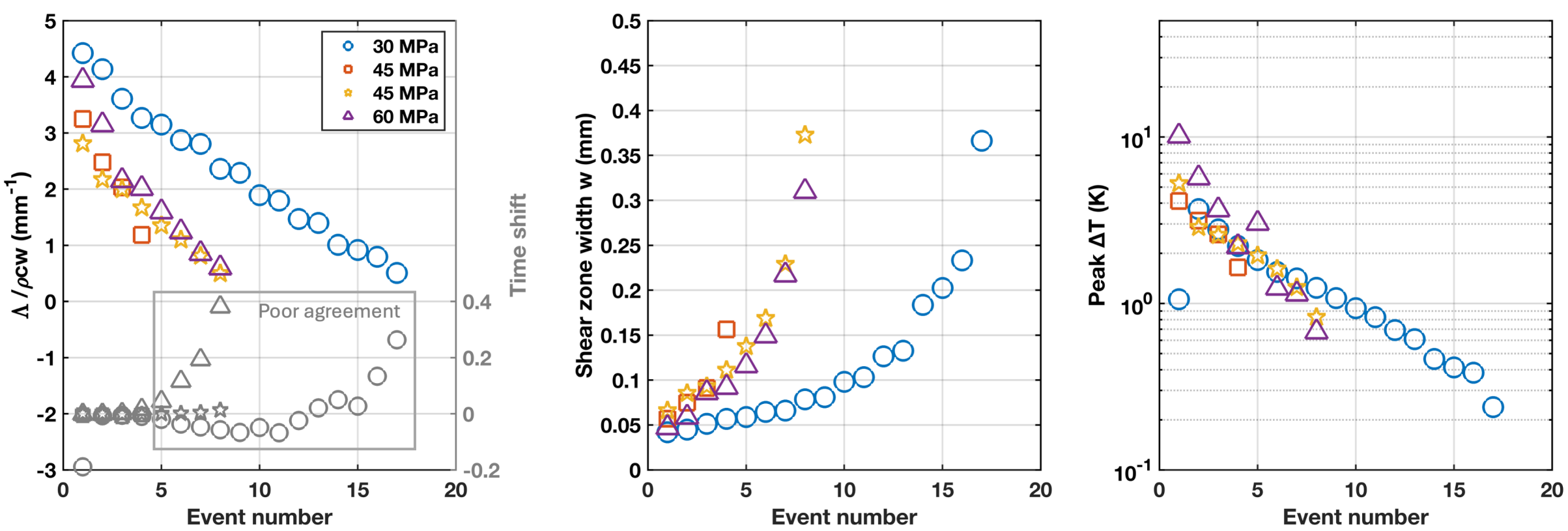

Figure 10. Event-by-event evolution of fitted $\Lambda/(\rho c w)$ (color symbols) and corresponding time-shift, together with estimates of effective shear zone width *w* and temperature change $\Delta T$ for all events with co-seismic *Pp* rise. Increasing time shifts means poor agreement between inverted *Pp* and measured *Pp* evolution, especially for slow

events. Symbols denote experiments conducted at different effective confining pressures (Star symbols corresponds to the high-permeability sample WGF8).

### 4.3 Constraints on co-seismic near-fault dilatancy

Stick-slip events exhibiting co-seismic *Pp* drop require an increase in pore volume during rupture. Such pressure decreases cannot be produced by elastic effects alone and instead imply dilatant deformation associated with porosity creation. Here, rather than reproducing the full *Pp* time history, we estimate the magnitude of plastic porosity change required to explain the observed co-seismic *Pp* drop during dilatant strengthening (DS) events.

We adopt a simplified undrained and adiabatic framework without hydraulic or thermal diffusion, appropriate for the co-seismic window defined by the slip-velocity pulse (Section 4.2). Elastic porosity effects are incorporated in the thermal pressurization factor Λ and are not treated explicitly (Rice, 2006). The governing balance equations are therefore

$$\begin{cases} \frac{\partial p}{\partial t} = \Lambda \frac{\partial T}{\partial t} - \frac{1}{\beta} \frac{\partial \phi_{pl}}{\partial t} \\ \frac{\partial T}{\partial t} = \frac{\tau V}{\rho c w} \end{cases} \tag{2}$$

where $\phi_{pl}$ denotes plastic porosity, $\beta$ is the storage capacity ($MPa^{-1}$), which is initially inverted (Table 1) based on *Pp* transducers' calibration. Because the effective shear zone width cannot be independently constrained for DS events, we adopt the value inferred for the last TP-type event and treat it as fixed. Under this assumption, the TP-induced *Pp* rise represents a potential contribution, and the inferred plastic porosity change therefore constitutes an upper bound on dilatancy. In the opposite limiting case where TP is neglected, the measured *Pp* drop would be entirely attributed to dilatancy, providing a lower bound. Together, these two end-member assumptions define physically reasonable bounds on the magnitude of dilatancy required by the data. In the following, we focus on constraining the upper bound. Figure 11 illustrates the inversion procedure for a representative fast DS event (#110) in experiment WGF9. The dilatancy-induced *Pp* drop accounting for TP is inferred from the observed *Pp* decrease and the estimated TP contribution based on shear stress, slip velocity, and the assumed shear zone width obtained from last TP event (Figure 11a). The resulting *Pp* drop is converted into plastic porosity change $\Delta\phi_{pl}$. Therefore, applying this approach to all DS events we could obtain each co-seismic $\Delta\phi_{pl}$ for each slip. As fault $\phi_{pl}$ is an analogue of damage zone, the total co-seismic $\phi_{pl}$ captures the evolution of fault damage versus cumulative slip, increasing from $\sim 10^{-5}$ to $10^{-2}$, i.e., at most on the order of 3% (Figure 11b). Since the post-seismic compaction ($\phi_{pl}$ recovery) might exist (Segall & Rice, 1995) but is neglected here, the total co-seismic $\phi_{pl}$ provides an upper bound. With increasing cumulative slip, $\phi_{pl}$ exhibits a superlinear trend for all experiments with a power-law exponent ($n$) larger than 1 (see inset). Therefore, damage might accumulate continuously, without clear saturation over the experimental slip range. In addition, the exponent has a decreasing trend with increasing effective confining pressure, indicating an inhibiting effect of pressure on plastic porosity accumulation and dilatancy.

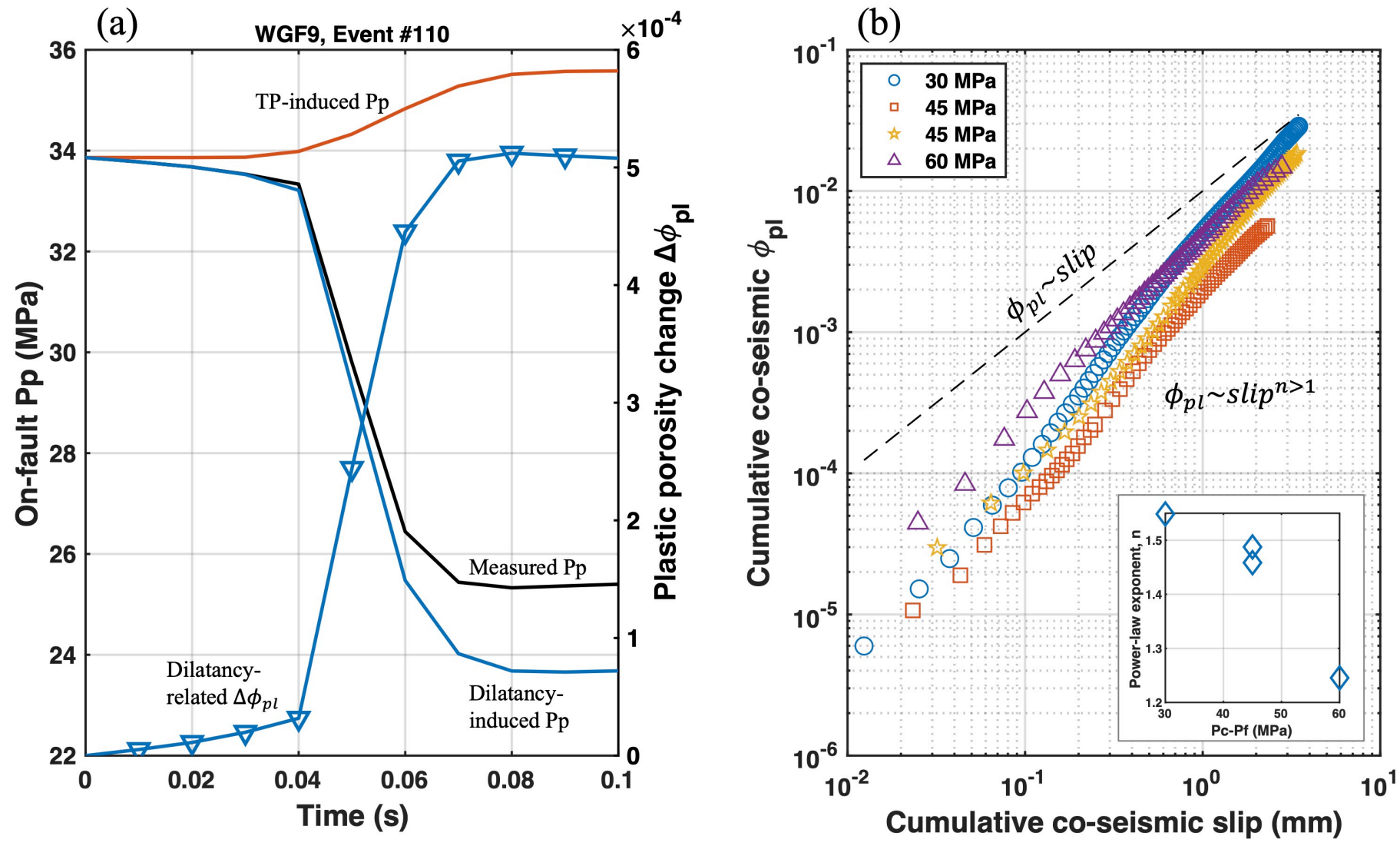

Figure 11. Inversion of plastic porosity change for dilatant strengthening (DS) events. (a) Recorded on-fault pore pressure ($Pp$) drop (black line) during the co-seismic stage of event #110 in experiment WGF9 ($P_c - P_f = 30$ MPa), together with upper bound of dilatancy-induced $Pp$ drop (blue line) with consideration of a potential thermal pressurization (TP) contribution (red line). The dilatancy-induced $Pp$ drop is transformed into plastic porosity change $\Delta\phi_{pl}$. (b) Power-law scaling of cumulative plastic porosity $\phi_{pl}$ of all DS events as a function of cumulative co-seismic fault slip. All experiments exhibit superlinear growth (n > 1). Inset shows the dependence of the fitted exponent n on effective confining pressure ($P_c - P_f$), indicating a decreasing trend with pressure. Symbols denote experiments conducted at different effective confining pressures (Star symbols corresponds to the high-permeability sample WGF8).

# 5. Discussion

### 5.1 Rupture transition from TP- to DS-dominated behavior

We classify co-seismic on-fault $Pp$ responses as thermal pressurization (TP) when $Pp$ rises, and dilatant strengthening (DS) when $Pp$ drops. An alternative explanation for the $Pp$ rise due to shear-induced compaction, which could locally reduce pore volume and raise fluid pressure cannot be strictly excluded without direct measurements of normal displacement and shear-zone temperature, but several lines of evidence argue against it. On-fault and off-fault $Pp$ increase nearly synchronously prior to the first stick-slip event, indicating that most of the shear zone's compaction capacity is already consumed by monotonic loading (on-fault exceeds off-fault by only ~1.5 MPa in WGF10); during rupture itself, off-fault $Pp$ rise stays below 0.5 MPa while on-fault $Pp$ reaches several MPa, so bulk compaction cannot explain the signal, and the rapid stress-reduction path during rupture does not favor further compaction-driven pressurization. Laboratory gouge-shearing studies show initial compaction typically occurs at low slip rates before localization drives progressive dilatancy (Marone et al., 1990; Faulkner et al., 2018), with clay-rich gouges compacting at low velocity (Ujiie & Tsutsumi, 2010; Ujiie et al., 2013) while quartz- or illite-rich gouges dilate at the higher velocities relevant here (French et al., 2014; Yao et al., 2016). Stick-slip experiments on simulated glass-bead gouge similarly show that volumetric response is microstructure-dependent, small spherical grains can compact co-seismically with inter-seismic dilation (Johnson et al., 2008; Scuderi et al., 2014, 2015), while larger or differently shaped grains dilate co-seismically instead (Hu et al., 2023), and bare-rock experiments show smooth surfaces compact less than rough ones, which dilate via asperity breakdown (Morad et al., 2022). Critically, in these cases co-seismic compaction typically follows inter-seismic dilation that replenishes pore volume, whereas our samples are already pre-compacted by the first event, leaving little capacity for further collapse. We therefore interpret the co-seismic *Pp* rise as primarily thermal-pressurization-driven.

A progressive transition from on-fault *Pp* rise to *Pp* drop, accompanied by a shift from fast to slow and back to fast stick-slip events, was observed in our experiments. This transition tracks progressive fault-zone thickening (Section 4.2). Slip initially occurs on a smooth surface, allowing thermal pressurization to operate efficiently; the subsequent increase in shear-zone thickness, evidenced by SEM observations of gouge patches (Section 3.5), distributes frictional work over a larger volume and reduces the efficiency of thermal weakening (Rempel & Rice, 2006; Rice, 2006), consistent with prior experimental work (Scholz, 1987; Boneh et al., 2014; Tesei et al., 2017; Noël et al., 2023). This *Pp* evolution is summarized in the conceptual model shown in Figure 12. Consistently, recent experiments on initially rough faults observed only co-seismic *Pp* drops, without an *Pp*-rise behavior (Brantut et al., 2026), further supporting the role of fault structure in controlling the co-seismic *Pp* response.

With continued deformation, growing gouge patches provide a physical basis for velocity-dependent dilatancy (Morrow & Byerlee, 1989; Marone et al., 1990; Samuelson et al., 2009b, 2011; Affinito et al., 2025), and the development of a near-fault damage zone, again seen in SEM images (Section 3.5), enhances inelastic porosity generation, consistent with the progressively stronger dilatancy observed during the DS-dominated stage.

TP- and DS-dominated events also differ in post-seismic behavior. TP events are followed by thermal and hydraulic diffusion into the surrounding bulk, well reproduced by a simple diffusion model (Supporting Information Text S4). DS events instead potentially show additional post-seismic compaction within the shear zone, which prevents a straightforward diffusion-based inversion. This contrast underscores the fundamentally different hydromechanical processes underlying TP- and DS-type events.

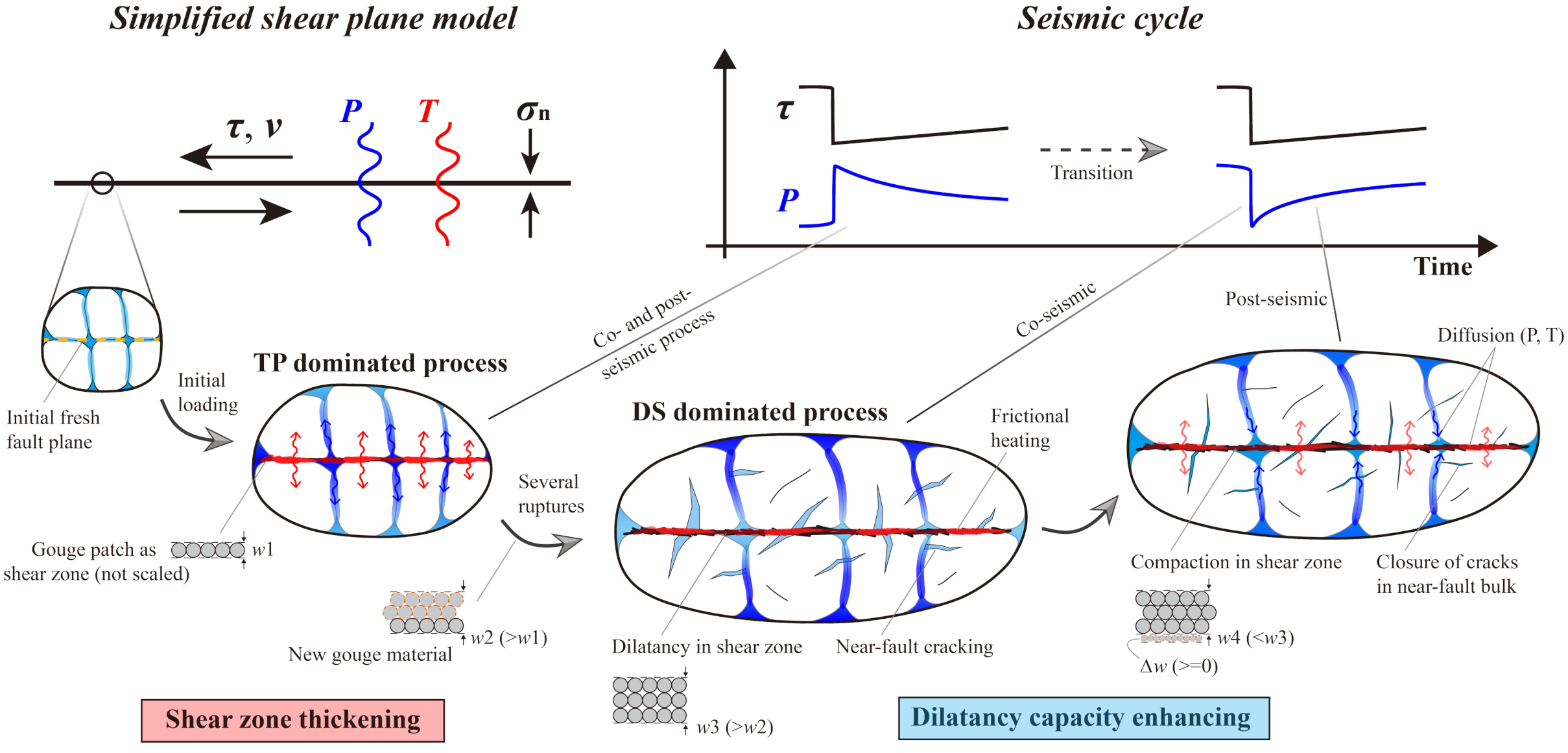

Figure 12. Conceptual model of fault structural evolution, including shear-zone thickening and near-fault damage development, driving the transition from pore pressure rise dominated by thermal pressurization (TP) to drop dominated by dilatant strengthening (DS) during stick-slip rupture.

### 5.2 Experimental stiffness constraint on slip instability

The transition in *Pp* behavior is explained above by fault structural evolution; the parallel transition in slip stability can be interpreted through fault stiffness. In a rate-and-state spring-slider framework, a critical stiffness $k_c$ sets the instability condition via $\kappa = k/k_c$ (Ruina, 1983; Leeman et al., 2016): unstable slip occurs when $\kappa < 1$, with slower or faster slip depending on how far $\kappa$ falls below unity. Since total loading stiffness is approximately constant across our experiments (~42 MPa/mm, Figure S20), the observed fast–slow–fast transition implies an evolving $k_c$.

Rather than compute $k_c$ directly, we use a co-seismic weakening stiffness $k_w$ (the initial slope of shear stress versus slip during weakening, see Figure S22) and track $k/k_w$ as an experimental proxy for $k/k_c$. This ratio decreases monotonically from ~1.5 to 0.1 with increasing slip velocity, with a clear threshold at $k/k_w \approx 0.8$ separating slow and fast events, although one experiment exhibits a lower transition value of 0.6. For the three low-permeability experiments ($P_c - P_f$ = 30, 45, 60 MPa), which show fast–slow–fast transitions, $k/k_w$ first increases then decreases, crossing this threshold twice; the high-permeability experiment shows only fast events and a $k/k_w$ that never crosses the threshold, consistent with the observed evolution of shear stress drop.

This evolution of $k/k_w$ tracks the TP-to-DS transition: toward the late DS-dominated stage, $k/k_w$ decreases (indicating enhanced weakening) even as co-seismic *Pp* drops, hence dilatancy, grow larger. Since dilatancy is generally expected to stabilize slip, this coincidence of stronger dilatancy with faster, more unstable rupture suggests that pore pressure evolution alone does not control rupture stability, and that additional weakening processes operate in the late-stage DS regime.

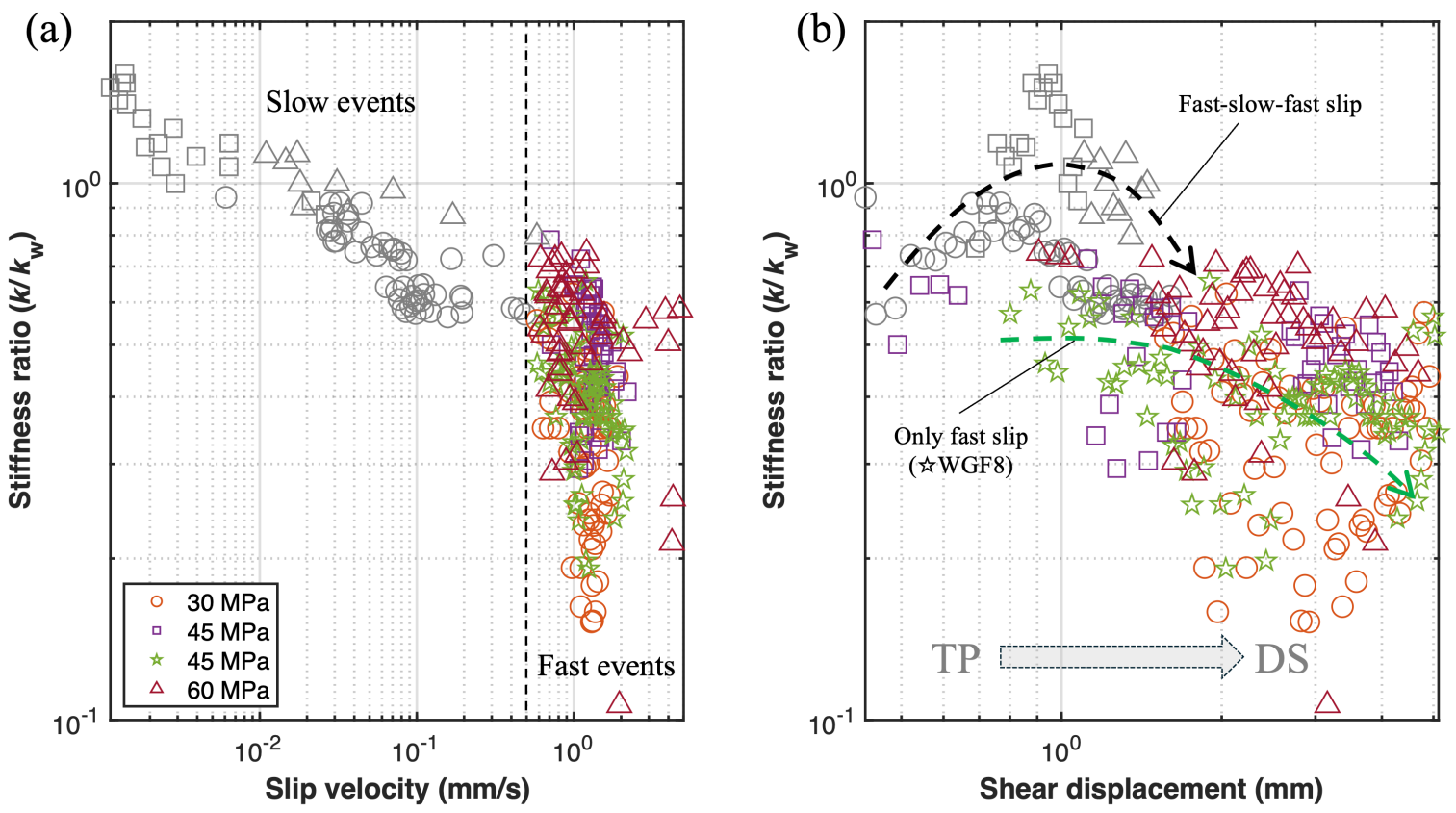

Figure 13. Experimental stiffness ratio ($k/k_w$) as a function of slip velocity (a) and shear displacement (b), illustrating the fast–slow–fast instability transition during the TP-to-DS evolution. Symbols denote experiments conducted at different effective confining pressures (Star symbols corresponds to the high-permeability sample WGF8).

### 5.3 Fracture energy

Fracture energy $G$ (or breakdown work Wb) is the energy dissipated during rupture propagation and generally scales with co-seismic slip (Wong, 1982; Cocco et al., 2023). We estimate $G$ using the classical linear slip-weakening approximation, $G = {}^{1}/_{2}\,(\tau_p - \tau_r)D_c$ (Ida, 1972), consistent with prior laboratory work (Passelègue et al., 2016a); an integral-based estimate accounting for nonlinear stress evolution gives comparable scaling (Figure S23).

Fracture energy for slow events ranges from ~1–10 J/m², and for fast events from ~10–1000 J/m²; despite this difference in magnitude, slow and fast events follow a similar energy–slip scaling, largely independent of whether co-seismic *Pp* rises or drops, though *Pp*-rise events tend toward slightly lower $G$. Our results (slip range $10^{-5}$ to $2\times10^{-4}$ m) are consistent with previous measurements on saw-cut granite (Ohnaka, 2003; Passelègue et al., 2016a) and with compiled seismological and laboratory estimates (Cocco et al., 2023), indicating that the observed scaling is robust across experimental approaches.

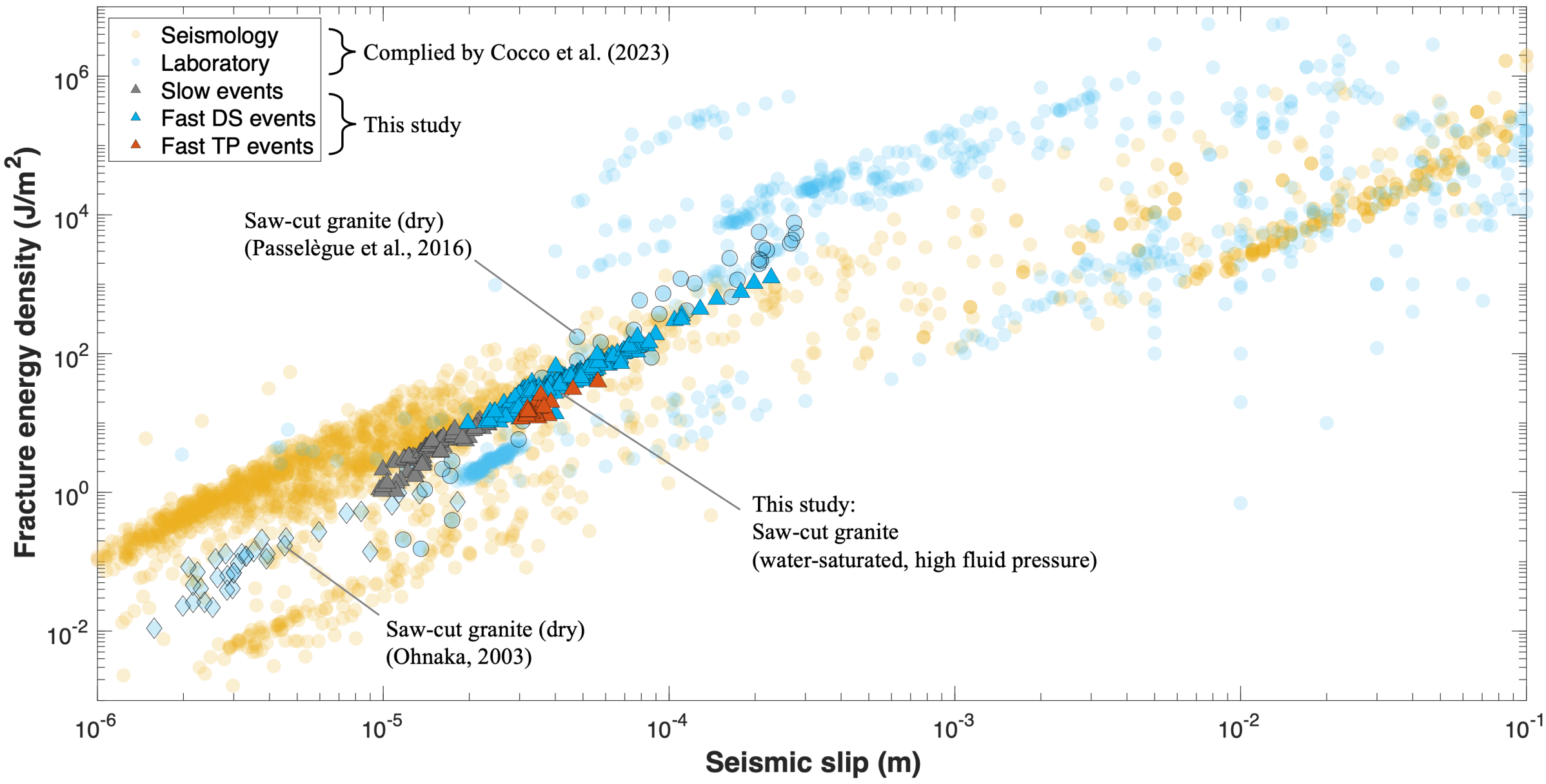

Figure 14. Comparison of fracture energy density versus seismic slip from this study with seismologically inferred and laboratory estimates compiled by Cocco (2023) over a comparable slip range. Data from Passelègue et al. (2016a) and Ohnaka (2003), compiled by Cocco (2023), are highlighted with distinct symbols for direct comparison with the same saw-cut granite configuration. Filled triangles represent fracture energy estimated using the linear slip-weakening approximation in this study. Gray symbols denote slow events, blue symbols fast DS-type events, and red symbols fast TP-type events.

### 5.4 Limitations and perspectives

Mixed *Pp* responses during transitional events (e.g., events #9–11 in WGF10) show *Pp* rising rapidly above the pre-slip level with an initial *Pp* drop (Figure 5a), either late in the co-seismic stage or shortly after slip stops — a pattern that classical TP alone, in which *Pp* should track heating and peak during slip itself, does not directly predict. Critically, this rise occurs within the same event rather than being inherited from an earlier one: the eventual peak *Pp* reached in these mixed events still follows the same systematic background trend seen in ordinary TP-dominated events, despite an initial co-seismic *Pp* drop. We interpret this as dilatancy and thermal pressurization acting simultaneously within a single event: dilatancy dominates the earliest, fastest part of slip and produces the initial drop, while frictional heating continues to raise near-fault pressure throughout the velocity pulse. Because the transducer's finite fluid volume buffers and delays its response, this ongoing TP signal only becomes visible at the sensor once dilatancy's effect wanes, near the end of slip or shortly after. This is why the eventual peak still matches the TP-related *Pp* background trend despite the early drop. Post-seismic compaction and near-fault diffusion may also contribute to the recovery, but consistency with the TP background trend is the strongest evidence for a residual, delayed TP contribution as the primary driver.

A second limitation is that our analysis relies on spatially averaged on-fault *Pp*, whereas the four on-fault transducers reveal spatial heterogeneity (Figure S24) that develops after the first rupture and persists across subsequent cycles, likely reflecting rupture-induced structural changes such as gouge development and localized dilatancy/compaction. Distinguishing between these possibilities requires independent constraints on damage evolution and is beyond the scope of this study.

# 6. Implications for fault structural evolution

Our experiments provide direct measurements of transient *Pp* during laboratory earthquakes: rapid rises of several MPa followed by a shift to pressure decrease. Such large, rapid *Pp* variations are rarely documented experimentally despite their central role in fluid–fault interaction models (Rice, 2006; Viesca & Garagash, 2015). The measured >5 MPa on-fault *Pp* rise is well reproduced by TP under undrained, adiabatic assumptions (Figure 9), consistent with reports of MPa-scale *Pp* rise in rotary-shear experiments (Yao et al., 2023; C.-C. Hung et al., 2025), and provides direct experimental verification of TP theory.

However, only a few *Pp*-rise events were observed, and only during early, low-stress loading; all subsequent events show *Pp* drops even as background stress increases. This challenges the view that TP universally dominates large, fast earthquakes (Viesca & Garagash, 2015), and suggests instead that structural evolution (e.g., damage zone) in the near-fault region increasingly inhibits TP as loading progresses — i.e., structural evolution on a simple fault need not favor TP dominance at later, higher-stress stages. This is consistent with modeling showing that damage-related permeability enhancement can reduce TP efficiency (Toffol et al., 2026).

This structural framework also bears on fault maturity, though the comparison to natural faults should be drawn carefully. Field and laboratory observations show that slip in mature faults localizes into narrow principal slip zones within broader damage zones (Beeler et al., 1996; Mitchell & Faulkner, 2009; Perrin et al., 2016; Scuderi et al., 2017). Rather than treating our experimental sequence as a single fault simply maturing over time, we note that the early, mirror-polished-surface events may themselves be representative of one class of natural fault, one with highly localized slip within cemented fault rock. In contrast, the later events, with their more developed gouge and distributed damage, may better represent a structurally different, more damaged natural fault, rather than a later stage of the same one. Comparisons with natural fault sequences also require caution because our experiments neglect interseismic healing and sealing of the fault core and fracture network, processes that in nature would limit the dilation available during subsequent ruptures. With that caveat, transient *Pp* responses during rupture may still serve as a useful indicator of fault structural state.

These results also bear on rupture stability. Although dilatancy is generally expected to stabilize slip, the TP-to-DS transition here does not prevent dynamic rupture: strong dilatancy coexists with increasingly fast rupture, unlike triaxial experiments on fractured granite where sustained dilatancy stabilizes slip (Aben & Brantut, 2021). We interpret this in light of dynamic damage generation near the fault (Passelègue et al., 2016b; Okubo et al., 2019), consistent with our microstructural observations: dilatancy and its associated *Pp* drop primarily reflect the evolving damage zone behind the rupture front rather than directly controlling stability, whose dependence on the competition between dilatancy, frictional weakening, and TP is well established (Segall et al., 2010; Garagash & Rudnicki, 2003a, 2003b).

Within a rate-and-state framework, structural evolution modulates rupture behavior by altering the hydromechanical properties that set both pore pressure feedback and effective critical stiffness (Ruina, 1983; Leeman et al., 2016; Mclaskey & Yamashita, 2017). Overall, our results suggest that the classical TP–dilatancy competition framework should be extended to explicitly incorporate fault structure and its evolution, particularly near-fault damage-zone development (Okubo et al., 2019; Jara et al., 2021; Ferry et al., 2025; Almakari et al., 2026).

## 7. Conclusion

Across triaxial stick-slip cycles at varying effective confining pressure (30, 45, 60 MPa), we document a systematic transition in co-seismic on-fault pore pressure (*Pp*) from rise to drop, with a complex, mixed-response middle phase, accompanied by a spectrum of fast and slow slip events coexisting on the same evolving fault. Early cycles show *Pp* rise, reproduced by a TP model under undrained, adiabatic conditions; later cycles show progressively larger *Pp* drops consistent with dilatancy. The magnitude of *Pp* rise decreases over early events while *Pp* drop and rupture intensity both grow with continued loading, indicating strengthening dilatancy despite more intense ruptures.

To our knowledge, this is the first experimental demonstration of both opposing *Pp* responses and a fast–slow rupture spectrum on a single evolving fault under continuous loading. These results suggest a systematic transition from thermal pressurization (TP) to dilatant strengthening (DS) as the fault structurally evolves, and show that transient *Pp* signals provide a direct window into evolving fault hydromechanical properties. In mature fault zones, fault weakening may thus be governed largely by co-seismic dilatancy, though fast, strong ruptures can still occur when dilatant strengthening is insufficient to arrest an ongoing rupture.

## Conflict of Interest

The authors declare there are no conflicts of interest for this manuscript.

## Data Availability Statement

The experimental data from the present study are available in Zenodo (Fan et al., 2026).

## Acknowledgments

CF acknowledges China Scholarship Council for providing funding to perform this study at ENS. HSB acknowledge the European Research Council for its support of this work through the PERSISMO Grant (865411). CF expresses his appreciation to Teng-fong Wong, Daniel Faulkner, Di Toro Giulio and François Passelègue for fruitful discussions regarding this work. CF also thanks Ankit Gupta, Suli Yao, Yishuo Zhou and Wei Feng for meaningful discussions. ChatGPT is used to correct grammatical mistakes in the manuscript.

Supporting information for

# Competition between thermal pressurization and dilatant strengthening in laboratory ruptures

Caiyuan Fan[1], Gang Lin[1], Jérôme Aubry[2], Damien Deldicque[1], Carolina Giorgetti[1], Harsha S. Bhat[1], Alexandre Schubnel[1]

[1]Laboratoire de Géologie, École Normale Supérieure/CNRS UMR 8538, PSL University, Paris, France,

[2]Université Savoie Mont Blanc, Université Grenoble Alpes, CNRS, IRD, University Gustave Eiffel, Le Bourget du Lac, France

## Contents of this file

## Text S1. Pp transducer calibration and hydromechanical parameter inversion

Under target confining pressure, we controlled boundary fluid pressure $P_f$ by fluid pump to have stepwise change to get responses of strain gauges on Pp transducers in different pressure level (Figure S1a, WGF10 as an example). Then, extracting stable value at each level, a linear relationship between them could be obtained with calibrated factor (Figure S1b):

$$P_f = a\varepsilon + b$$

where *a* and *b* are constants, fitted by data.

Since under controlled pressure source at the boundary, the fluid pressure diffuses to inside the sample and finally diffusion finishes when pressure is balanced everywhere inside the sample. Such a process satisfies diffusion equation

$$\frac{\partial p}{\partial t} = a_{hy}\frac{\partial^2 p}{\partial y^2}$$

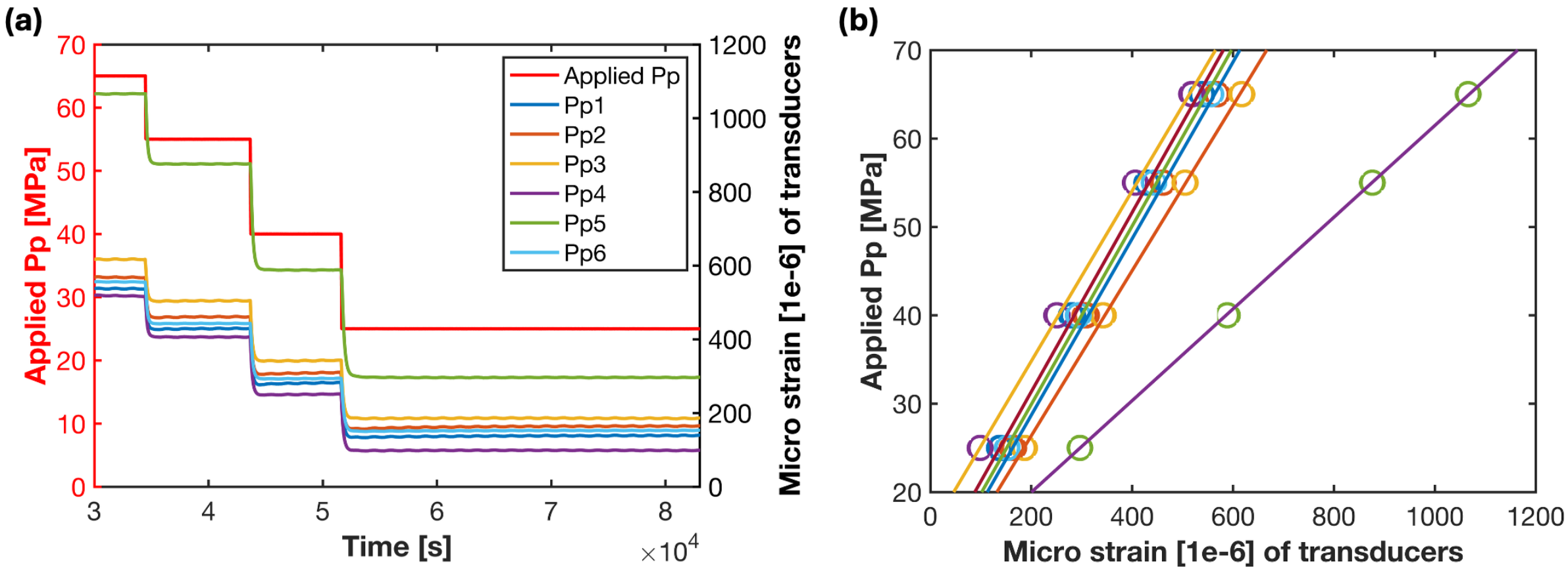

Figure S1. Calibration of internal pore pressure sensors. (a) Original stepwise boundary applied fluid pressure change with responses of Pp transducers. (b) the linear response of strain value of transducers with applied fluid pressure at the boundary.

Build the finite difference form of the equation

$$\frac{p_j^{n+1} - p_j^n}{\Delta t} = a_{hy} \frac{p_{j-1}^n - 2p_j^n + p_{j+1}^n}{(\Delta y)^2}$$

Then,

$$p_j^{n+1} = r\left(p_{j-1}^n - 2p_j^n + p_{j+1}^n\right) + p_j^n$$

Where r is

$$r = \frac{a_{hy}\,\Delta t}{(\Delta y)^2}$$

A 1D diffusion model is built with 25~35 mm length, the left boundary ($y = 0$) is set Dieterich' boundary condition, provided by fluid pump pressure data. The right side ($y = L$) is Neumann (zero-flux) boundary condition. That is

$$\frac{\partial p}{\partial y}\Big|_{y=L} = 0$$

By using fixed $\Delta t = 1$ s and satisfying $r = 0.4$ (it should be <0.5 to satisfy the stability condition of numerical calculation in explicit method of finite difference method), we automatically calculate a $\Delta y$ by giving a potential $a_{hy}$ to fit the fluid pressure curve during calibration. Eventually, we obtained the $a_{hy}$ with minimum L2 error of fitting. The fitting result could be seen in Figure S2 and shown in Table 1 in the main text.

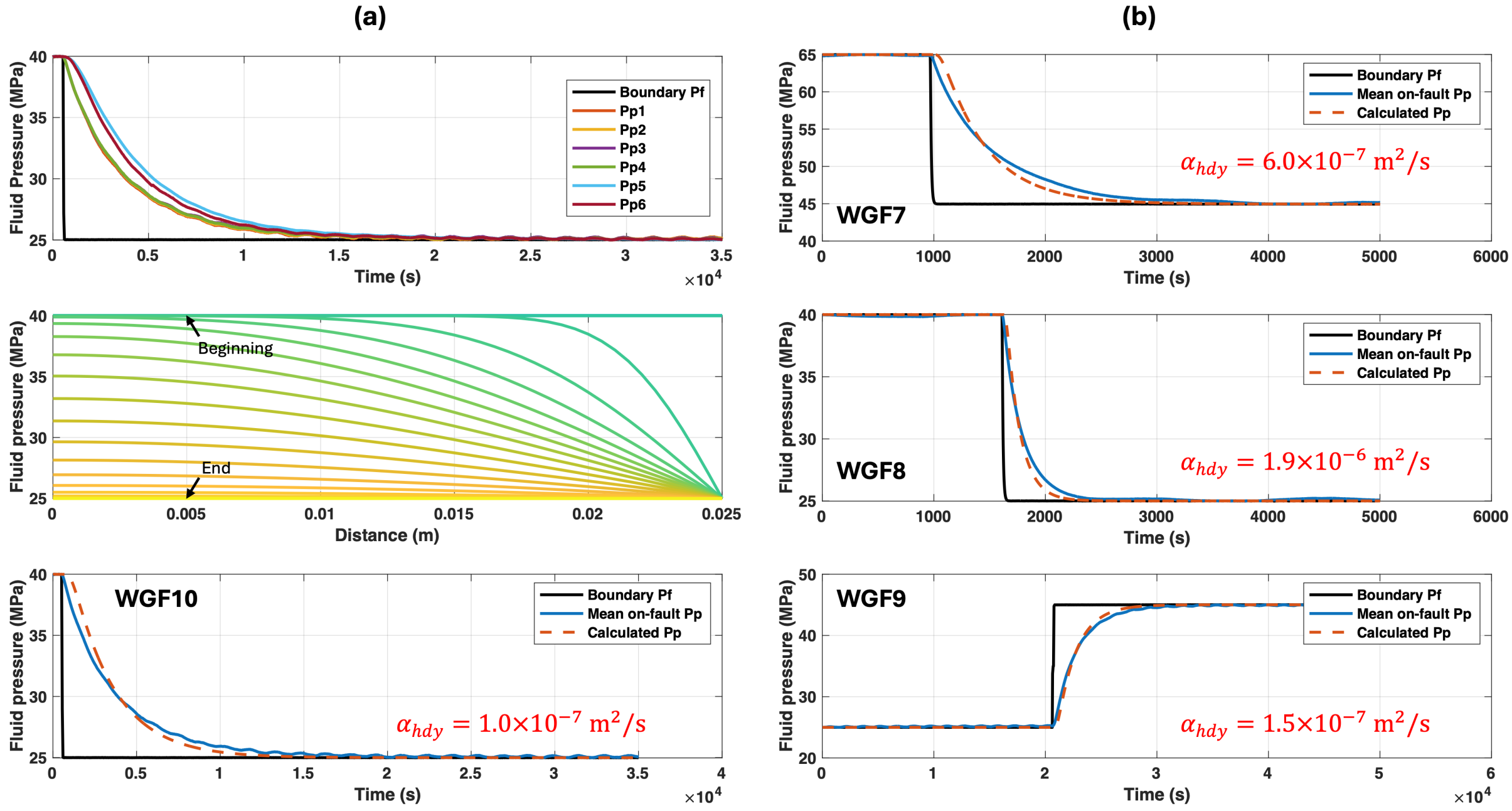

Figure S2. Pore pressure diffusion from boundary to internal Pp transducers for all four samples. (a) Initial boundary fluid pressure and internal Pp data, inverted Pp diffusion snapshot and inverted mean on-fault Pp curve for WGF10, (b) Inverted result of samples WGF 7~9.

Furthermore, hydraulic diffusivity can be calculated by

$$a_{hy} = \frac{k}{\eta\beta}$$

Where k is permeability ($m^2$), $\eta$ is viscosity (Pa*s), $\beta$ is storage capacity ($Pa^{-1}$).

Since k is been measured before Pp transducer's calibration at target fluid pressure (see in Table 1) and viscosity depends on fluid temperature and pressure while at current stage temperature is regarded as room temperature and it does not change too much with pressure in the range of 25 MPa to 45 MPa (our target fluid pressure at the boundary), yield about $10^{-9}$ Pa*s. Therefore, storage capacity can be calculated by equation and yield $1\sim6\times10^{-5}$ $MPa^{-1}$ (Table 1) which is similar to some reference.

Furthermore again, we also know initial porosity of about 1.5~2.5 % at room condition which could be regarded as an upper bound for the experimental condition with effective confining pressure. Using a porosity of 2% and fluid compressibility of $4\times10^{-5}$ $MPa^{-1}$ and general express of storage capacity

$$\beta = \phi(\beta_f + \beta_\phi)$$

We could then get $\beta_\phi$ yield $0.3\sim3\times10^{-3}$ $MPa^{-1}$.

## Text S2. Slip velocity pulse calculated by LVDT slip

Slip velocity is calculated as the time derivative of the fault slip. To reduce noise amplification during differentiation, the slip time series is first smoothed using a first-order Savitzky–Golay (SG) filter and then differentiated using cubic spline interpolation. Because fast and slow events occur on different timescales, different SG window lengths are applied. For fast events (peak velocity > 0.5 mm $s^{-1}$), a narrow window (3 points) is used to preserve the short-duration velocity pulse (left panel in Figure S3). For slow events, a wider window (50 points) is adopted to suppress long-period fluctuations, followed by light smoothing of the derived velocity (right panel in Figure S3). Moderate variations in smoothing parameters do not affect event classification or peak velocity within measurement uncertainty.

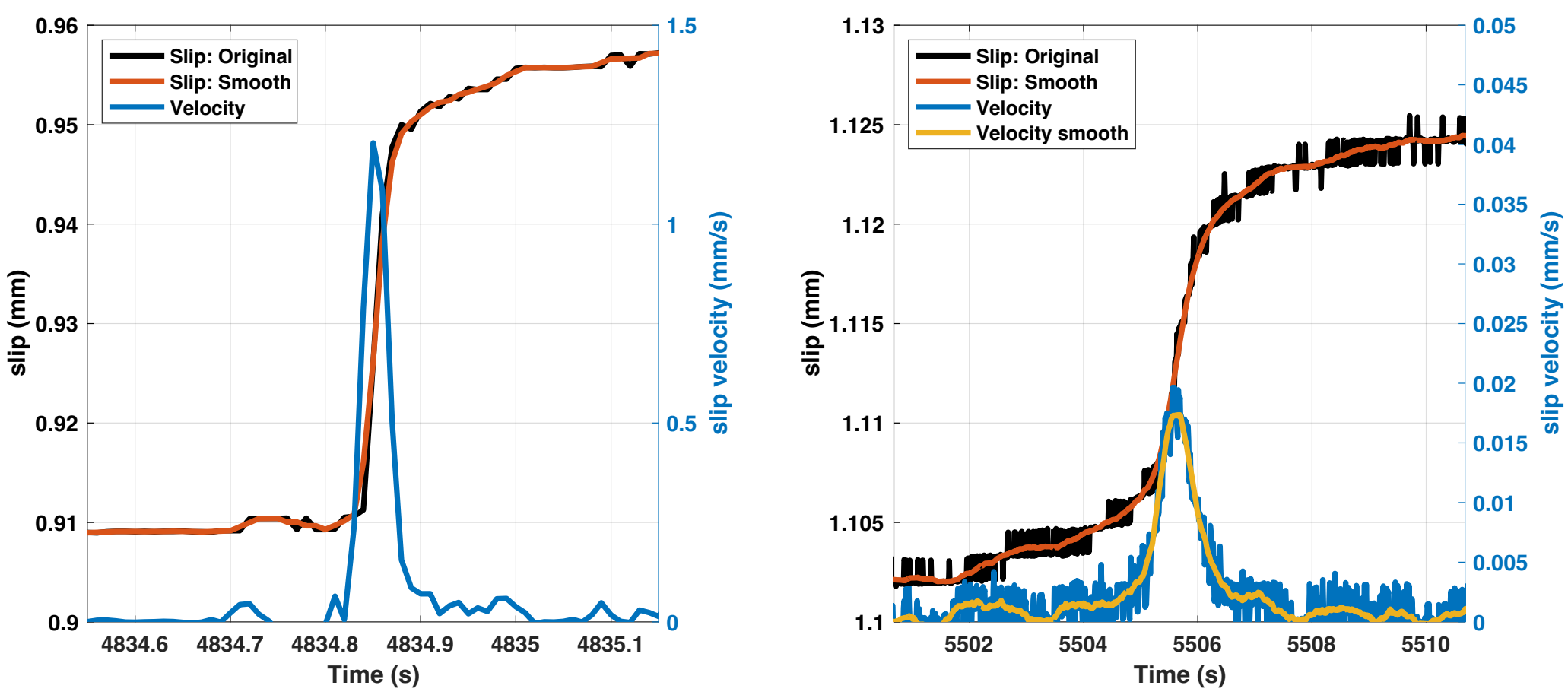

Figure S3. Example of slip-velocity pulse extraction for a representative fast event (left) and slow event (right). Black curves show original fault slip, red curves indicate Savitzky–Golay–smoothed slip, and blue curves represent the derived slip velocity. For slow events, additional smoothing (yellow curve) is applied to the velocity to suppress differentiation noise. Note the two-order-of-magnitude contrast in peak velocity between fast and slow ruptures.

## Text S3. Microscopic image and crack density calculation

This section describes the method used to quantify crack density from cross-sectional SEM images. Figure S4 illustrates an example with well-developed near-fault cracks. The original grayscale image is converted into a binary image using a color threshold (0–100), in which cracks are identified as white pixels. Although some fine cracks are not fully captured, the dominant near-fault damage is well resolved. The binary image is then divided into square grids of 5 μm. For each grid, crack density is defined as the ratio of crack pixel area to total grid area, yielding values between 0 and 1. This metric represents a 2D crack density, which effectively captures the spatial distribution of damage both normal to and along the fault. Note that the images are cropped with the upper boundary corresponding to the fault surface. However, due to spatial variations in surface roughness and local damage, a small portion of the region above the slip interface (<~5 μm) may be included. This effect is minor and does not significantly influence the estimation of damage zone thickness.

The resulting spatial distribution of crack density (Figure S4c) reproduces the main features of the original image. Depth profiles are obtained by averaging crack density along each row, using both pixel-scale and grid-averaged approaches. The latter provides a smoother representation of damage evolution. In this example, the damage zone thickness is defined as the depth at which crack density decreases to a background level (~0.04), yielding a value of ~40 μm. Local variations are observed, reflecting non-uniform crack development and preferential crack orientations subparallel to the fault. Along the fault-parallel direction, horizontal profiles are obtained by averaging each column. These profiles reveal spatial heterogeneity, with localized zones of higher crack density, indicating that damage is not uniformly distributed along the fault.

The same processing method is applied, with profile calculations adapted to the large image scale: fault-parallel crack density is computed over 100 µm × 100 µm regions, whereas depth profiles are derived from 1 µm–thick strips averaged over ~1.5 mm segments or over the full image length (~75 mm) to obtain local and mean profiles, respectively. Background crack density is estimated at depths of 150–200 µm (~0.05). The mean depth profile yields an average damage zone thickness of ~60 µm, while local profiles show strong variability, ranging from ~10 to 160 µm. This variability reflects heterogeneous damage development along the fault, including localized zones of concentrated cracking that persist over the entire imaged length (~75 mm).

Figure S5 shows the location of the gouge patch presented in Figure 6d of the main text. Mineral phases are identifiable in the SEM image; however, no systematic interpretation is made regarding preferential crack development within specific minerals, as this is beyond the scope of the present analysis.

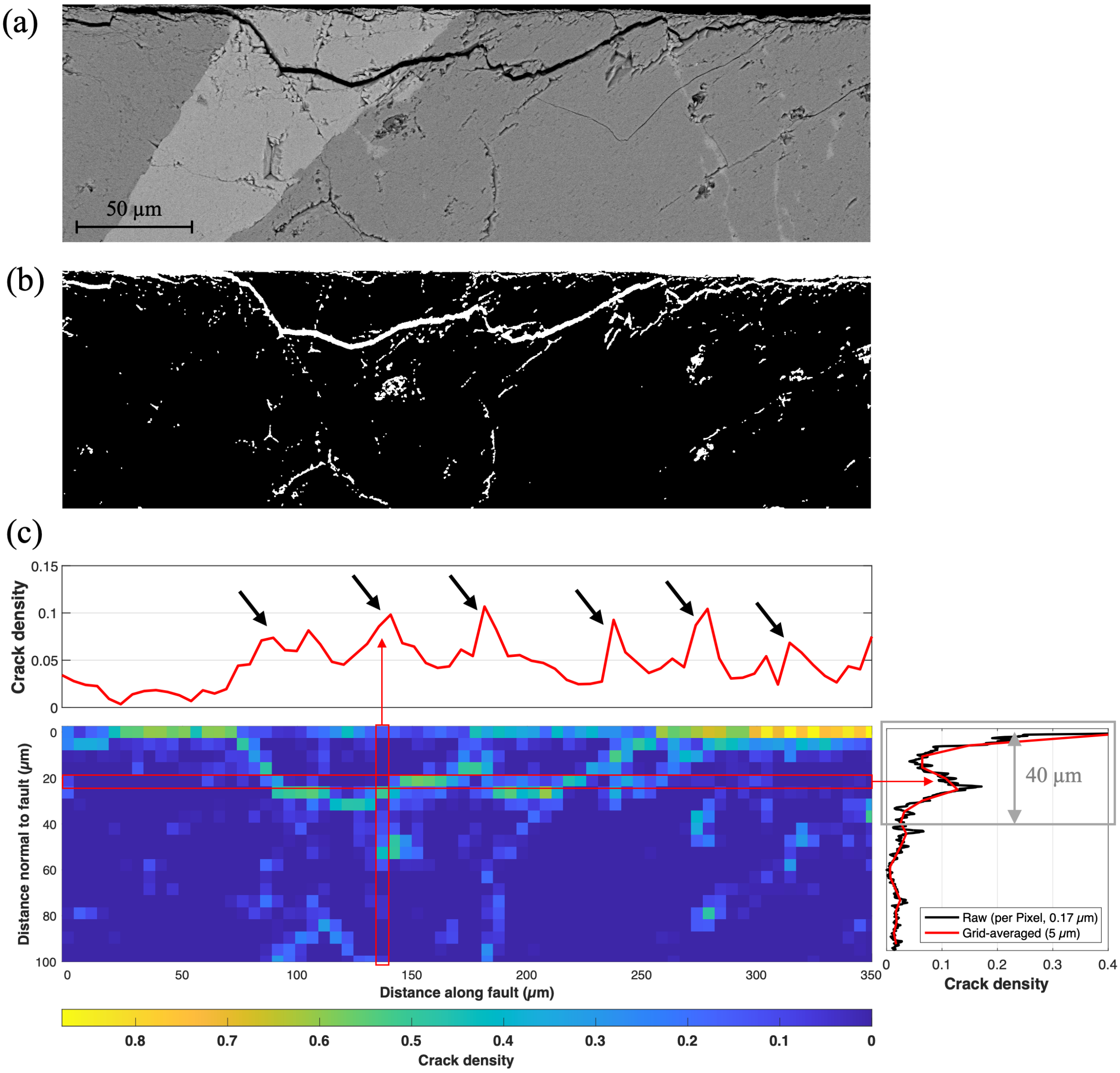

Figure S4. Cross-sectional SEM images perpendicular to the fault plane into the bulk, illustrating the crack density calculation procedure.

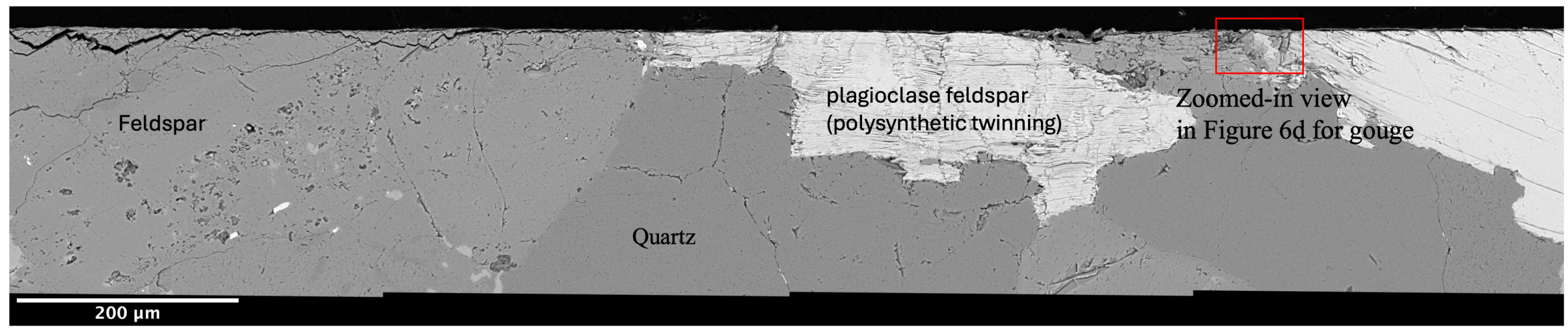

Figure S5. Cross-sectional SEM images perpendicular to the fault plane into the bulk, showing crack development in relation to mineral phases and a gouge patch (see Figure 6d in the main text for a zoomed-in view).

# Text S4. Numerical analysis on thermal pressurization (TP)

### S4.1 Theoretical description

Thermal pressurization (TP) is governed by the coupled conservation of thermal energy and fluid mass in a porous medium. Following Rice (2006), pore fluid diffusion obeys Darcy's law, and heat transport by fluid advection is neglected. The full governing equations for temperature $T$ and pore pressure $p$ evolution in a deforming porous medium are derived in Rice (2006), including the effect of inelastic porosity change. For the present study, we adopt the simplified formulation of Rice (2006) under the following assumptions (see Figure S6): 1) Physical properties (thermal conductivity, heat capacity, permeability, viscosity, storage capacity) are spatially uniform and constant. 2) Shear strain is homogeneously distributed within a shear zone of width $w$, such that $\dot{\gamma} = V/w$. 3) Diffusion is considered only in the direction perpendicular to the shear zone (1D geometry). 4) Heat generation is uniformly distributed within the shear zone. 5) Inelastic porosity change is neglected for analyzing TP-type events. Under these assumptions, the governing equations reduce to

$$\frac{\partial T}{\partial t} - \frac{\tau V}{\rho c w} = \alpha_{th} \frac{\partial^2 T}{\partial y^2}$$

$$\frac{\partial p}{\partial t} - \Lambda \frac{\partial T}{\partial t} = a_{hy} \frac{\partial^2 p}{\partial y^2}$$

where $\alpha_{th}$ is thermal diffusivity, $a_{hy}$ is hydraulic diffusivity, and $\Lambda$ is the thermal pressurization coefficient. This reduced system forms the basis for the 1D finite-difference simulations described below. Note we directly assumed measured on-fault pore pressure by Pp transducers as fluid pressure in shear zone.

Because the measured pore pressure evolution includes both near-fault diffusion and transducer-scale buffering effects, the inferred diffusivity is generally lower than theoretical host-rock values. Accordingly, both the shear zone thickness and diffusivity derived here should be regarded as effective parameters describing the bulk thermo-hydraulic response at the measurement scale.

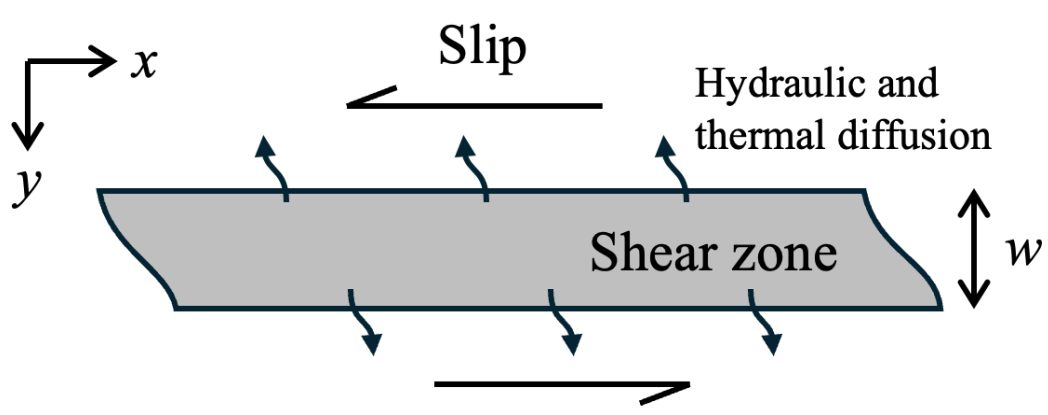

Figure S6. Finite-width shear zone model for coupled thermal and hydraulic diffusion during slip. Modified from Urata et al. (2015).

### S4.2 One-dimension coupled diffusion model with finite difference method

We solve these coupled TP equations using an explicit finite-difference method on a symmetric non-uniform grid that is refined near the fault (y=0) to resolve the strong thermal gradients. To balance accuracy near the shear zone and computational efficiency away from it, we generate a two–sided adaptive grid over [-L, +L]. Nodes are clustered near y=0 using an exponential stretching factor $\beta$, ensuring higher resolution where temperature and pressure gradients are largest. If $n_y$ is the total number of grid points (odd), the half-grid is constructed as:

$$y_{\mathrm{half}}(i) = L\,\frac{e^{\beta\, i/(N_{\mathrm{half}}-1)} - 1}{e^{\beta} - 1}, \qquad i = 0, \dots, N_{\mathrm{half}} - 1.$$

The full symmetric grid is $y = \left[-\mathrm{flip}\left(y_{\mathrm{half}}(2\!:\!end)\right),\, y_{\mathrm{half}}\right]$. This produces a smoothly varying spacing $\Delta y_i = y_{i+1} - y_i$, which is smallest near the fault and increases toward the boundaries, as shown in the grid-structure figure S7.

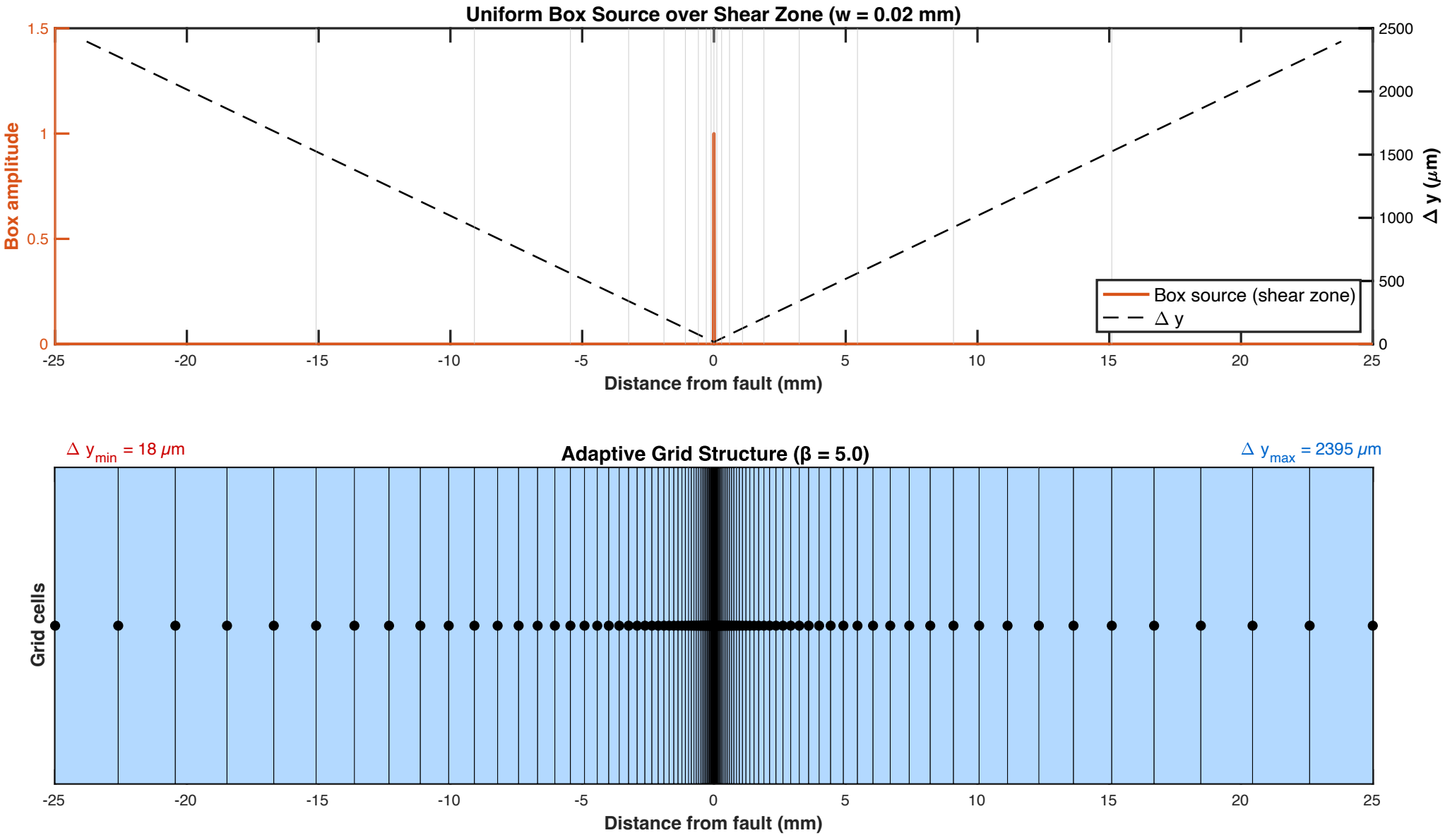

Figure S 7. Adaptive exponentially stretched grid used for the 1-D coupled thermal–hydraulic diffusion simulations.

Because the grid is non-uniform, standard central differences are not applicable. We adopt the second-order accurate finite-difference approximation for non-uniform spacing. The expression for temperature is

$$\left.\frac{\partial^2 T}{\partial y^2}\right|_i = \frac{2}{\Delta y_{i-1} + \Delta y_i}\left[\frac{T_{i+1} - T_i}{\Delta y_i} - \frac{T_i - T_{i-1}}{\Delta y_{i-1}}\right].$$

The same expression is used for pressure. This formulation preserves accuracy while allowing arbitrary grid stretching.

For the time stepping, an explicit forward-Euler scheme is used to update temperature and pressure. The form for temperature is

$$T_i^{n+1} = T_i^n + \Delta t\, \alpha_{th} \left.\frac{\partial^2 T}{\partial y^2}\right|_i^n + Q_i^n \Delta t,$$

where the shear heating source term is $Q = \frac{\tau(t)V(t)}{\rho_c w}$.

It is applied uniformly within the shear zone thickness:

$$Q_i = \begin{cases} Q, & |y| \leq \frac{w}{2} \\ 0, & |y| > \frac{w}{2} \end{cases}$$

The update of fluid pressure is, as a discrete form of equation xx.

$$p_i^{n+1} = p_i^n + \Delta t\, a_{hy} \left.\frac{\partial^2 p}{\partial y^2}\right|_i^n + \Delta t \left(\Lambda \frac{T_i^{n+1} - T_i^n}{\Delta t}\right).$$

Thus, the pressure is thermally driven by the instantaneous temperature increment.

The Neumann (zero-flux) boundary conditions are used at both ends:

$$\frac{\partial T}{\partial y}|_{\pm L} = 0, \qquad \frac{\partial p}{\partial y}|_{\pm L} = 0.$$

These are implemented as:

$$T_1 = T_2, \qquad T_{end} = T_{end-1},$$

$$p_1 = p_2, \qquad p_{end} = p_{end-1}.$$

This corresponds physically to a thermally and hydraulically insulated far field.

To ensure the calculation stability, the explicit scheme requires CFL<0.5. Since the space is defined in advance, the time step $\Delta t$ should satisfy $\Delta t_{\text{stable}} \leq 0.5 \min_i \frac{\Delta y_i^2}{a_{max}}$, where $a_{max} = \max(\alpha_{th}, a_{hy})$. In practice, the minimum $\Delta y$ occurs at the fault, so the time-step is automatically tied to the required heat-resolution scale. The algorithm selects $\Delta t = \min(\Delta t_{\text{exp}}, \Delta t_{\text{stable}})$ where $\Delta t_{\text{exp}}$ is the sampling time interval (0.01 s) in the experiments.

In total, the algorithm updates temperature first (since the heating term appears only in the T equation), then computes pressure using the updated temperature to guarantee $p^{n+1}$ depends on $T^{n+1}$, which correctly preserves causality of thermal expansion–induced pore pressure change. This also stabilizes the explicit scheme by avoiding temporal lag.

**Case 1 – Co-seismic fitting under near-undrained conditions**

In this case, we examine whether the finite-difference solver reproduces the analytical undrained solution derived in Section 4.2. To approximate undrained conditions, very small thermal and hydraulic diffusivities ($1 \times 10^{-10}$ m²/s) are assigned. The measured shear stress and slip velocity are directly prescribed as the shear-heating source term, ensuring that the numerical model is driven by the experimentally observed rupture dynamics.

For the first fast event of WGF10 (effective Pc = 60 MPa), the analytical undrained inversion yielded a shear zone thickness of w = 47 μm. Using this value in the numerical model, the characteristic diffusion time

$$t_d = \frac{w^2}{4a_{hy}}$$

is approximately 5.5 s, much larger than the co-seismic duration (~0.1 s), indicating that diffusion should be minimal during slip.

However, when diffusion is explicitly included, even with such small diffusivity, slight thermal leakage occurs during heating. As a result, the modelled pore pressure rise is marginally smaller than the purely undrained analytical prediction. To reproduce the observed co-seismic amplitude, the effective shear zone thickness must therefore be slightly reduced to w = 38 μm.

Amont the shear zone thickness, the spatially averaged simulated pore pressure,

$$p_{\text{sim}}(t) = \frac{\int_{-w/2}^{w/2} p(y,t)\ dy}{\int_{-w/2}^{w/2} dy},$$

matches the measured co-seismic pressure increase very closely (Figure S8). The small difference between the analytical (47 μm) and numerical (38 μm) values reflects the fact that the analytical solution represents a strict undrained upper bound, whereas the numerical model always permits limited diffusion during heating.

Importantly, even under these near-undrained conditions, pore pressure begins to decay immediately after the co-seismic peak, demonstrating that diffusion becomes significant during the post-seismic stage. This confirms that the undrained adiabatic assumption is appropriate only for the co-seismic interval and tends to overestimate the true shear zone thickness while underestimating peak temperature rise.

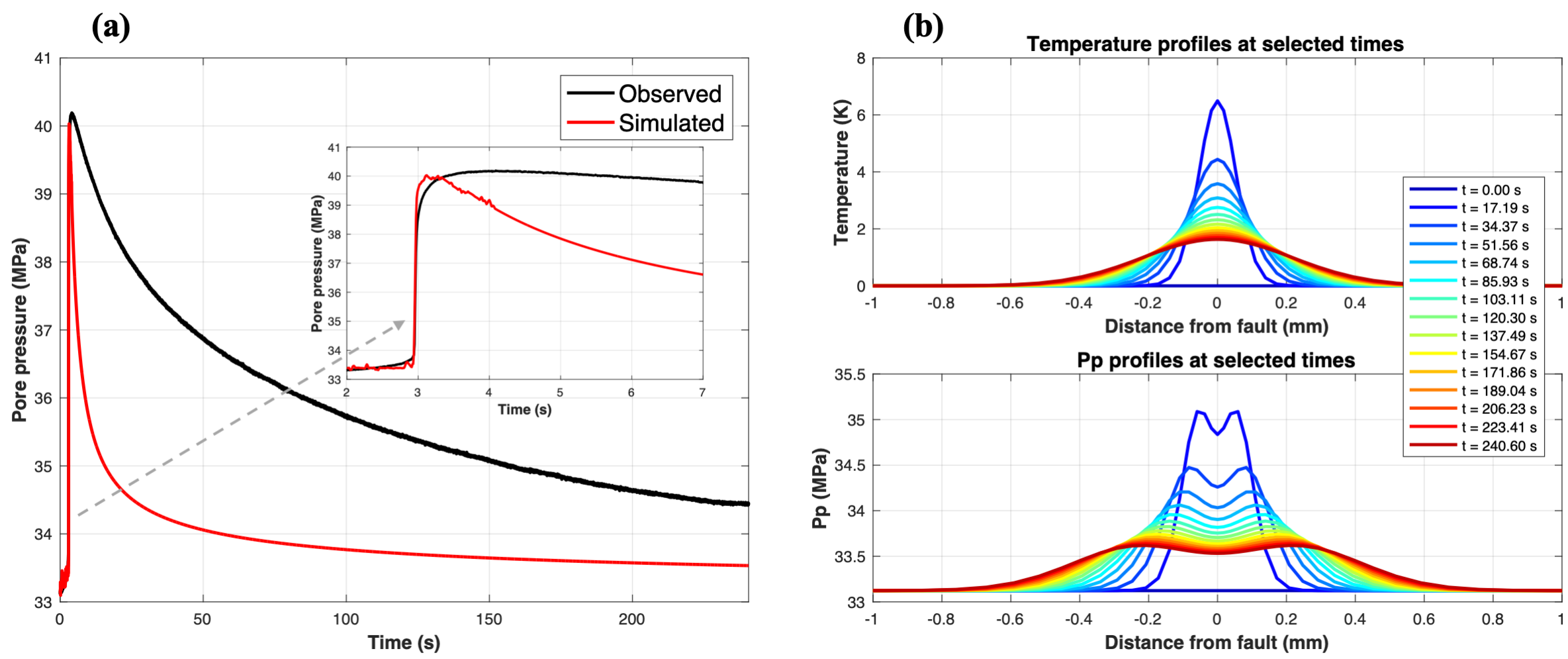

Figure S8. Co-seismic and post-seismic evolution of pore pressure and temperature simulated using the 1-D coupled thermal–hydraulic diffusion model under near-undrained conditions ($a_{hy}, a_{th} = 1 \times 10^{-10}$ m²/s, $w = 38$ μm). (a) Comparison between observed and simulated on-fault pore pressure. (b) Temperature and pore pressure profiles across the shear zone at selected times.

**Case 2 – Co-seismic fitting under realistic diffusion**

In this case, both thermal and hydraulic diffusivities are set to $1 \times 10^{-7}$ m²/s. The hydraulic diffusivity corresponds to the initial value of WGF10 (Table 1), and the thermal diffusivity is consistent with commonly adopted rock values. We then examine the co-seismic pore pressure evolution without imposing undrained conditions.

As shown in Figure S9, the model predicts a very sharp co-seismic pore pressure increase, followed immediately by rapid decay once shear heating ceases. Most of the thermally generated pressure is lost within a short time due to diffusion. In contrast, the measured pore pressure continues to increase gradually and decays much more slowly. This discrepancy indicates that the observed co-seismic Pp evolution is strongly influenced by measurement-scale buffering associated with the transducer volume and spatial averaging. The true shear zone response is likely more transient and more localized than recorded by the sensors. Importantly, this does not invalidate the undrained analytical fitting in the main text. The undrained framework provides a consistent and comparable way to track the evolution of effective shear zone thickness across events. However, because the measured Pp includes buffering effects, the inverted shear zone thickness under undrained assumptions should be regarded as an effective upper bound.

The rapid numerical response under realistic diffusivity suggests that the true co-seismic shear zone thickness is likely smaller than the undrained estimate. In contrast, the post-seismic stage, where pressure evolves over longer timescales, may better reflect diffusion-controlled equilibration within the shear zone. This motivates a third case focusing on post-seismic fitting independent of the co-seismic peak.

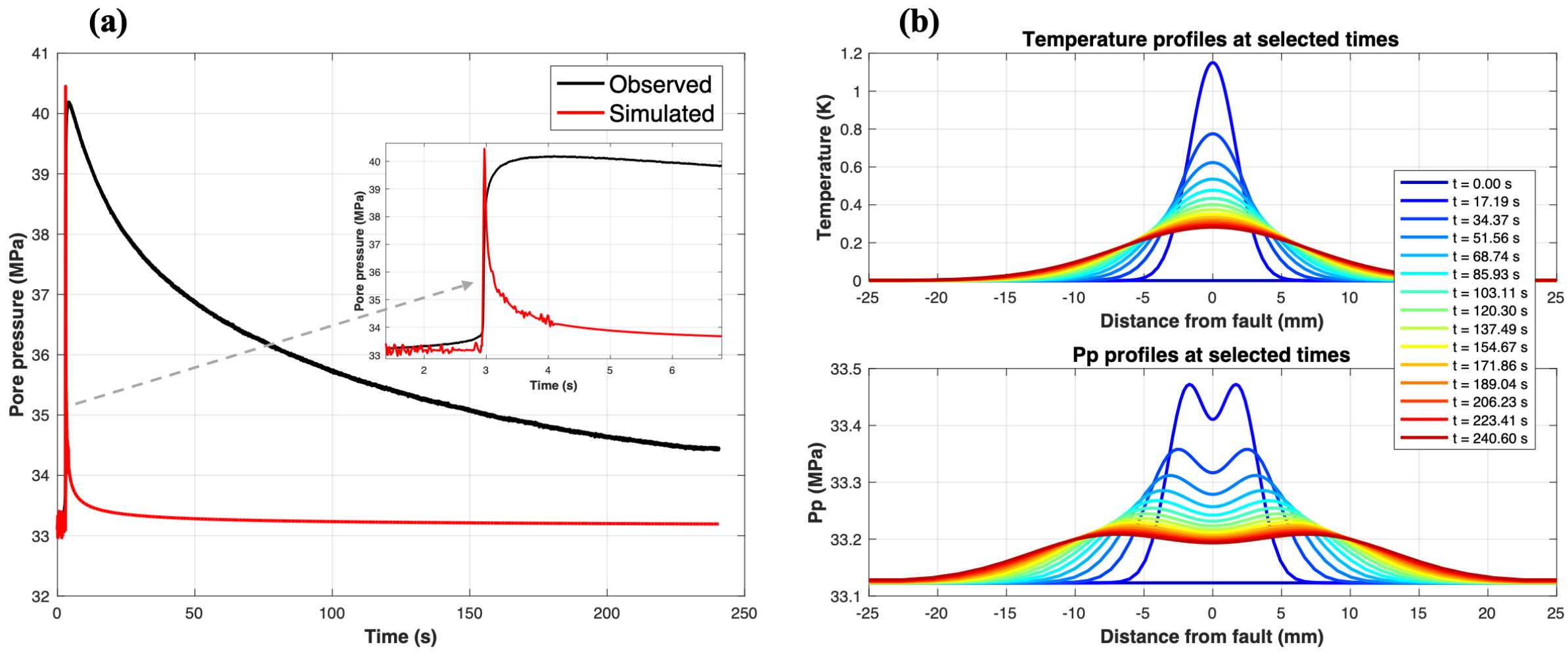

Figure S9. Co-seismic and post-seismic evolution of pore pressure and temperature simulated using the 1-D coupled thermal–hydraulic diffusion model under realistic diffusion ($a_{hy}, a_{th} = 1 \times 10^{-7}$ m²/s, $w = 2.2$ μm). (a) Comparison between observed and simulated on-fault pore pressure. (b) Temperature and pore pressure profiles across the shear zone at selected times.

**Case 3 – Post-seismic fitting under realistic diffusion**

After slip stops, pore pressure relaxes over timescales of seconds to minutes, primarily governed by thermal diffusivity, hydraulic diffusivity, and shear zone thickness. In this regime, diffusion dominates the pressure evolution. Here we fix the thermal diffusivity ($\alpha_{th} = 10^{-7}$ m$^2$/s) and jointly invert hydraulic diffusivity $\alpha_{hy}$ and shear zone thickness w by fitting only the post-seismic pore pressure decay. The co-seismic peak is intentionally excluded from fitting, as it is strongly influenced by rapid heating and measurement buffering, whereas the post-seismic stage more faithfully reflects diffusion-controlled relaxation.

To account for the finite sensing volume of the transducers, the simulated pore pressure is spatially averaged over ±0.5 mm around the fault plane,

$$p_{\text{sim}}(t) = \frac{\int_{|y|\leq 0.5 \text{ mm}} p(y,t)\ dy}{\int_{|y|\leq 0.5 \text{ mm}} dy}.$$

corresponding to the effective sampling region of the sensor with 1 mm diameter. Only times $t \geq t_{\text{peak}} + 10$ s are used for inversion to ensure that primarily diffusive relaxation is captured.

Because diffusion timescale scales as $t_d \sim w^2/4a_{hy}$, the inversion reveals a pronounced trade-off between $w$ and $a_{hy}$: multiple parameter pairs yield nearly identical post-seismic decay curves (Figure S10). The misfit landscape exhibits a valley-shaped structure, reflecting that the data mainly constrain the ratio $w^2/4a_{hy}$ rather than each parameter independently. Consequently, increases in shear zone thickness can be compensated by decreases in hydraulic diffusivity, and vice versa.

Importantly, the inversion for successive events accounts for the residual pore pressure distribution remaining at the end of each previous event. For event #1, the best-fitting solution leaves a non-uniform residual pressure profile (Figure S10b). This residual profile is then used as the initial condition for event #2, and the procedure is repeated for events #3–6 (Figure S11). If the initial profile is artificially reset to a uniform state before each event, the quality of post-seismic fitting progressively deteriorates. This demonstrates that pressure memory between events plays a significant role in controlling subsequent diffusion behavior.

Fitting a sequence of TP-type events (events #1–6, including fast events #1–3 and slower events #4–6) reveals a systematic shift of the trade-off valley toward larger shear zone thickness with event number, indicating gradual structural thickening. In contrast, hydraulic diffusivity remains within the constrained range of $10^{-8}{\sim}10^{-7}$ m²/s and does not exhibit significant evolution. Compared with the undrained co-seismic inversion, the post-seismic fitting yields smaller absolute shear zone thickness values, reflecting that the latter is controlled by diffusion over sensor-scale distances rather than strictly by localized heating. Importantly, despite differences in absolute magnitude, the inferred evolutionary trend of shear zone thickening is consistent with the undrained analysis in Section 4.2, demonstrating the robustness of the structural interpretation across different modeling assumptions.

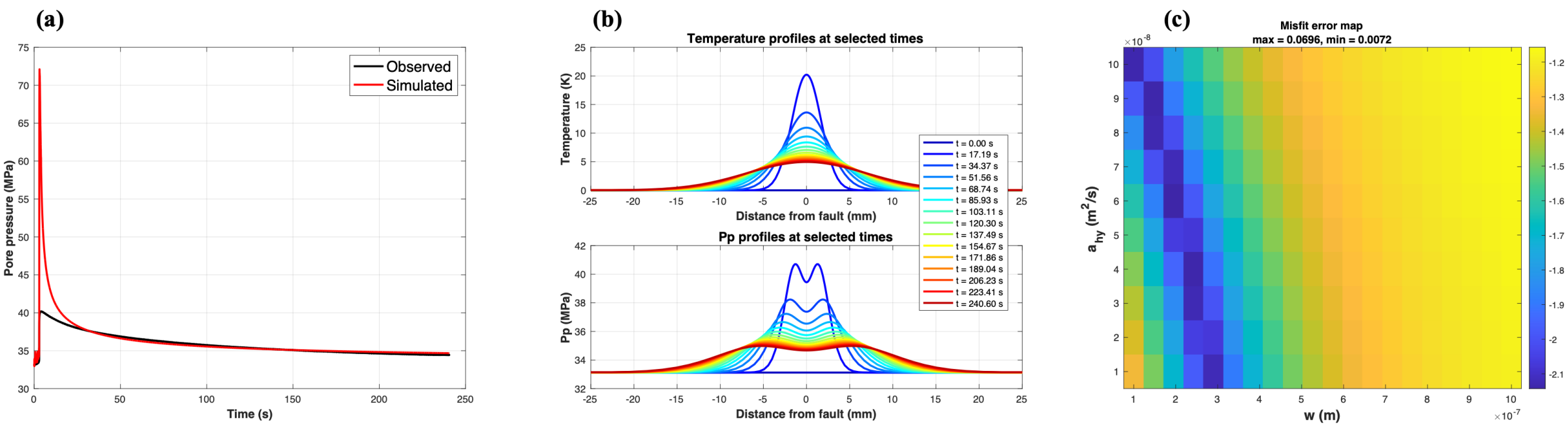

Figure S10. Post-seismic fitting of event #1 of WGF10 under effective Pc = 60 MPa. (a) Comparison between observed and simulated pore pressure decay; the red curve represents the spatially averaged pressure over |y|<0.5 mm corresponding to the transducer sensing volume. (b) Temperature and pore pressure profiles across the shear zone at selected times. (c) Trade-off relationship between hydraulic diffusivity and shear zone thickness based on L2 misfit.

(a) Event #2

(b) Event #3

(c) Event #4

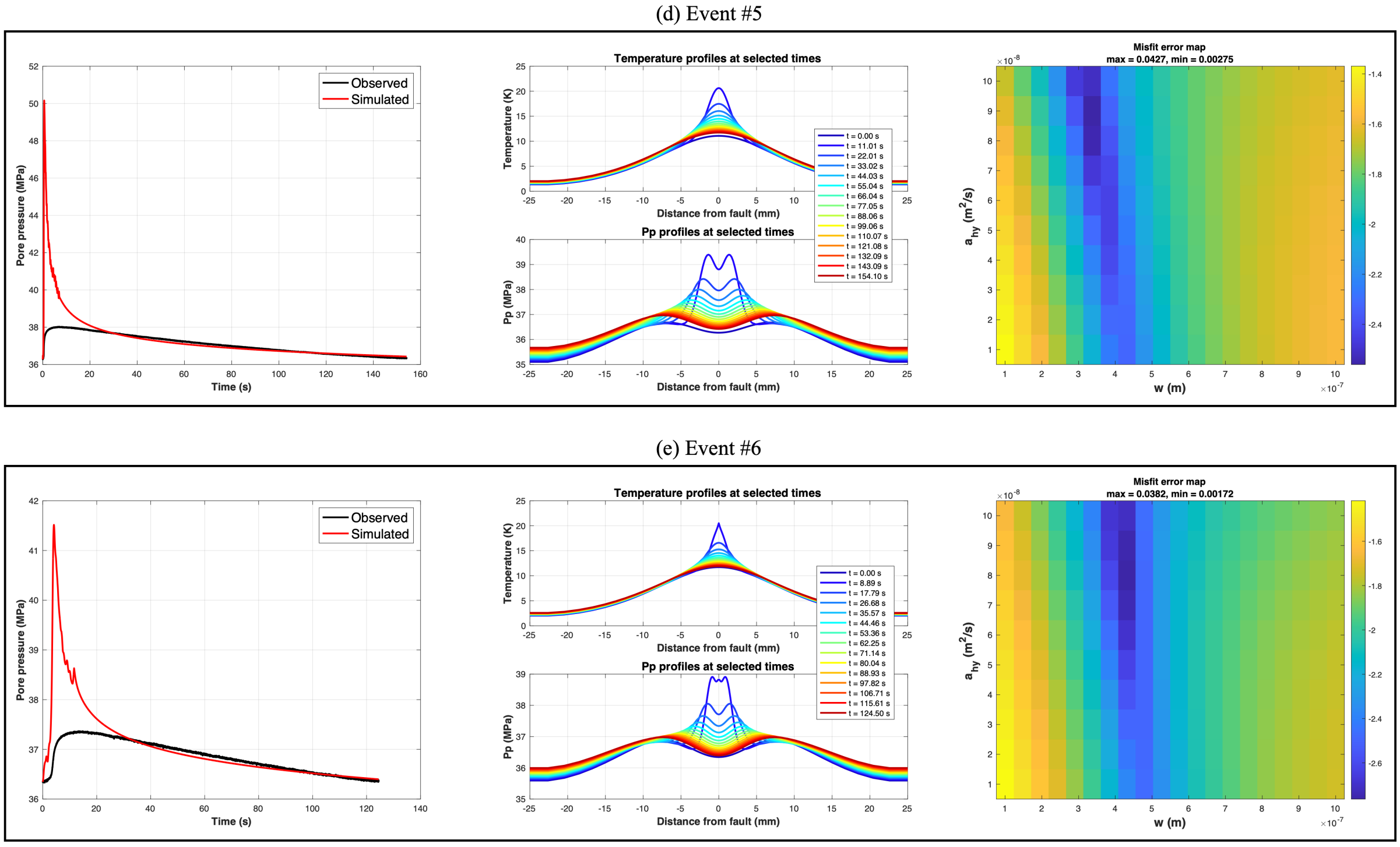

Figure S11. Post-seismic fitting results for events #2–6 of WGF10 under effective Pc = 60 MPa, following the fitting of event #1 in Figure S10. For each event: (left) comparison between observed and spatially averaged simulated pore pressure decay; (middle) temperature and pore pressure profiles at selected times; (right) L2 misfit maps showing the trade-off between hydraulic diffusivity and shear zone thickness. The progressive shift of the trade-off valley toward larger w indicates gradual shear zone thickening across the event sequence. Residual pressure profiles from the preceding event are used as initial conditions for each simulation.

### S4.3 Summary: Upper and lower bounds of shear zone pore pressure evolution

To synthesize the numerical analyses, we combine the results from Case 2 (co-seismic fitting with realistic diffusion) and Case 3 (post-seismic fitting with spatial averaging) to constrain the plausible range of shear zone pore pressure evolution. Event #1 of WGF10 (a representative fast TP-type event) is used as an illustrative example (Figure S12). In Case 2, using realistic thermal and hydraulic diffusivities, the modeled shear zone response exhibits a sharp and transient pore pressure increase followed by rapid decay once shear heating ceases. Because diffusion efficiently removes pressure from the localized shear zone, this solution provides a lower bound for the persistence of shear zone pressurization under finite diffusion conditions. For estimating shear zone thickness, the undrained adiabatic analysis in Section 4.2 yields the largest estimate of w, while Case 1 reproduces the near-undrained pore pressure evolution using the finite-difference solver. In Case 3, fitting only the post-seismic decay while accounting for sensor-scale spatial averaging yields a slower pressure relaxation. The simulated pressure is averaged over ±0.5 mm to mimic the transducer response, producing a response that reflects the effective pore pressure evolution at the sensor scale. When parameters derived from this post-seismic fitting are used to reconstruct the co-seismic stage, the predicted peak pressure is larger than the measured value and can be regarded as an upper bound of plausible shear zone pressurization.

The measured co-seismic pressure rise is likely underestimated due to spatial averaging and buffering by the transducers. The true shear zone Pp evolution therefore likely lies between the diffusive shear zone response (Case 2) and the reconstructed response constrained by post-seismic fitting (Case 3). The exact trajectory depends on the coupled evolution of shear zone thickness and hydraulic diffusivity (assuming constant thermal diffusivity) and cannot be uniquely determined with the present measurements.

Importantly, although the absolute amplitudes differ among modeling approaches, all scenarios consistently indicate structural evolution of the shear zone. The bounding analysis therefore refines the magnitude of inferred parameters while preserving the robustness of the evolutionary trend identified in the main text.

The same buffering effect also applies to DS-type events, implying that the measured co-seismic pressure drop is likewise a lower-bound estimate of the true pressure decrease within the shear zone.

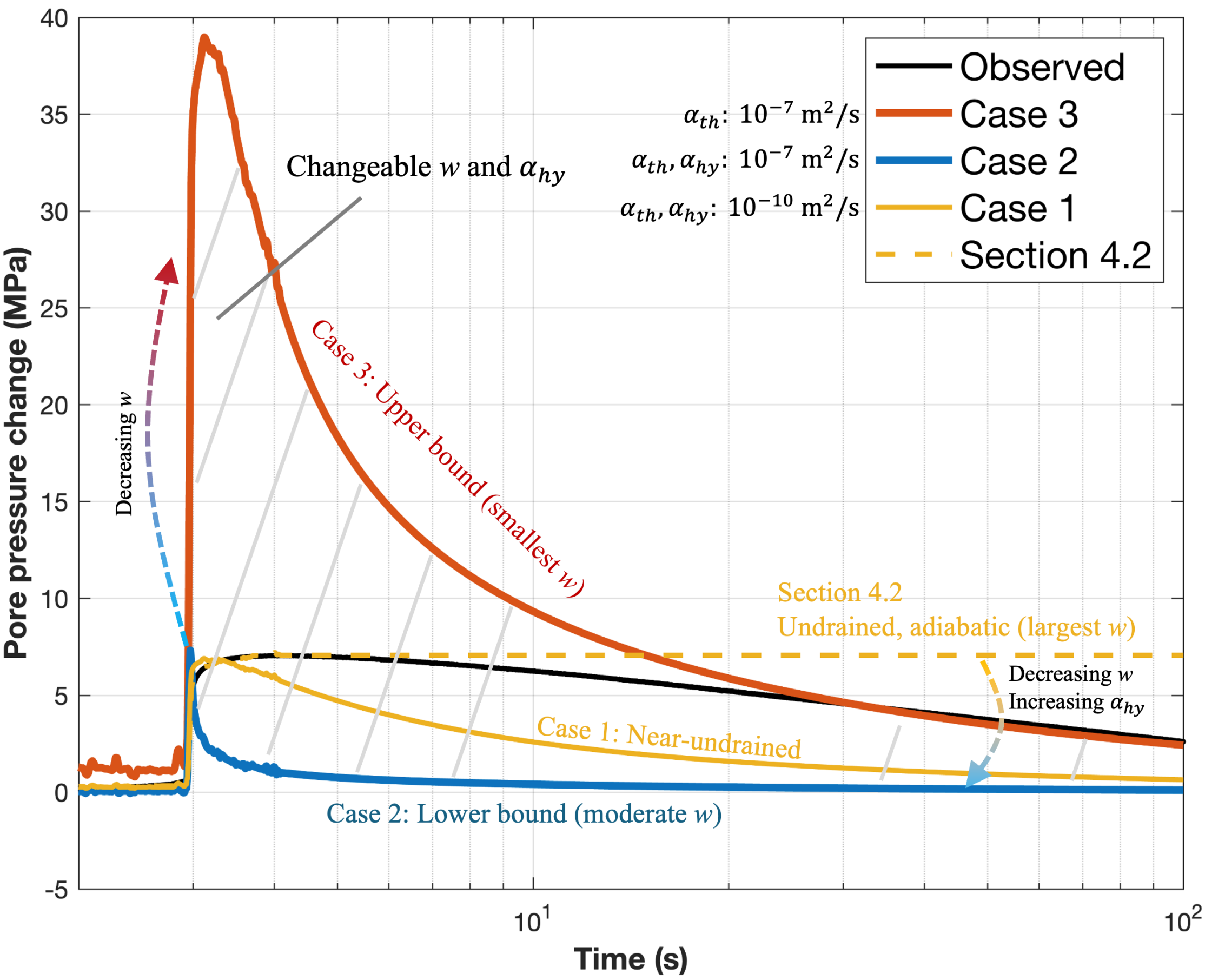

Figure S12. Bounding analysis of on-fault pore pressure evolution for event #1 of WGF10. Measured pore pressure change (black) is compared with numerical solutions from undrained–adiabatic conditions (Section 4.2), near-undrained conditions (Case 1), and realistic diffusion (Cases 2–3). The undrained–adiabatic solution provides the largest shear zone thickness, while Case 2 (co-seismic fitting) and Case 3 (post-seismic fitting with spatial averaging) define lower and upper bounds of plausible shear zone Pp evolution.

## S5. Supplementary figures

Figure S13–S16 show the original time series of shear stress, slip, and pore pressure for all four experiments. These data form the basis of the analyses presented in Sections 3–5.

Figure S17 presents the evolution of the apparent effective friction coefficient during the experiments.

Figure S18 shows the laser-vibrometer velocity waveform.

Figure S19 shows off-fault pore pressure responses in experiment WGF10 as a representative example.

Figure S20 shows the relationship between shear stress drop and slip for all events, used to estimate the effective loading stiffness.

Figure S21 compares stick-slip event parameters between WGF7 and WGF8, highlighting the role of hydraulic diffusivity in controlling rupture style.

Figure S22 shows the method used to determine weakening stiffness.

Figure S23 compares fracture energy estimated using two methods.

Figure S24 shows the complete dataset from the pore pressure transducers, indicating heterogeneity in on-fault Pp.

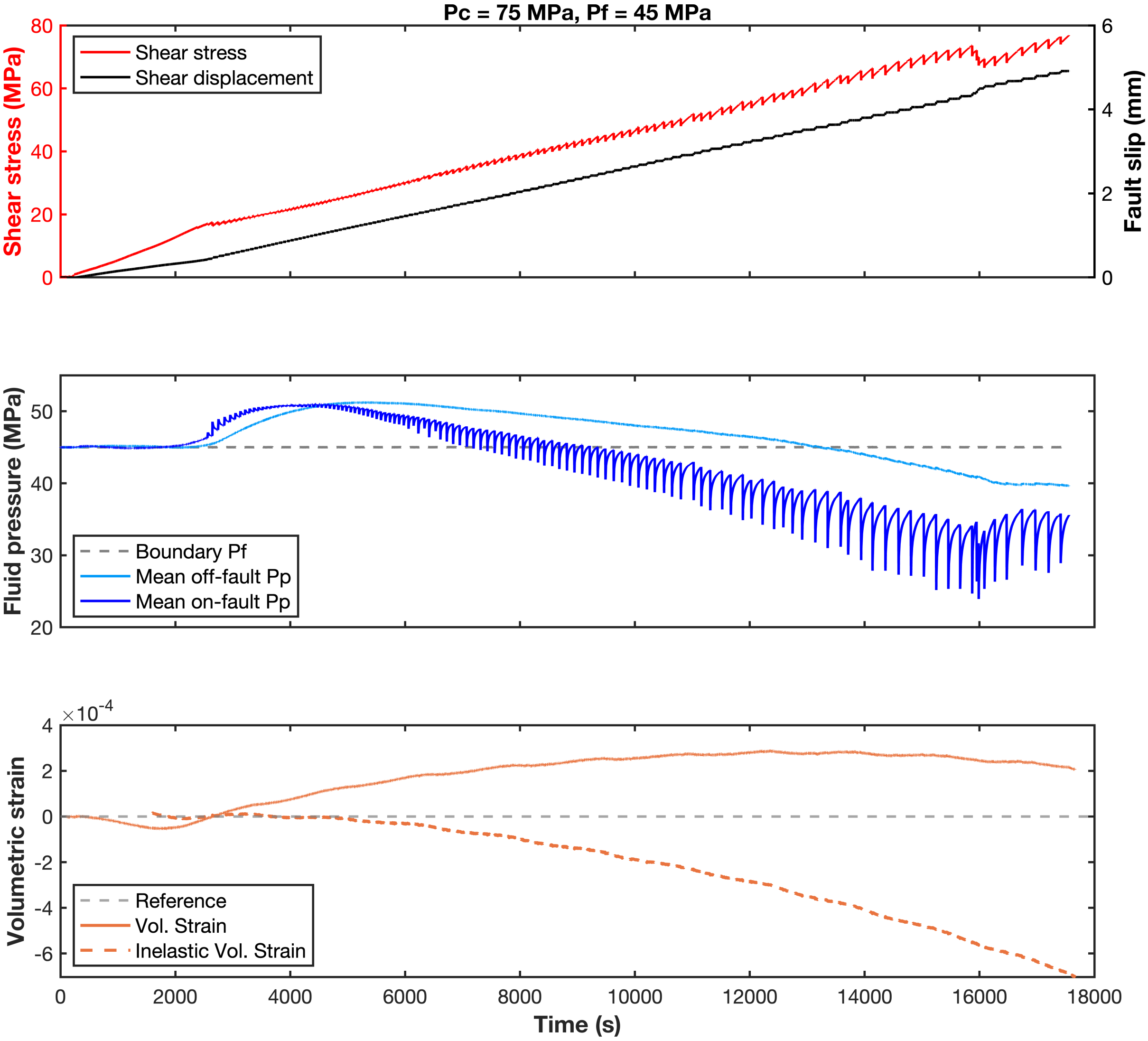

Figure S13. Evolution of shear stress, cumulative fault slip, mean on- and off-fault pore pressure (Pp), and volumetric strain during experiment WGF9 conducted at $P_c - P_f = 30$ MPa. Mean on-fault Pp is averaged from transducers No. 1–4 and mean off-fault Pp from transducers No. 5–6. Volumetric strain is calculated from fluid pump volume change normalized by the sample volume, and the inelastic volumetric strain is obtained by removing the linear elastic trend.

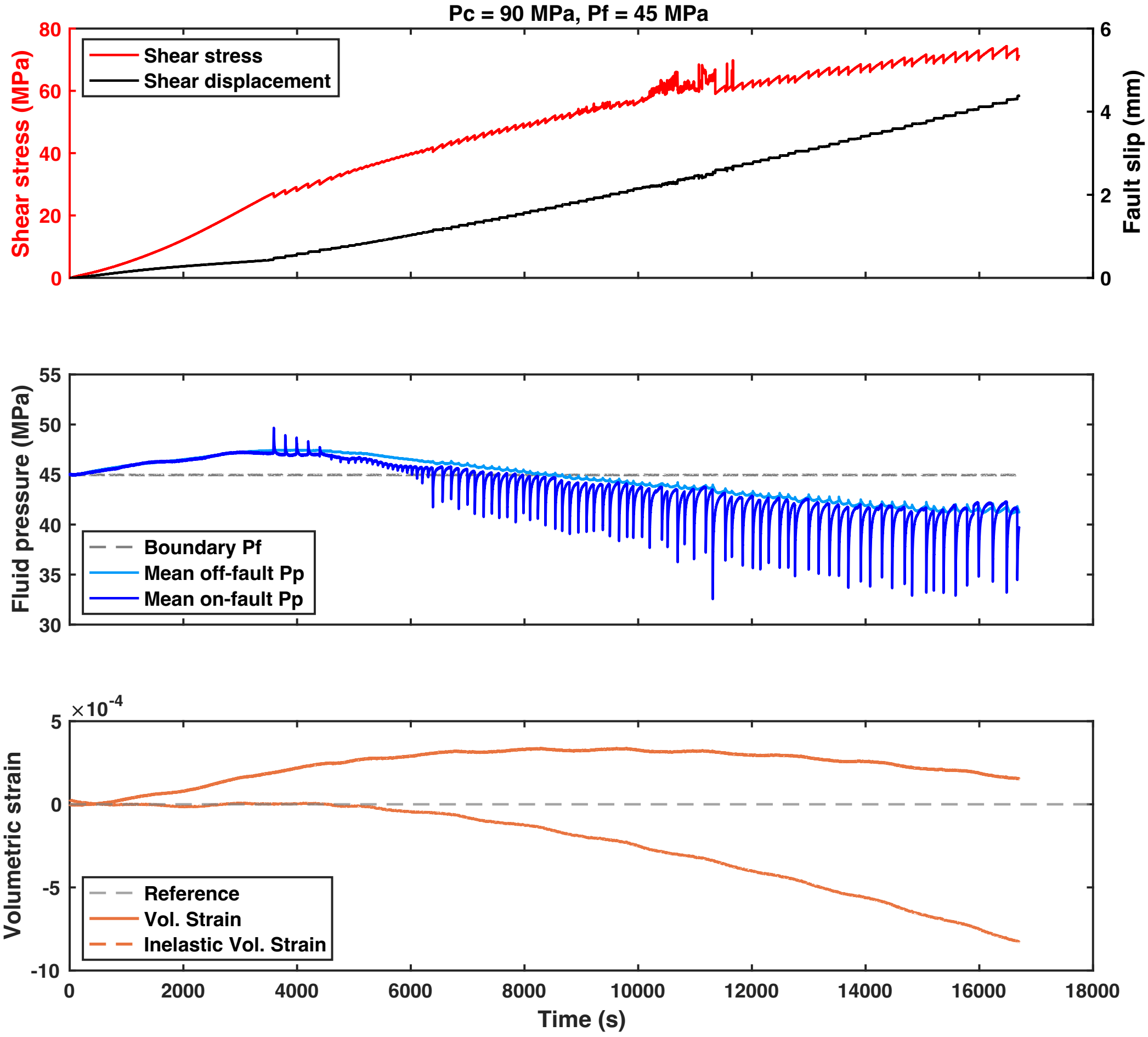

Figure S14. Evolution of shear stress, cumulative fault slip, mean on- and off-fault pore pressure (Pp), and volumetric strain during experiment WGF7 conducted at $P_c - P_f = 45$ MPa. Plotting conventions and data processing are the same as in Figure S13.

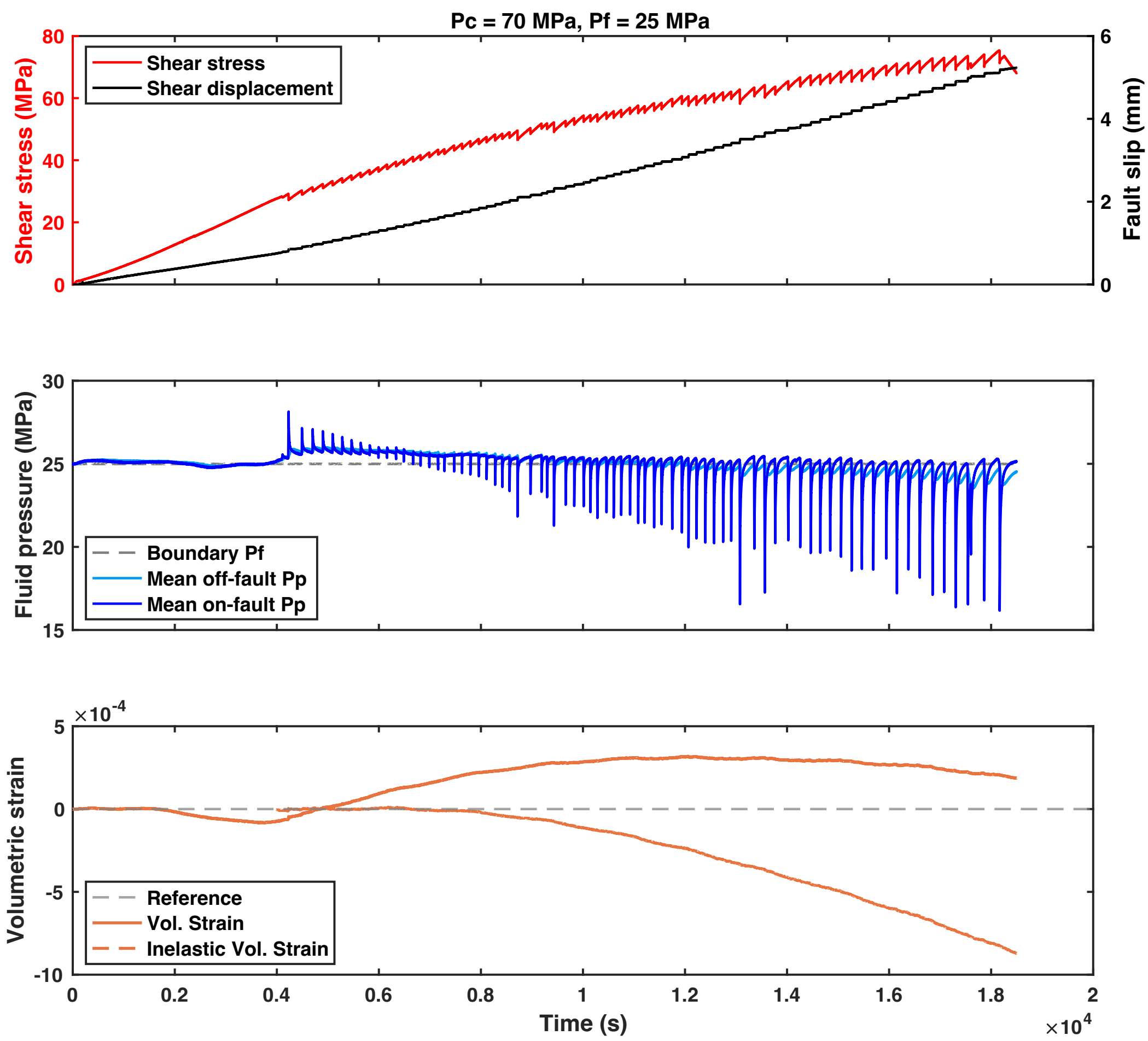

Figure S15. Evolution of shear stress, cumulative fault slip, mean on- and off-fault pore pressure (Pp), and volumetric strain during experiment WGF8 (with higher permeability) conducted at $P_c - P_f = 45$ MPa. Plotting conventions and data processing are the same as in Figure S13.

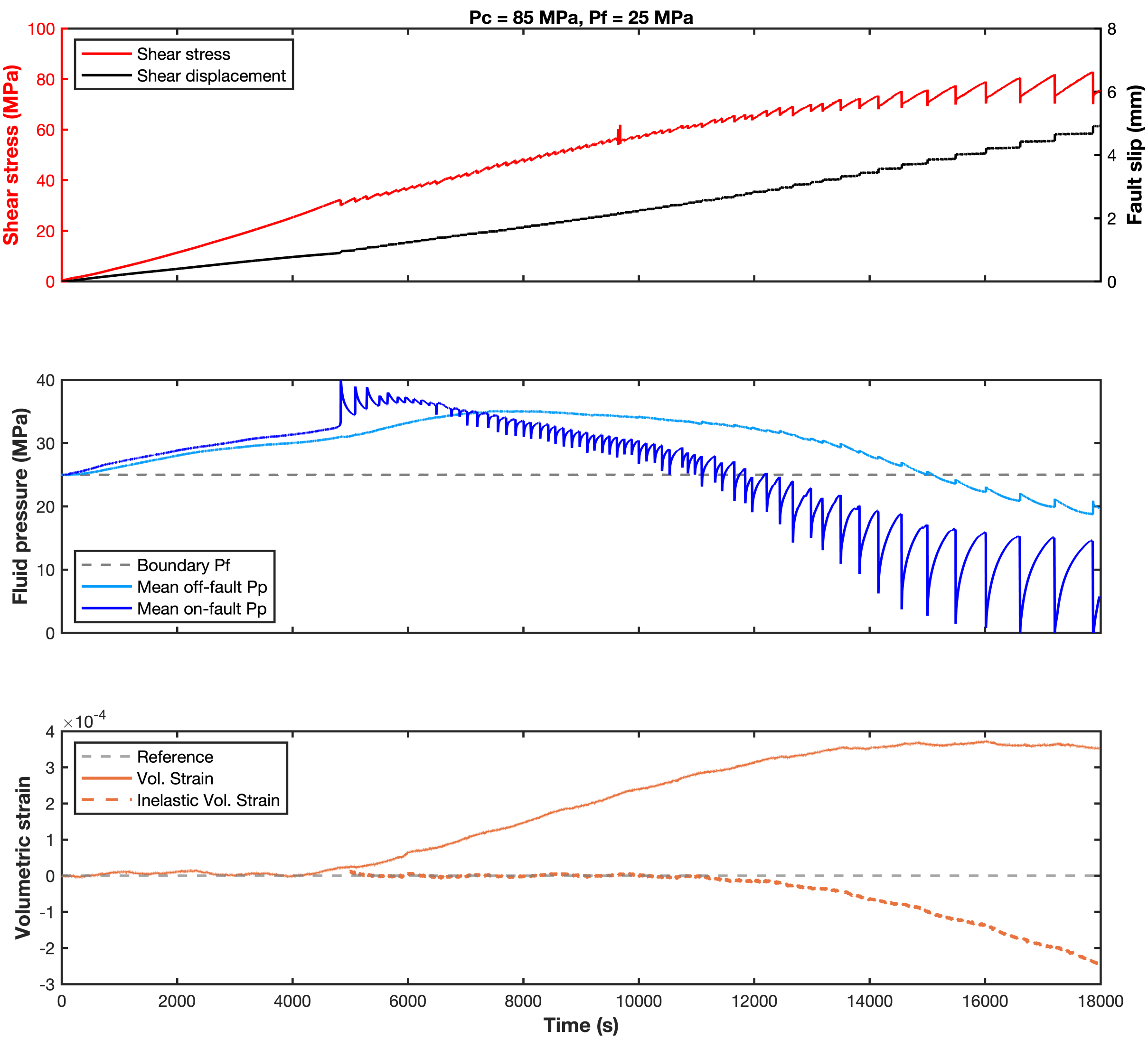

Figure S16. Evolution of shear stress, cumulative fault slip, mean on- and off-fault pore pressure (Pp), and volumetric strain during experiment WGF10 conducted at $P_c - P_f = 60$ MPa. Plotting conventions and data processing are the same as in Figure S13.

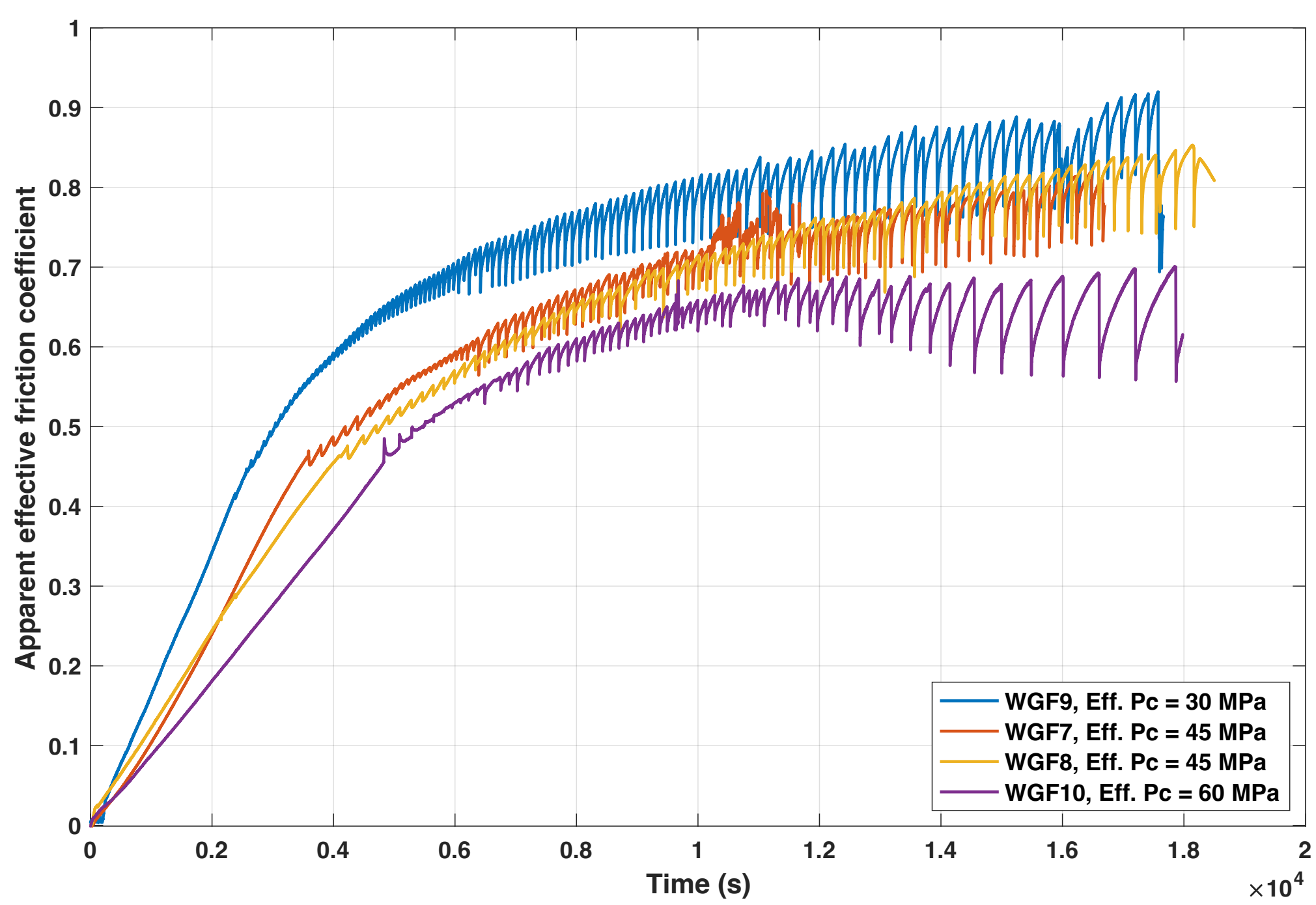

Figure S17. Temporal evolution of apparent effective friction coefficient for all experiments under different effective confining pressures. The effective friction coefficient is calculated as $\mu = \tau/(\sigma_n - P_{on-fault})$, where $\tau$ is shear stress, $\sigma_n$ is normal stress resolved on the fault plane, and $P_{on-fault}$ is the mean on-fault pore pressure. Under higher effective confining pressure conditions, the long-term effective friction level progressively decreases, consistent with Passelègue et al. (2019). In experiment WGF10, the apparent increase in friction during early stick–slip events results from rapid co-seismic Pp rise, which reduces effective normal stress more strongly than the accompanying shear stress drop.

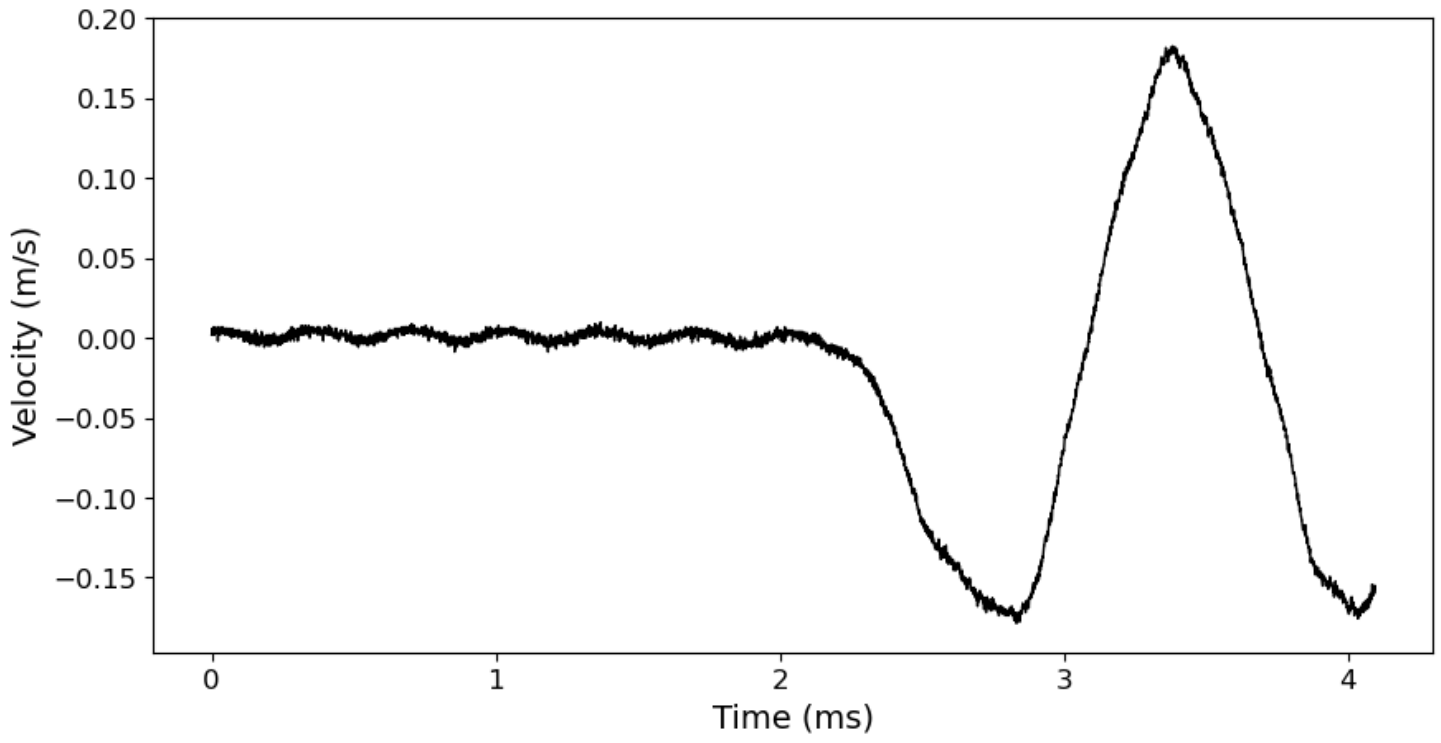

Figure S18. Representative laser-vibrometer velocity waveform of a fast stick-slip event, as a constraint on slip velocity of fast events. Laser signals were obtained only for the relatively fast events in our data set, which generally have LVDT-derived peak slip velocities greater than 0.5 mm/s. The waveform illustrates peak dynamic slip velocities reaching the order of 0.1 m/s.

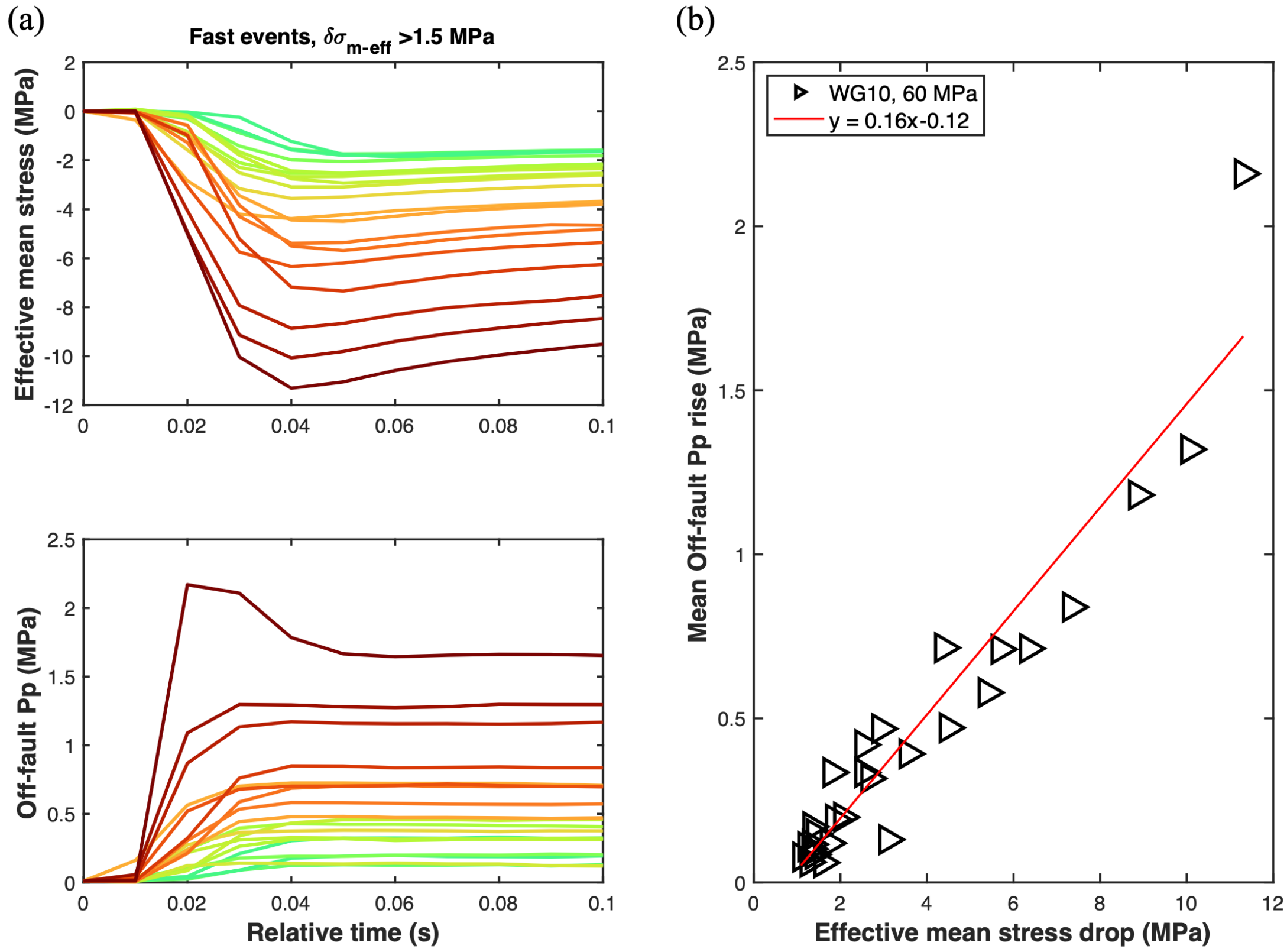

Figure S19. (a) Co-seismic evolution of effective mean stress and off-fault pore pressure (*Pp*) for events with larger effective mean stress drop (>1.5 MPa) in experiment WGF10, as a representative example. Both y-axes are offset for clarity. Resolvable co-seismic off-fault Pp increases are observed. The off-fault Pp drop reaches up to ~2.2 MPa for the largest events. (b) Peak off-fault *Pp* rise as a function of total effective mean stress drop (with >1 MPa) for fast events in WGF10, showing a linear growth relationship. Off-fault Pp variations are substantially smaller than the corresponding on-fault signals (see main text). Those events with show smaller effective mean stress drop (<~1.5 MPa for WGF10) mainly consists of slow events and have no resolvable off-fault Pp change above the background noise level (<0.05 MPa).

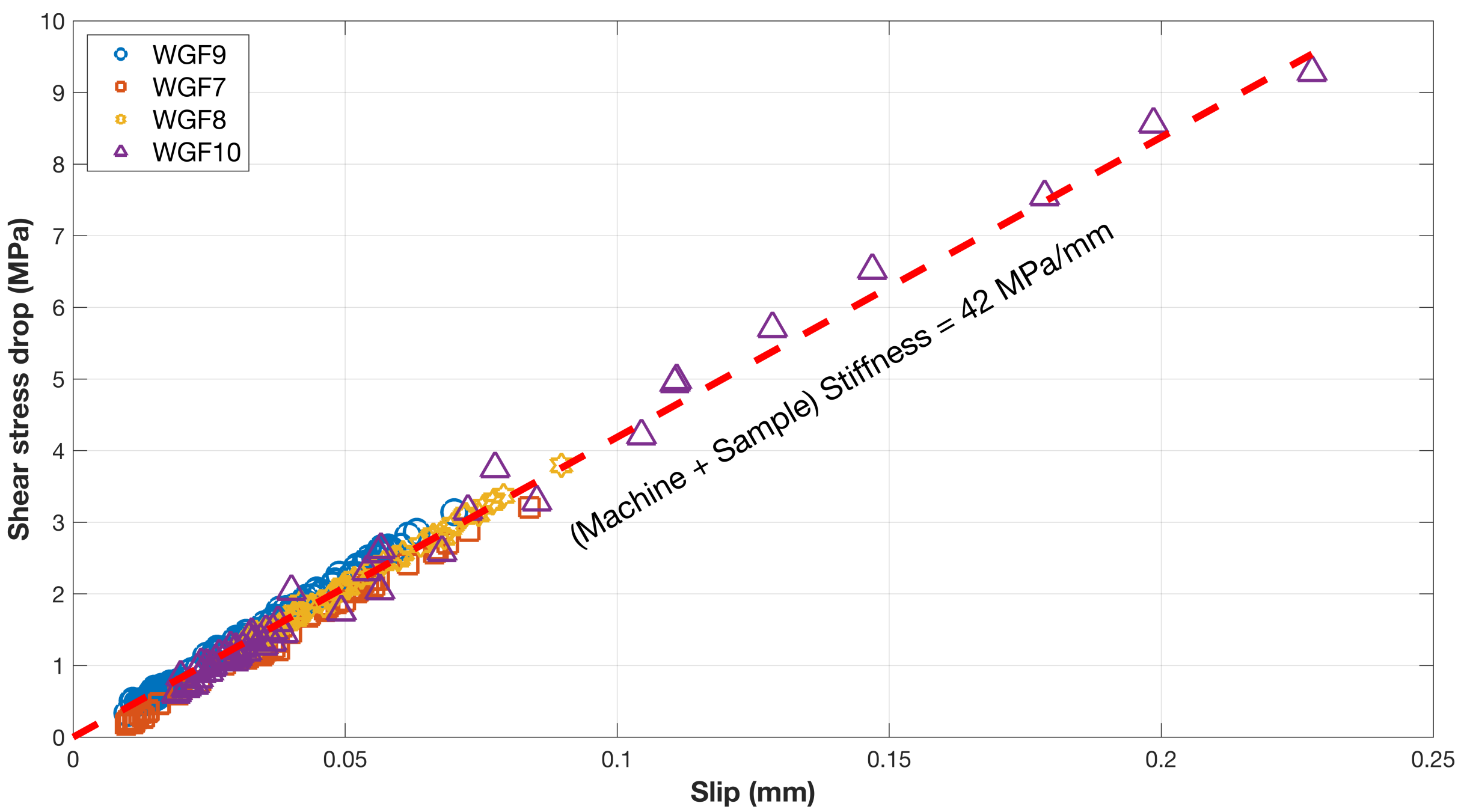

Figure S20. Static stress drop as a function of fault slip for all events of the four triaxial experiments (Pc-Pf = 30, 45 and 60 MPa). The dataset provides the basis of estimating the effective loading stiffness *K* (~42 MPa/mm) of the machine–sample system. The stiffness is estimated by linear regression constrained to pass through the origin, consistent with the physical expectation that zero slip corresponds to zero stress change, following Aubry et al. (2018).

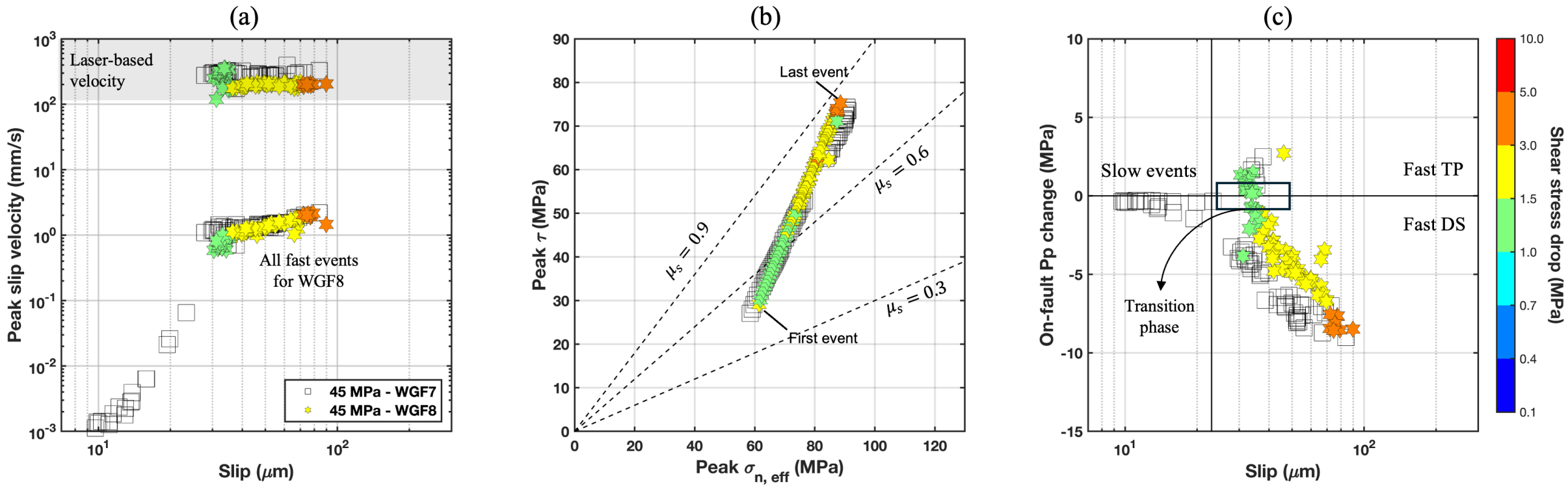

Figure S21. Stress state and co-seismic rupture characteristics of stick-slip events for two experiments conducted at effective confining pressures of 45 MPa on samples with different initial permeability (WGF7: $8 \times 10^{-21}$ m$^2$; WGF8: $1 \times 10^{-19}$ m$^2$). Symbol color denotes shear stress drop magnitude using the same color scale as Figure 8. WGF8 is color-coded, while WGF7 is shown as open symbols for reference (see Figure 8 for the colored version). (a) Peak slip velocity versus total slip. Slip velocities exceeding 100 mm s$^{-1}$ are derived from laser vibrometer measurements and correspond to fast events only. (b) Peak shear stress versus peak effective normal stress immediately prior to stress drop, illustrating the stress state at rupture onset. (c) Maximum co-seismic on-fault Pp change during rupture. A transition phase box is roughly shown in panel c for WGF8 based on on-fault Pp co-seismic response even though no slow events exist in this experiment.

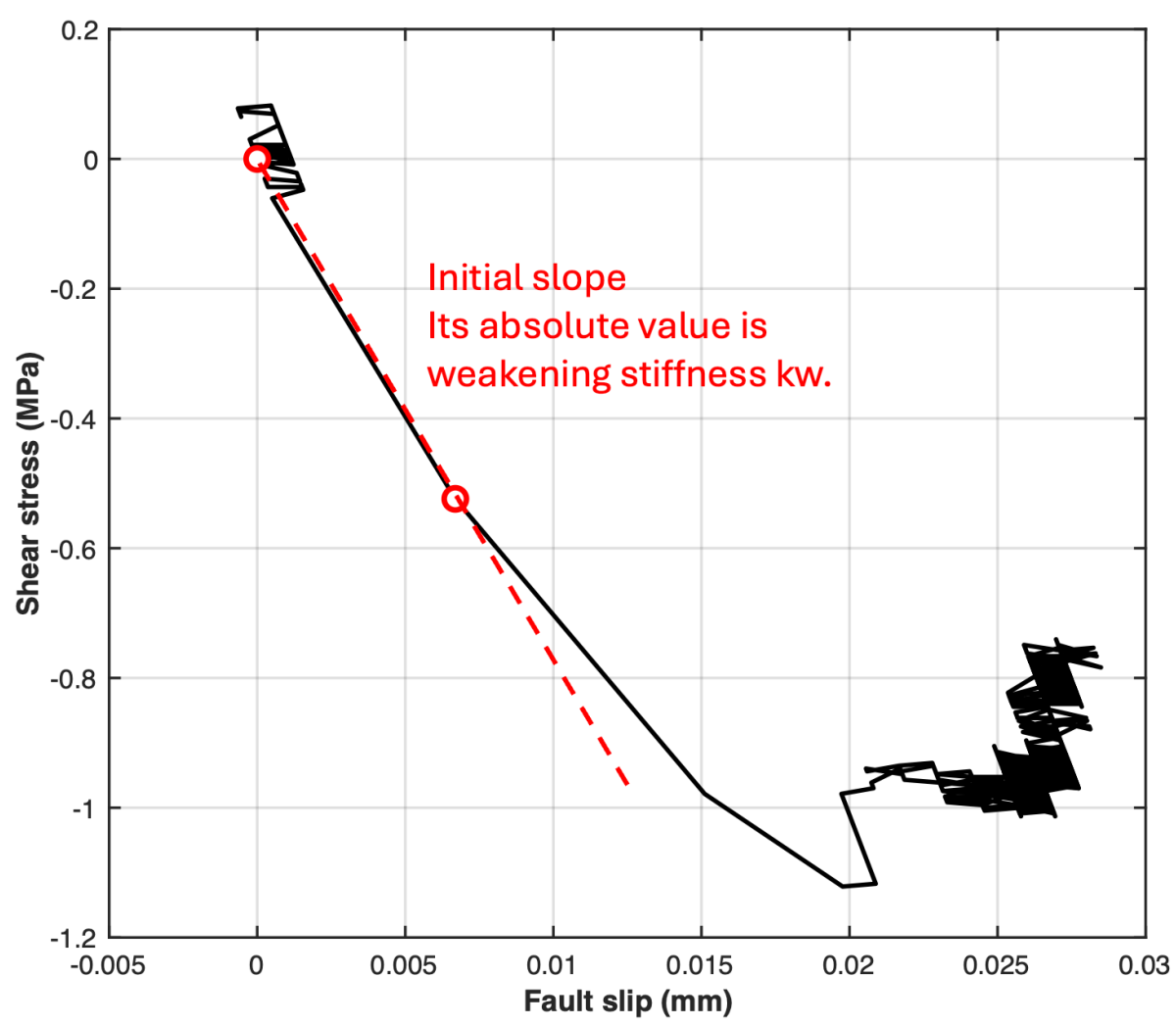

Figure S22. Illustration of weakening stiffness (kw) estimation from the initial slope of the shear stress–fault slip curve.

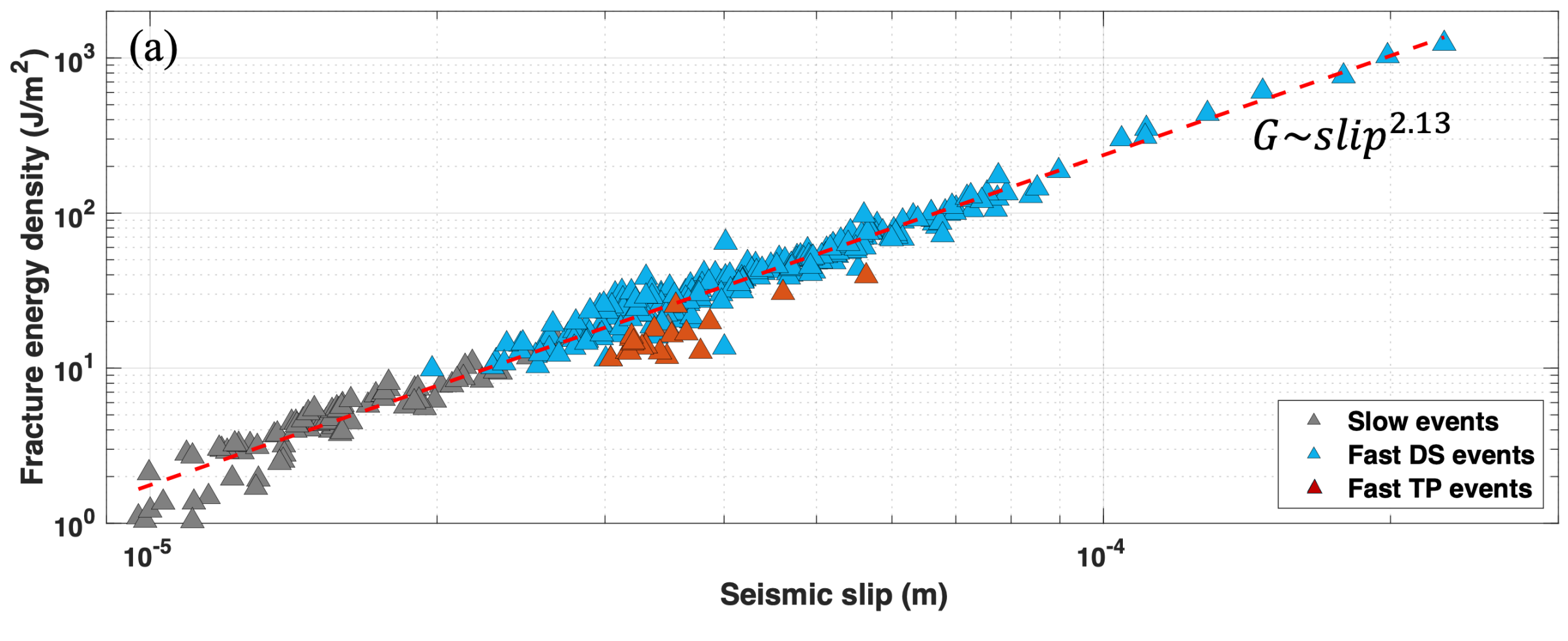

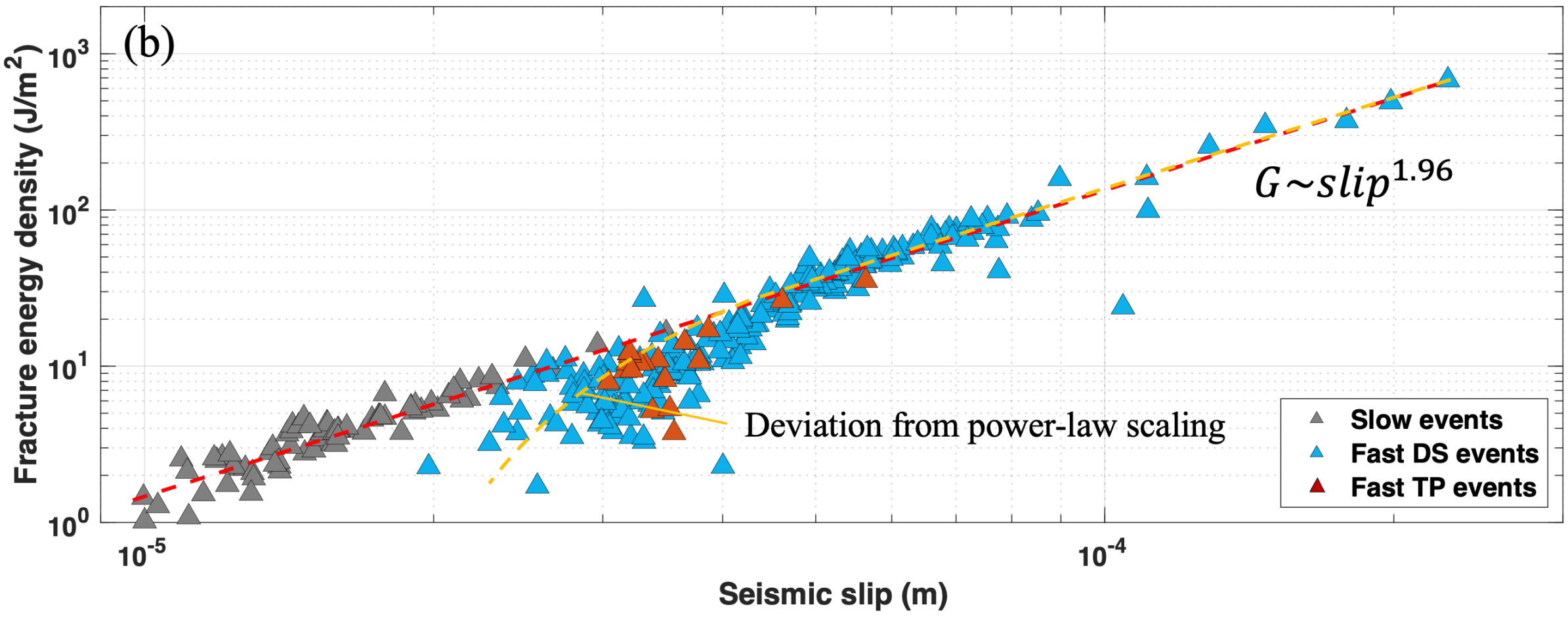

Figure S23. Comparison of fracture energy density as a function of seismic slip estimated using two approaches: (a) the linear slip-weakening approximation (same dataset as Figure 14 in the main text) and (b) direct integration ($G = \int_0^{\delta_*} [\tau(\delta) - \tau_r(\delta_*)] d\delta$, e.g., Wong, 1982; Ohnaka, 2003; Wang & Dresen, 2025) of the measured shear

stress–slip curve during dynamic weakening Both approaches yield similar power-law scaling, with exponents of 2.13 and 1.96 for the linear and integral methods, respectively. Integral estimates are generally lower, with the largest deviations occurring for fast events near a slip of ~2.5 × $10^{-5}$ m (25 μm).

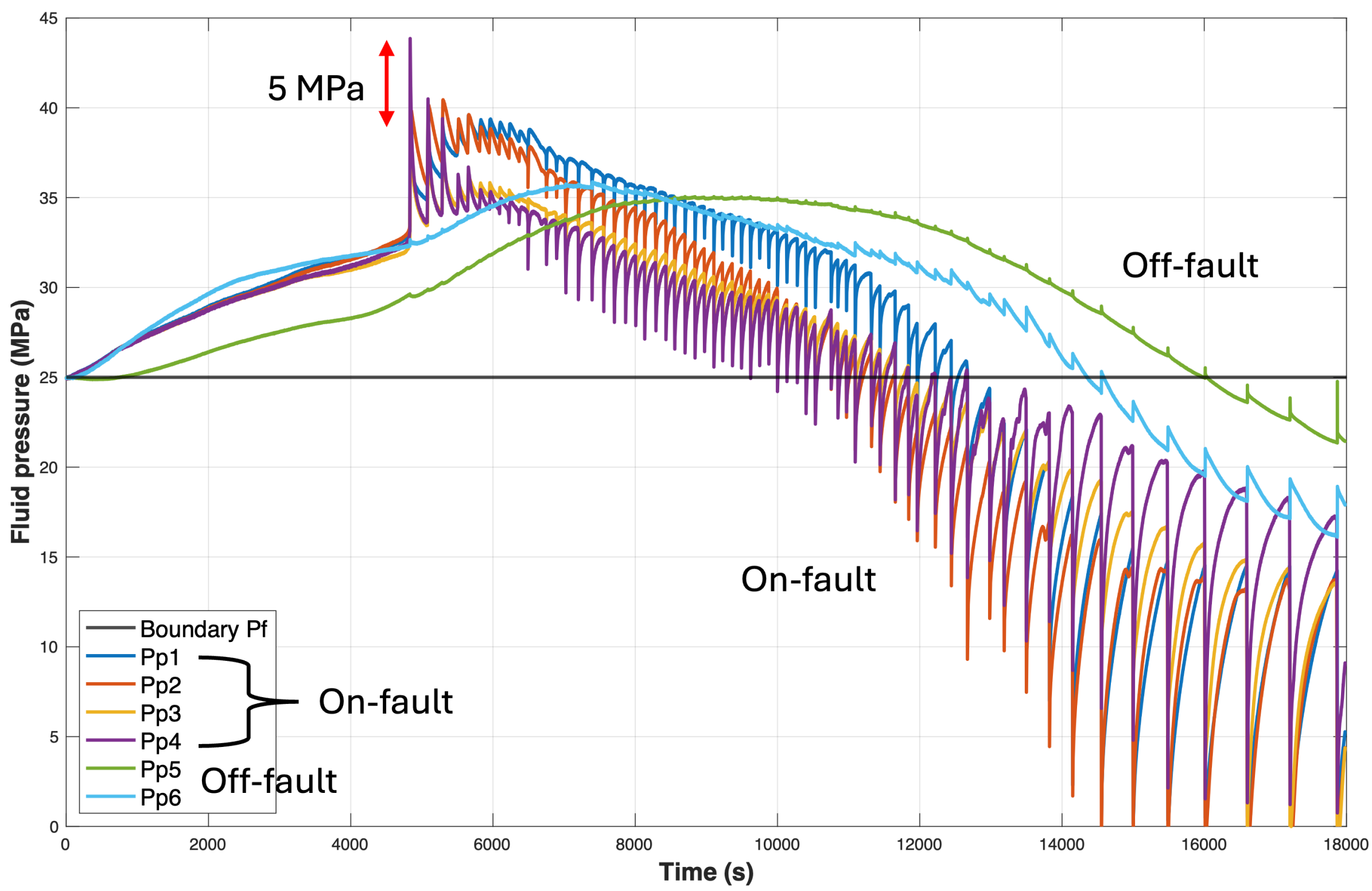

Figure S24. Complete pore pressure record for WGF10 at $P_c - P_f = 60$ MPa, showing heterogeneity in on-fault pore pressure. The figure shows the full-time series of all six pore pressure transducers (four on-fault and two off-fault) together with the boundary fluid pressure. The data illustrate the emergence of spatial heterogeneity in on-fault pore pressure following the first rupture event.

## References From the Supporting Information